\documentclass[useAMS,usenatbib]{mn2e}

\usepackage{amssymb,amsmath}
\usepackage{graphics}
\usepackage{subfigure}
\usepackage[dvips]{graphicx}
\usepackage{textcomp}
\usepackage{pdflscape}
\usepackage{epstopdf}
\usepackage{float}
\usepackage{xcolor}
\usepackage[percent]{overpic}
\usepackage{pifont}

\newcommand{\htwo}{\textsc{H}_2}
\newcommand{\co}{\textsc{CO}}

\newcommand{\h}{\textsc{H}}
\newcommand{\hp}{\textsc{H}^+}
\newcommand{\cp}{\textsc{C}^+}

\newcommand{\kms} {\,{\rm km\,s}^{-1}}
\newcommand{\cm} {\,{\rm cm}^{-3}}
\newcommand{\pc} {\,{\rm pc}}
\newcommand{\kpc} {\,{\rm kpc}}

\newcommand{\mo}{\,{\rm M}_\odot}
\newcommand{\yr}{\,{\rm yr}}
\newcommand{\Myr}{\,{\rm Myr}}

\newcommand{\mopc}{\,{\rm M}_{\odot}\ {\rm pc}^{-2}}

\newcommand{\gsim}{\lower.7ex\hbox{$\;\stackrel{\textstyle>}{\sim}\;$}}
\newcommand{\lsim}{\lower.7ex\hbox{$\;\stackrel{\textstyle<}{\sim}\;$}}

\newcommand{\rthr}{\rho_{\rm sink}}
\newcommand{\nthr}{n_{\rm sink}}
\newcommand{\raccr}{r_{\rm accr}}

\newcommand{\SNIa}{Type\ {\sc I}a\ }

\newcommand{\sfrinst}{{\rm SFR_{\rm inst}}}

\newcommand{\sfrob}{{\rm SFR_{\rm OB}}}

\newcommand{\sfrsd}{{\rm \Sigma_{\rm SFR}}}
\newcommand{\sfrsdinst}{{\rm \Sigma_{\rm SFR_{\rm inst}}}}
\newcommand{\sfrsdob}{{\rm \Sigma_{\rm SFR_{\rm OB}}}}

\newcommand{\sfrsdks}{{\rm \Sigma_{\rm SFR_{\rm KS}}}}
\newcommand{\vinf}{\ensuremath{v_{\rm wind}}}
\newcommand{\vesc}{\ensuremath{v_{\rm esc}}}
\newcommand{\Teff}{\ensuremath{T_{\rm eff}}}
\newcommand{\kmsw}{\ensuremath{{\rm km}\,{\rm s}^{-1}}}

\title[SILCC III. Regulation of star formation by winds and SNe]
{The SILCC project: III. Regulation of star formation and outflows by
  stellar winds and supernovae} 

\author[A. Gatto, et al.]
{Andrea Gatto,$^{1}$
Stefanie Walch,$^{2}$\thanks{e-mail: walch@ph1.uni-koeln.de} Thorsten Naab,$^{1}$ Philipp Girichidis,$^{1}$\newauthor Richard W\"{u}nsch,$^{3}$ Simon C. O. Glover,$^{4}$ Ralf S. Klessen,$^{4,7}$ Paul C. Clark,$^{5}$\newauthor Thomas Peters,$^{1}$ Dominik Derigs,$^{2}$ Christian Baczynski$^{4}$ and Joachim Puls$^{6}$\\
$^{1}$Max-Planck-Institut f\"{u}r Astrophysik, Karl-Schwarzschild Strasse 1, D-85748 Garching, Germany\\ 
$^{2}$I. Physikalisches Institut, Universit\"{a}t K\"{o}ln, Z\"{u}lpicher Strasse 77, D-50937 K\"{o}ln, Germany\\
$^{3}$Astronomick\'{y} \'{U}stav, Akademie v\u{e}d \u{C}esky Republiky, Bo\u{c}n\'{i} ́II 1401, C-14131 Praha, Czech Republic\\
$^{4}$Universit\"{a}t Heidelberg, Zentrum f\"{u}r Astronomie, Institut f\"{u}r Theoretische Astrophysik, Albert-Ueberle-Str. 2, D-69120 Heidelberg, Germany\\
$^{5}$School of Physics and Astronomy, Cardiff University, 5 The Parade, Cardiff CF24 3AA, UK\\
$^{6}$LMU Munich, Universit\"{a}ts-Sternwarte, Scheinerstrasse 1, D-81679 M\"{u}nchen, Germany\\
$^{7}$Universit\"at Heidelberg, Interdisziplin\"ares Zentrum f\"ur WIssenschaftliches Rechnen (IWR), D-69120 Heidelberg, Germany
}

\begin{document}

\date{Accepted --. Received --; in original form --}

\pagerange{\pageref{firstpage}--\pageref{lastpage}} \pubyear{2015}

\maketitle

\label{firstpage}

\begin{abstract}
We study the impact of stellar winds and supernovae on the multi-phase interstellar medium using three-dimensional hydrodynamical simulations carried out with {\sc FLASH}. The selected galactic disc region has a size of $(500 \pc)^2 \times \pm 5 \kpc$ and a gas surface density of $10 \mopc$. The simulations include an external stellar potential and gas self-gravity, radiative cooling and diffuse heating, sink particles representing star clusters, stellar winds from these clusters which combine the winds from individual massive stars by following their evolution tracks, and subsequent supernova explosions. Dust and gas (self-)shielding is followed to compute the chemical state of the gas with a chemical network. We find that stellar winds can regulate star (cluster) formation. Since the winds suppress the accretion of fresh gas soon after the cluster has formed, they lead to clusters which have lower average masses ($10^2 - 10^{4.3} \mo$) and form on shorter timescales ($10^{-3} - 10$ Myr). In particular we find an anti-correlation of cluster mass and accretion time scale. Without winds the star clusters easily grow to larger masses for $\sim$ 5 Myr until the first supernova explodes. Overall the most massive stars provide the most wind energy input, while objects beginning their evolution as B type stars contribute most of the supernova energy input. A significant outflow from the disk (mass loading $\gtrsim 1$ at 1 kpc) can be launched by thermal gas pressure if more than 50\% of the volume near the disc mid-plane can be heated to $T > 3\times 10^5$ K. Stellar winds alone cannot create a hot volume-filling phase. The models which are in best agreement with observed star formation rates drive either no outflows or weak outflows. 
\end{abstract}
\normalfont
\begin{keywords}
galaxies: ISM -- ISM: evolution -- structure -- kinematics and dynamics -- clouds -- methods: numerical
\end{keywords}

\section{Introduction}\label{sec:introduction}
\let\thefootnote\relax\footnotetext{The SILCC project:
  \textbf{\texttt{www.astro.uni-koeln.de/silcc}}} 

The life cycle of the interstellar medium (ISM) is tightly connected with the star formation activity of a galaxy. Cold and dense molecular gas can partly undergo gravitational collapse leading to star formation. Eventually, newly formed massive stars (with mass $>8 \mo$) strongly impact the surrounding ISM by ionizing radiation
\citep[e.g.][]{Peters+10,Peters+11,Dale+12,Walch+12-2,Walch+13,Dale+14,Geen+15},
radiation pressure
\citep[e.g.][]{KrumholzMatzner09,Murray+10,KrumholzThompson12},
stellar winds
\citep[e.g.][]{Wunsch+08,Pellegrini+11,ToalaArthur11,Wunsch+11,Dale+12,RogersPittard13,Mackey+15, Klassen2016},
and supernova explosions
\citep[e.g.][]{MacLow+05,Dib+06,IffrigHennebelle15,Martizzi+15,Gatto+15,Li+15,WalchNaab15,Haid2016}. These processes - termed 'feedback' in astrophysical slang - locally heat and disperse the surrounding ISM, but may also
compress some fraction of the gas and trigger the formation of new stars.   

Stellar feedback may drive supersonic turbulent motions in the ISM gas \citep[see e.g.][]{Klessen2016}. As an example, observations of broad $\h$ and $\co$ emission lines show that warm, cold, and molecular gas are shaped by supersonic turbulent motions with a typical  velocity dispersion from few to $\approx 10\kms$. \citep{Larson81,Goodman+98,HeilesTroland03, PetricRupen07, Tamburro+09,Caldu+13,Ianjamasimanana+15}.    

Further, it has been proposed that feedback from (massive) stars can locally limit the fraction of gas mass that is converted into stars, i.e. the star formation efficiency, $\epsilon_{\rm SF}$. In the Milky Way, star formation is inefficient with $\epsilon_{\rm SF}\sim$ 1\% \citep{ZuckermanEvans74,MacLowKlessen04}. The inefficiency of star formation has been confirmed for a large number of star-forming galaxies at local and high redshifts $z\approx0-2$ \citep{Leroy+08,Genzel+10,Tacconi+13}. The importance of stellar feedback for the regulation of star formation relative to other
processes, such as large scale shear flows around spiral arms \citep{DobbsPringle13}, is still a matter of debate.  

Stellar feedback influences the thermal and kinetic pressures of the gas at larger scales \citep{Ostriker+10,Girichidis+15}. In particular, supernova explosions can create a hot ionized medium \citep{CoxSmith74,McKeeOstriker77} with high volume-filling factors \citep{Ferriere01,KalberlaDedes08, Walch+15}, which may launch powerful outflows from galactic discs \citep[e.g.][]{Oppenheimer+10,Creasey+13,Hopkins+14,Marinacci+14,Girichidis+15, Peters2015}. Galactic outflows remove gas that would otherwise be available for star formation and hence might regulate galaxy evolution on global scales. In this context, the fundamental role of massive stars for the evolution of star-forming galaxies with a large range of masses has been emphasised in many recent numerical studies \citep[see e.g.][]{Agertz+13,Hopkins+14, Somerville2015}. These simulations test different feedback processes, but suffer from limited spatial and/or mass resolution and thus, cannot capture many physical processes regulating the ISM on small and intermediate scales. 

To understand the non-linear interaction between the interstellar matter and the young stellar population and to investigate the multitude of the relevant physical processes, many authors have carried out studies of the ISM in representative pieces of isolated, stratified, galactic discs using (magneto-)hydrodynamic (MHD) simulations. They investigate the structure of the ISM that is stirred by supernova (SN) feedback
\citep[e.g.][]{deAB04,deAB07,JoungMacLow06,Joung+09,Hill+12,ShettyOstriker12,KimOstriker+13},
with self-gravity (\citealp{G2013}a; \citealp{Gent+13a}b; \citealp{HennebelleIffrig14}; \citealp{KimOstriker2015}), and e.g. with different cooling
functions \citep[][]{Gent+13a}. 

In \citet[][hereafter Paper I]{Walch+15}, we demonstrated how the positioning of SN explosions relative to the cold and dense gas in the disc affects the multi-phase temperature (from $\sim 10$ to $10^8 $K) and chemical structure ($\htwo$, H, H$^+$) of the ISM. With a fixed SN rate, which is informed by the Kennicutt-Schmidt \citep[KS;][]{Kennicutt98a} relation and connecting the gas surface density to a SN rate using a standard IMF \citep[e.g. ][]{Kroupa2002, Chabrier03}, we evolved the simulation using MHD, gas self-gravity, a chemical network and radiative transfer of diffuse radiation to model the formation of molecular gas in the form of $\htwo$ and CO. We showed that SNe located at random positions lead to a bubbly ISM with a high volume-filling fraction of hot gas on the one hand, and at the same time help self-gravity to drive the formation of $\htwo$ in filaments and clumps. In these runs, the thermal feedback is strong enough to launch galactic fountain flows that have a multi-phase structure \citep[see][hereafter Paper II]{Girichidis+15}. SNe which explode within dense gas have a low heating efficiency and produce low $\htwo$ mass fractions. In the case where all SNe are associated with dense gas, we obtained a very low volume-filling fraction of hot gas \citep[see also][]{Gatto+15} and there were no outflows from the disc. However, these proof of concept studies lack a direct connection between dense gas and the formation of new stars.   

A recent study by \cite{HennebelleIffrig14}, uses sink particles to more self-consistently model the star formation in such a stratified galactic disc \citep[see also][for similar work using a periodic box]{Slyz+05}. Their SN rate is not fixed but correlated in space and time with the sink particle positions and accretion rate. The energy from SN explosions is injected right after the formation of each massive star \citep[see also e.g.][]{KimOstriker+13, KimOstriker2015}, thus neglecting the time delay of the explosions corresponding to the stellar lifetime of single stars (typically 5 - 40 Myr). They show that instantaneous SN feedback can significantly lower the star formation rate (SFR) by a factor of 20--30. A complication in this context is that it remains unclear whether in a more realistic setup the SN explosions remain the SFR limiting factor when other pre-supernova feedback processes and realistic SN delay time distributions are assumed.  

In this paper we improve on earlier studies (Paper I and II) by studying the mutual influence of the three-phase ISM, self-consistent star formation, and feedback from massive stars in the form of stellar winds and supernovae with realistic delay time distributions. The feedback is associated with accreting sink particles that represent young star clusters. We follow the evolution of each single massive star using the latest Geneva stellar evolution tracks by \cite{Ekstrom+12} and study the relative impact of stellar winds and SNe on the structure of the ISM, the star formation rate, and the onset of galactic outflows. We will argue that the inclusion of stellar winds (and possibly other pre-supernova injection processes of massive stars not investigated here, like ionizing radiation and radiation pressure) qualitatively change the timing and the regulation mechanisms for star cluster formation. 

The manuscript is organised as follows: in section \ref{sec:method} we describe our model and we list the important parameters and simulations. In section \ref{sec:results1} we present our qualitative results, with a more detailed discussion on the wind and SN feedback regulation processes in section \ref{sec:results2}. The effects on disk outflows are presented in section \ref{sec:results3}, and we conclude in section \ref{sec:conclusions}.




\section{Numerical Method}
\label{sec:method}
We use the Eulerian, adaptive mesh refinement (AMR), MHD code \texttt{FLASH 4} \citep{flash,Dubey+08,Dubey+13} with the directionally split, Bouchut HLL5R solver \citep{Bouchut+07,Waagan09,Bouchut+10,Waagan+11} to simulate the ISM in a stratified disc. The size of the vertically elongated box is $500 \pc \times 500 \pc \times \pm\ 5 \kpc$. We set periodic boundary conditions in $x$ and $y$ direction and use outflow boundary conditions in the $z$ direction. Near the disc mid-plane the resolution is $\Delta x \simeq 3.9 \pc$ and above and below $z=1$ kpc we use $\Delta x \simeq 7.8 \pc$. We solve the ideal MHD equations and additionally include  
\begin{itemize}
\setlength{\itemindent}{0.3cm}
 \item a static potential to model the old stellar component in the disc (sec. \ref{sec:gravity}),
 \item gas self-gravity (sec. \ref{sec:gravity}),
 \item radiative cooling and diffuse heating by a smooth interstellar radiation field (ISRF) with $G_0 = 1.7$ (sec. \ref{sec:cooling}),
 \item dust and gas (self-)shielding (sec. \ref{sec:cooling}),
 \item a chemical network to explicitly follow $\h,\hp,\htwo,\co,\cp$ (sec. \ref{sec:cooling}),
 \item star cluster sink particles (sec. \ref{sec:sinks}) with a sub-grid prescription that models the formation and evolution 
   of the massive stars in the star cluster using stellar tracks (sec. \ref{sec:sub-grid}), and 
 \item stellar winds and/or SN feedback from the star cluster sink particles (sec. \ref{sec:sub-grid}).
\end{itemize}
In this paper we do not include the impact of ionizing radiation from the massive stars on the ISM.  We also do not include galactic shear unlike e.g. \cite{KimOstriker2015} but for our particular setup the influence of shear is probably negligible (see section 7.4 in Paper II for an estimate of the Rossby number). Below, we briefly describe our numerical method, but also refer to Paper I for more details. 

\subsection{Gravity}\label{sec:gravity}
Three terms contribute to the gravitational acceleration of the gas: self-gravity, the static background potential caused by old stars in the disc, and newly forming sink particles:  
\begin{equation}
\mathbf{g} = \mathbf{g}_\mathrm{sg}+\mathbf{g}_\mathrm{ext}+\mathbf{g}_\mathrm{sinks}.
\end{equation}

The gravitational acceleration due to self-gravity, $\mathbf{g}_\mathrm{sg}$, is computed by solving Poisson's equation
for the gas in three dimensions using a tree-based method described in detail in Paper I and W\"{u}nsch et al.~(2016, in prep.).  

We neglect dark matter but consider the external potential generated by the old stellar component in the galactic disc, which we assume to follow the distribution \citep{Spitzer42}  
\begin{equation} \label{eq:extpot}
\rho_*(z) = \rho_*(0)\ \mathrm{sech}^2(z/2z_\mathrm{d})\ ,
\end{equation}
where $\rho_*$ is the density of stars at height $z$. We take $\rho_*(0) = 0.075 \mo \pc^{-3}$, which corresponds to a total stellar surface density of $30 \mopc$ with a scale height $z_\mathrm{d}=100 \pc$. We then integrate the one-dimensional Poisson equation along the $z$-direction for $\rho_*$ to get the external acceleration: $\mathbf{g}_\mathrm{ext}(z)$.    

The contribution of sink particles to the gravitational acceleration of the gas, $\mathbf{g}_\mathrm{sinks}$, is taken into
account. Following \cite{Federrath+10-2}, outside the accretion radius (see section \ref{sec:sinks}) this involves a direct summation for all computational cells and all particles. Within the accretion radius, a cubic spline gravitational softening scheme is applied to avoid diverging accelerations at close distances. The sink particles are advanced using a Leapfrog time integration scheme with sub-cycling. The according forces are computed from particle-particle as well as gas-particle interaction \citep[for more details see][]{Federrath+10-2}. In addition, we include the force due to the
external gravitational potential.  

\subsection{Cooling, heating and chemistry}\label{sec:cooling}
We include heating and cooling processes using a simplified chemical network based on\cite{GloverMacLow07a,GloverMacLow07b}, and \cite{NelsonLanger97} to follow the abundances of seven chemical species: $\h,\hp,\htwo,\co$, and $\cp$, as well as free electrons and atomic oxygen, which are tracked utilising conservation laws. The rate equations for $\htwo$ and \co\ include the effect of dust shielding and molecular (self-)shielding \citep{Glover+10}. The total, $\htwo$, and CO column densities, which are necessary to compute the shielding coefficients are estimated using the \texttt{TreeCol} algorithm of \citet{Clark+12}, which we implemented into \texttt{FLASH} 4. For further details see Paper I and W\"{u}nsch et al. (2016, in prep.).  
 
Cooling of gas with $T>10^4$ K is modelled with the cooling rates of \cite{GnatFerland12}, which assumes collisional ionisation equilibrium. For lower temperatures, non-equilibrium cooling rates for the respective chemical abundances as well as heating by the photo-electric effect, cosmic rays, X-rays, and UV radiation from a diffuse interstellar radiation field with $G_{\mathrm 0} = 1.7$ \citep{Habing68,Draine78} is included \citep{Glover+10,GloverClark12}. We assume a cosmic ray ionisation rate of $\zeta = 3 \times 10^{-17} {\rm s}^{-1}$,  and X-ray ionisation and heating rates based on\cite{Wolfire+95}. For simplicity, we assume that the ISRF is constant everywhere in the computational domain. However, it is attenuated in shielded regions, where the shielding depends on the column densities (total, $\htwo$, and CO), which are determined through \texttt{TreeCol}.

For all simulations the gas has solar metallicity with abundances $x_{\mathrm{O, tot}} =3.16 \times 10^{-4}$, $x_{\mathrm{Si^+}} = 1.5 \times 10^{-5}$, and $x_{\mathrm{C, tot}} = 1.41 \times 10^{-4}$ \citep{Sembach+00}. The (constant) dust-to-gas mass ratio is set to $10^{-2}$. For further details we refer the reader to Paper I. 

\subsection{Sink particles}\label{sec:sinks}
We include the sink particles unit from the \texttt{FLASH 4} public release, described in \citet[][see also \citealt{Bate+95,Krumholz+04,Jappsen+05,2013MNRAS.430.3261H,BleulerTeyssier14} for details on the implementation of sink particles in other SPH and Eulerian codes]{Federrath+10-2}. In our models, collisionless sink particles provide the framework to model the formation of internally unresolved star clusters in dense regions undergoing gravitational collapse. Following \cite{Federrath+10-2}, a sink particle is created in a particular cell if   
\begin{itemize}
\setlength{\itemindent}{0.3cm}
 \item the gas density is higher than a user-defined density threshold
   $\rthr$,  
\item all cells within the accretion radius, $\raccr$, are at the highest
  refinement level, 
\item the cell represents a local gravitational potential minimum,
 \item the gas within $\raccr$ is Jeans unstable,
\item the gas within $\raccr$ is in a converging flow ($\mathbf{\nabla} \cdot \mathbf{v} < 0$),
 \item the gas within $\raccr$ is gravitationally bound, and 
 \item the sink's accretion radius does not overlap with that of another existing sink.
\end{itemize}

Once a sink particle is formed, it can accrete gas within $\raccr$ if the gas density exceeds $\rthr$. Additional checks are performed to ensure that only bound, collapsing gas is removed from the grid and added to the sink. We set the accretion radius to $\raccr=4\times \Delta x = 15.6$ pc, where $\Delta x = 3.9$ pc is the cell size at the maximum refinement level \citep[a typical value, see e.g.][]{Krumholz+04,HennebelleIffrig14}. This satisfies the Truelove criterion \citep{Truelove+97} and the more stringent criterion of isothermal MHD collapse found by \cite{Heitsch+01}. 
The choice of $\raccr$ determines $\rthr$, below which the gas can be considered Jeans-stable. Then we have 
\begin{equation} \label{eq:jeans}
\lambda_\mathrm{J} = \biggl(\frac{\pi c_\mathrm{s}^2}{G \rthr}\biggr)^{\frac{1}{2}} = 2\times \raccr \approx 31.2 \pc \ ,
\end{equation}
with $c_\mathrm{s} = (k_\mathrm{B}T/ m_\mathrm{p})^{1/2}$ the
isothermal sound speed of monoatomic gas.  
This gives
\begin{equation} \label{eq:rthr}
\rthr = \frac{\pi k_\mathrm{B}}{m_\mathrm{p}G} \frac{T}{(2\times \raccr)^2}\ .
\end{equation}
For a temperature of $T=300$ K, below which we consider the gas to be in the thermally stable, cold phase, the density threshold is $\rthr \approx 1.26 \times 10^{-22} \;{\rm g\; cm}^{-3}$ \citep[this is an order of magnitude lower than the sink density threshold used by][]{HennebelleIffrig14}. Often, we find even lower temperatures in the dense gas (down to 10 K), for which the Jeans length cannot be resolved with our choice of $\raccr$. However, we do not consider this to be a severe problem since the sink particles in our simulations do not represent individual stars, but are rather considered to be tracing star clusters with an internal stellar initial mass function (IMF, see section \ref{sec:sub-grid}). Therefore, we do not need to resolve the fragmentation limit with $\rthr$, but rather treat it as a free parameter. We present simulations with different sink formation thresholds, ranging from $\rthr=2\times10^{-22}\;{\rm g\; cm}^{-3}$ or a particle density of $\nthr \approx 10^2\; {\rm cm}^{-3}$ to $\rthr=2\times10^{-20}\;{\rm g\; cm}^{-3}$ or $\nthr \approx 10^4\; {\rm cm}^{-3}$. 


\begin{figure}
 \includegraphics[width=0.49\textwidth]{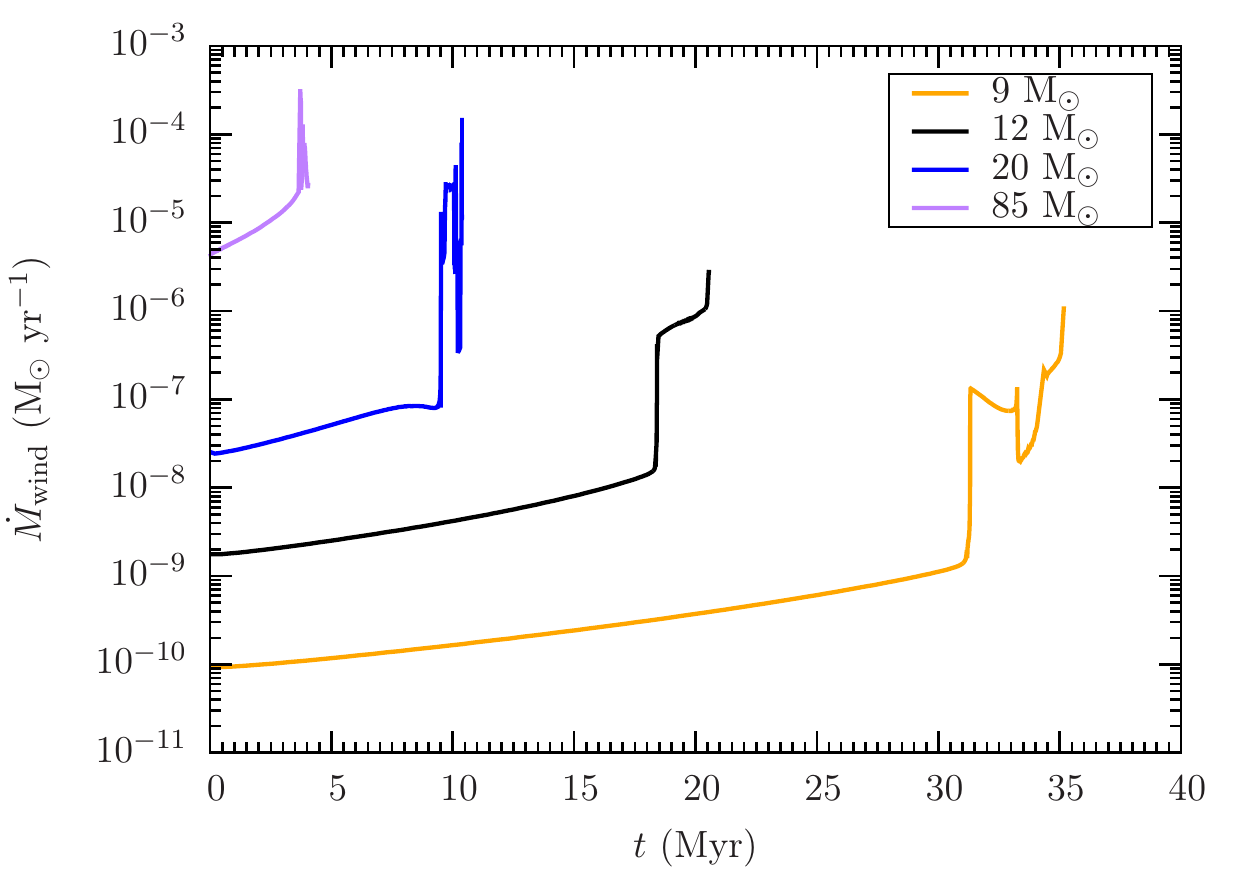}
 \includegraphics[width=0.49\textwidth]{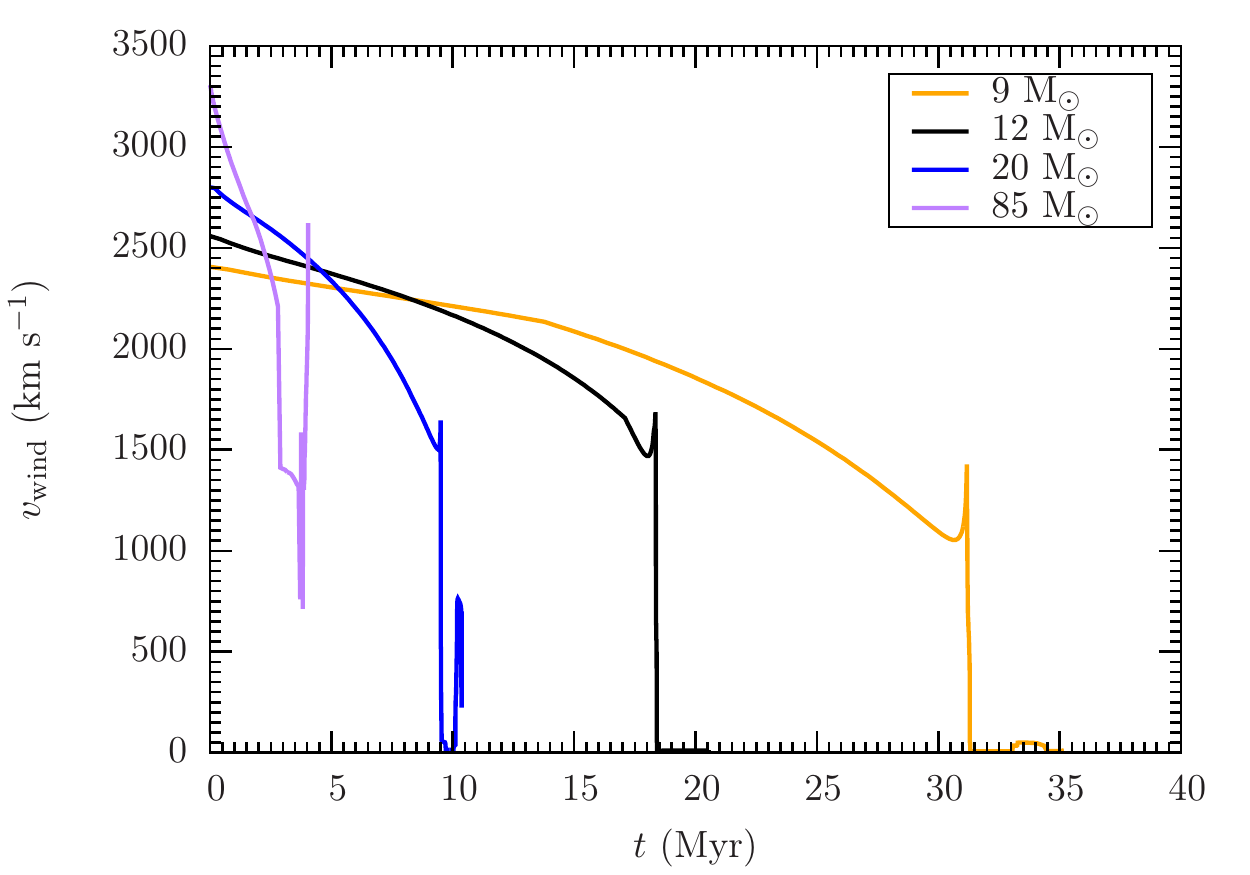}
 \includegraphics[width=0.49\textwidth]{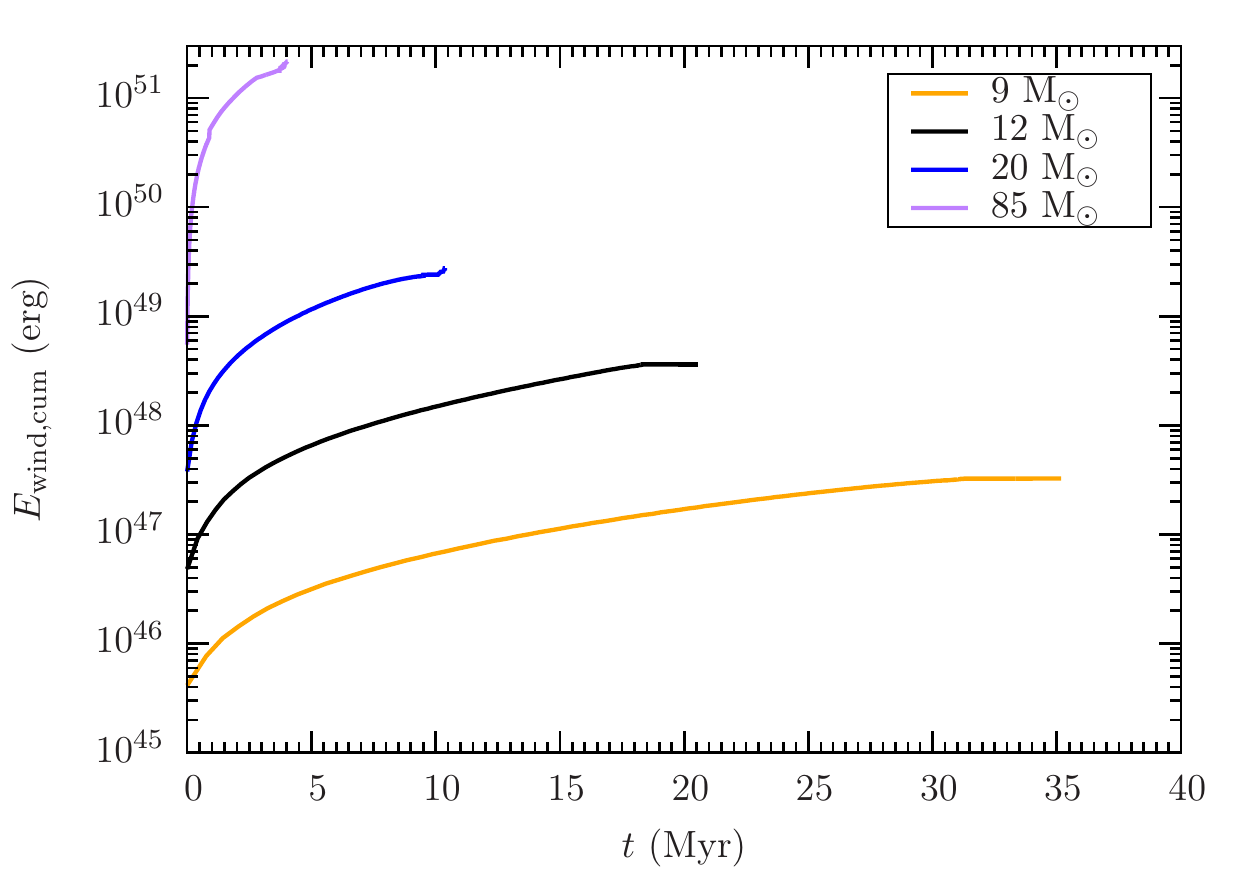}
\caption{Mass-loss rates (top panel), wind terminal velocities (middle panel) and cumulative energies (bottom panel) for the stellar winds of four different massive stars with initial masses of $9$ (orange), $12$ (black), $20$ (blue) and $85$ (purple) $\mo$.}\label{fig:tracks}   
\end{figure}

\subsection{Sub-grid model for cluster-sink particles}\label{sec:sub-grid}
The sink particles formed in our simulations have masses of $M_\mathrm{sink} \sim 10^2-10^{5.3} \mo$, i.e. they are groups of stars (star clusters). We therefore call them {\it cluster sink particles} and implement a sub-grid model to follow the evolution of massive stars that are supposedly forming within them.  

We assume that all gas accreted onto a sink is converted into stars, which corresponds to a {\it cluster formation efficiency} of 100\%. This choice is numerically motivated and prevents gas from being artificially locked up inside the sink without the possibility to be heated or dispersed by stellar feedback or to eventually collapse into stars. We note that the {\it cluster formation efficiency} is a theoretical concept and is not equal to the {\it star formation efficiency}. The latter needs to be computed from the ratio of the star formation rate and the available mass in atomic and/or molecular hydrogen (as indicated when comparing our simulation results with recent observations in Fig. \ref{fig:obs}).\\   

  
{\sc Massive star content:} All of $M_\mathrm{sink}$ is available for star formation. We are only interested in following the evolution of individual massive stars that have significant stellar winds and explode as SNe (that is stars with mass $> 8 \mo$). Therefore, we have implemented the following model:  
\begin{itemize}
\setlength{\itemindent}{0cm}
\item One massive star is created for every $120 \mo$ of gas that is converted into or accreted onto a sink particle (star cluster).
\item The mass of every new-born star is randomly sampled from the Salpeter IMF \citep{Salpeter55} within a mass range of $9 - 120 \mo$.  
\item The rest of the mass is assumed to reside in low-mass stars, which are not followed individually.  
\item Not every massive star is created upon sink formation. Whenever enough mass (a mass unit of $120 \mo$) becomes available (it has been accumulated by gas accretion onto the sink), a new massive star is spawned.  
\item The number of massive stars within each sink, $N_\star$, is different for each sink and changes as a function of time. 
\end{itemize}

{\sc Stellar wind model:} The evolution of each massive star is followed using the latest Geneva stellar evolution tracks from the zero-age main sequence (ZAMS) to the Wolf-Rayet (WR)/pre-SN phase by \cite{Ekstrom+12}. We interpolate and store 112 tracks (for stars with $9$ to $120 \mo$, separated by $1 \mo$). We do not take into account a delay time due to star formation or a proto-stellar phase, but immediately start with the ZAMS evolution of the formed massive stars. Then, in each time-step, the age and initial mass of each star are used to determine the appropriate mass-loss rate and terminal velocity of the stellar wind.  

While the mass-loss rates can be directly taken from the tracks by \cite{Ekstrom+12} (for the corresponding scaling relations, see their section 2.6), we estimate terminal velocities (\vinf), which are not given in the tracks, according to their evolutionary status (defined from the surface abundances of the models, see \citealt{Georgy+12}):  

\begin{enumerate}
\setlength{\itemindent}{0cm}
\item For OB type stars and A supergiants, we use a slightly modified version of the scaling relations provided by \cite{KudPuls2000} and \cite{MarkovaPuls08}, namely \vinf\ = 2.45 \vesc\ for \Teff\ $> 2.3 \times 10^4$ K, \vinf\ = 1.3 \vesc\ for \Teff\ $< 1.8\times 10^4$ K, and a linear interpolation in between (the so-called bi-stability jump, see \citealt{Puls+08} and references therein). Here, \vesc\ is the photospheric escape velocity corrected for the radiative acceleration by electron-scattering and \Teff\ corresponds to the {\it corrected}, effective temperature as provided by the tracks.  

\item For WR stars, we adapt observational data compiled by \cite{Crowther07}, using linear interpolations. In particular, for WNL and WNE stars, we use \vinf\ = 700~\kmsw\ for \Teff\ $< 2\times 10^4$ K, and a linear inter/extrapolation between 700 and 2100~\kmsw\ for $2\times 10^4 <$ \Teff\ $< 5\times 10^4$ K, whilst for WC stars we use again \vinf\ = 700~\kmsw\ for \Teff\ $< 2\times 10^4$ K, and a linear inter/extrapolation between 700 and 2800~\kmsw\ for $2\times 10^4$ K $<$ \Teff\ $< 8\times 10^4$ K.  

\item For red supergiants, we follow \cite{vanLoon06}, with $\vinf \propto L^{0.25}$, normalised to \vinf\ = 10~\kmsw at a luminosity of $L = 3\times 10^4 {\rm L}_{\odot}$.  

\item Finally, the terminal velocities for objects in between blue and red supergiants (rather insecure) have been approximated by the geometric mean of the \vinf-values for the 'neighbouring' blue and red supergiants, resulting in typical values of \vinf\ $\approx$ 50~\kmsw\ for yellow supergiants.   
\end{enumerate}

Fig. \ref{fig:tracks} shows the stellar evolution tracks used in this work for four representative stars with initial masses of $9, 12, 20$, and $85 \mo$. The most massive stars show significantly higher mass-loss rates, wind terminal velocities, and wind luminosities but about an order of magnitude shorter lifetimes (only $\sim 4$ Myr for a star with $85 \mo$).  The bottom panel of Fig. \ref{fig:tracks} shows the cumulative wind energies, which depend strongly on the initial
mass of the star. Stars with relatively low masses ($\sim 9-20 \mo$) release only little wind energy, i.e. $\sim
10^{-2}-10^{-4} \times E_\mathrm{SN}$, where the typical energy released by a single SN event is $E_\mathrm{SN}=10^{51}$ erg. However, the most massive stars inject as much or even more energy in winds than in their final SN explosion. Following these tracks, it requires $\sim$ 6600 stars with $9 \mo$ each to produce the same wind energy as a single $85 \mo$ star.   

For single stellar populations, stars at the lower end of our considered mass range  (i.e.\ B type stars) are considerably more numerous and have longer lifetimes than the WR- and massive O-stars which produce the strongest stellar winds\footnote{Energetically, i.e. with respect to luminosities and winds, WR-stars dominate \citep[e.g.][]{Leitherer1992,Doran2013}.}. Therefore, stellar winds only dominate the energy budget during the early evolution of the stellar population (for the first $\approx 5-20 \Myr$).\\

{\sc Stellar wind feedback:} We apply the following prescription to model the wind energy input in our simulations:
\begin{itemize} 
\setlength{\itemindent}{0.3cm}
\item For each cluster-sink and at each time step, we calculate the total mechanical luminosity by adding up the contributions of all $N_\star$ stellar winds 
\begin{equation}
L_{\mathrm{tot}} = \frac{1}{2} \sum_{i=1}^{N_\star} \dot{M}_{\mathrm{wind,i}} \times v_{\mathrm{wind,i}}^2\;[{\rm erg \; s}^{-1}]
\end{equation}

\item The total mass lost by all winds in the cluster is,
\begin{equation}
\dot{M}_{\rm tot} = \sum_{i=1}^{N_\star} \dot{M}_{\mathrm{wind,i}}.
\end{equation}
Within each time step $\Delta t$, we add a total mass of $\dot{M}_{\rm tot}\times \Delta t$ to the injection region, which we set equal to $\raccr$. Per unit volume, the mass is evenly distributed amongst all the cells which overlap with the spherical injection region. Note that the mass of the cluster-sink is reduced accordingly (the net sink mass can still increase due to the accretion of fresh gas). 
\item The mass which is added to the injection region carries a certain amount of internal energy, which we take into account.
\item We inject the wind feedback in the form of kinetic energy, $e_\mathrm{inj}$, which we evenly distribute within $\raccr$. Thus, we have
\begin{eqnarray*}
e_\mathrm{inj} &=& \dot{e}_\mathrm{inj} \times \Delta t  \\
&=&  L_{\mathrm{tot}} \times \Delta t\\
& =&  \frac{1}{2} M_\mathrm{inj} v_r^2 ,
\end{eqnarray*}
where $M_\mathrm{inj} = M_\mathrm{inj, old} + \dot{M}_\mathrm{tot} \times \Delta t$ is the sum of the previously present gas mass within the injection region and the returned stellar wind material, and $v_r$ is the radial velocity. The wind is assumed to be spherically symmetric and we neglect possible cancellation effects within the cluster sink due to wind collisions. The radial velocity applied within the injection region is hence computed from 
\begin{equation}
v_r^2 = 2 \frac{L_{\mathrm{tot}}\times \Delta t}{M_\mathrm{inj}}. 
\end{equation}
 
\end{itemize}

{\sc Supernova feedback:} Once a star has reached the end of its lifetime, it is assumed to explode as a Type II SN. In our model, each SN releases an energy of $E_\mathrm{SN}$, which is typically injected in the form of thermal energy provided that the adiabatic phase of the SN remnant is resolved. If the density in the injection region is high, such that the Sedov-Taylor phase would be unresolved, we switch to a momentum input scheme \citep[see][ for a detailed description of the SN model]{Gatto+15}. The mass of the SN progenitor star is also added to the injection region. For simplicity, we do not account for stellar remnants, which are unresolved. In run {\it FSN-n1e2} where stellar winds are not included, we still follow the evolution of each star to model the supernova delay time.  

Each feedback event is centred on the position of the cluster sink. We do not account for runaway stars that are ejected from their parental star clusters \citep[see e.g.][for a discussion]{Li+15}. Moreover, we neglect the slow winds from stars with $M \leqslant 8 \mo$. \SNIa SNe originating from an old stellar population are also not included in our model.   

\begin{table}
\begin{tabular}{lccc}
\hline
\hline
Name                   & $\rthr$ &Wind   & SN     \\
			    &$[2\times10^{-24}\;{\rm g\;cm}^{-3}]$ & & \\
\hline
\hline
NoF-n1e2         & $ 10^2$ &no            &no   \\ 
FSN-n1e2        	& $10^2$ &no            &yes       \\ 
FW-n1e2            & $10^2$ &yes          &no        \\ 
FWSN-n1e2    & $10^2$ &yes          &yes        	\\ 
FWSN-n1e3	     	&$10^3$ &yes		& yes \\ 
FWSN-n1e4	      &  $10^4$&yes		& yes \\ 
\hline
\end{tabular}
\caption{Overview of all presented simulations. We list the run names (column 1), the sink density threshold $\rthr$ (column 2), and the included feedback mechanisms (stellar winds in column 3, SNe in column 4). }
\label{tab:simlist}
\end{table}

\subsection{Simulation setup}
\label{sec:simulations}
\subsubsection{List of simulations}\label{sec:runs}

We present a set of 6 simulations (see table \ref{tab:simlist}), with which we are able to show the effect of the different feedback mechanisms. For reference, we include run {\it NoF-n1e2}, which is a run with clustered star formation but without feedback. Then we switch on either wind feedback (run {\it FW-n1e2}) or supernova feedback (run
{\it FSN-n1e2}), or both (run {\it FWSN-n1e2}). As a second parameter, we increase the sink density threshold from $\rthr = 2\times 10^{-22}\;{\rm g}\;{\rm cm}^{-3}$ (all runs with ending {\it -n1e2}) to $\rthr = 2\times 10^{-21}\;{\rm g}\;{\rm cm}^{-3}$ (run {\it FWSN-n1e3}) and $\rthr = 2\times 10^{-20}\;{\rm g}\;{\rm cm}^{-3}$ (run {\it FWSN-n1e4}), respectively.  

\subsubsection{Initial conditions}\label{sec:initial}
The initial gas density profile (see Paper I and Paper II) is uniform in $x$ and $y$ but follows a Gaussian distribution in the $z$-direction
\begin{equation} \label{rhoini}  
\rho(z) = \rho_0 \mathrm{exp}\biggl[-\frac{1}{2}\biggl(\frac{z}{z_0}\biggr)^2\biggr]\ ,
\end{equation}
with a scale height of $z_0=30 \pc$ and a mid-plane density of $\rho_0=9\times 10^{-24}$ g $\cm$.  At large heights above the mid-plane, we truncate the Gaussian distribution at the background density of $\rho_\mathrm{bg}=10^{-28}$ g $\cm$. Altogether, the initial gas surface density of the disc is $\Sigma_\mathrm{gas}=10 \mopc$ and the total mass in the computational domain is $M_{0} = 2.55 \times 10^6 \mo$. 

We set the initial temperature within the disc mid-plane to $T=4500$ K and assume vertical pressure equilibrium to compute the temperature profile. Therefore the halo gas is hot with a temperature of $T=4\times 10^8$ K. According to the initial temperature profile, all hydrogen is initially atomic near the disc mid-plane and partially or fully ionized at larger scale heights. Carbon is fully ionized everywhere in the computational domain. 

To create inhomogeneities in the gas distribution and to partially support the disc against gravitational collapse, we initially drive turbulent motions in the disc. This is necessary as otherwise all gas would collapse towards the mid-plane and cause a strong burst of star formation. On the largest possible modes in the disc plane, $k = 1$ and $k = 2$ corresponding to the box size of 500 $\pc$ and half of the box size, the turbulent energy is injected with a flat power spectrum and a thermal mix of solenoidal (divergence-free) to compressive (curl-free) modes of 2:1. The energy input is adjusted such that the global, mass-weighted, 3D root-mean-square (rms) velocity remains constant at $v_\mathrm{3D, rms} \sim 10 \kms$. The turbulent energy input is evolved with an Ornstein-Uhlenbeck random process \citep{EswaranPope88} with a phase turnover time, which corresponds to the turbulent crossing time in the $x$ and $y$ directions of $\sim 50 \Myr$. The turbulence driving is switched off once the first sink particle has formed, which happens after $\gtrsim 9 \Myr$ ($t_\mathrm{sink, 0}=9$ Myr for the simulations with the lowest sink density threshold).


\begin{figure*}
 \centering
 \textbf{Run {\it FWSN-n1e3}, $t=45 \Myr$}\\
 \includegraphics[width=0.7\textwidth]{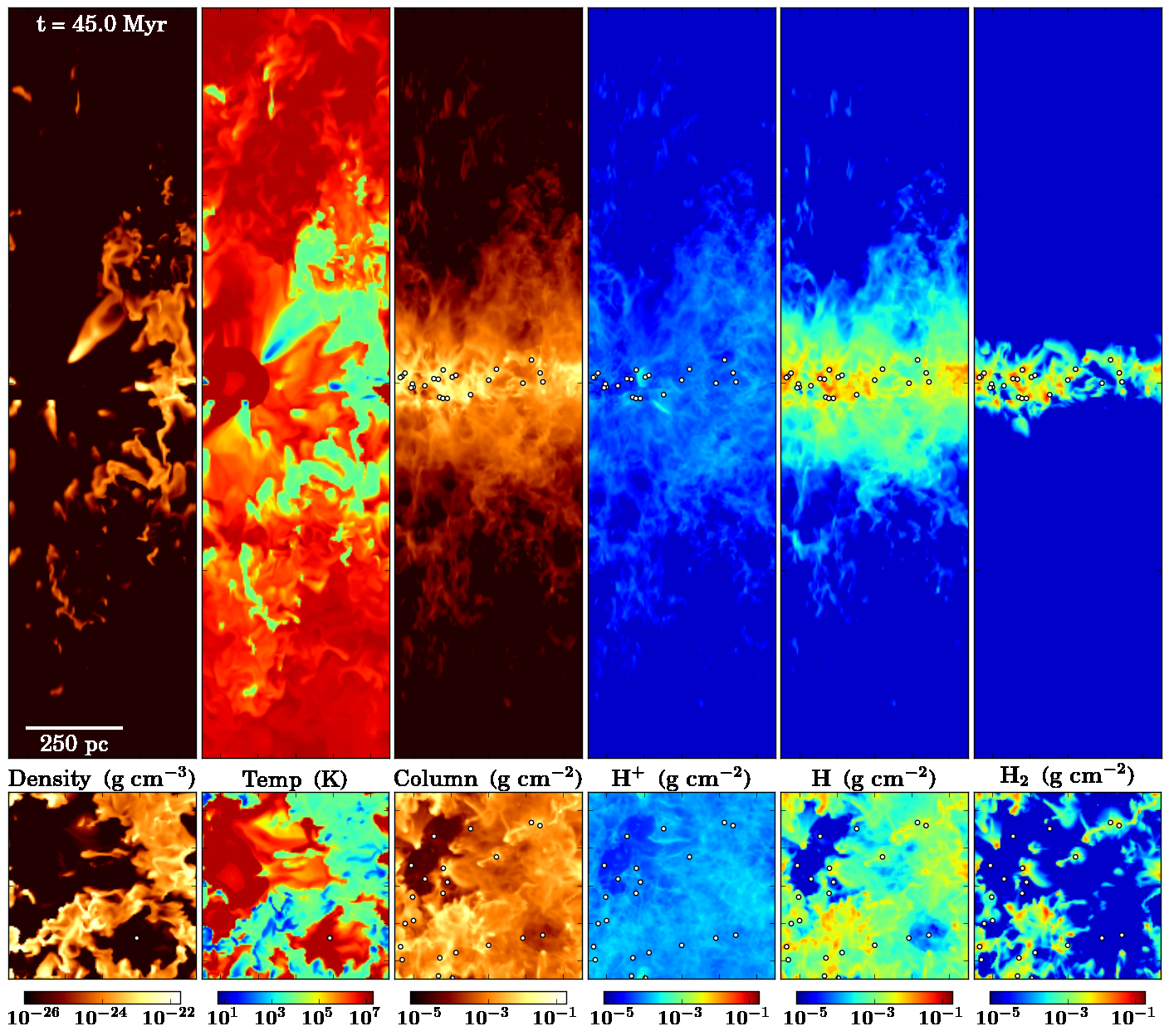}
\caption{Run {\it FWSN-n1e3} with stellar wind and supernova feedback
  from cluster sinks which are introduced above $\rthr = 2\times
  10^{-21}\;{\rm g\; cm}^{-3}$ at $t=45$ Myr as seen edge-on (upper
  panels) and face-on (lower panels). {\it From left to right:}
  density slice,  temperature slice, column density, and the column
  densities of $\hp,\ \h,\;{\rm and}\; \htwo$, respectively. The filled
  circles show the location of the cluster-sink particles.  
}\label{fig:snap1}
\end{figure*}

\begin{figure*}
\centering
\begin{tabular}{cc}
 \textbf{Run {\it NoF-n1e2}}  & \textbf{Run {\it FW-n1e2}}\\
 \includegraphics[width=0.5\textwidth]{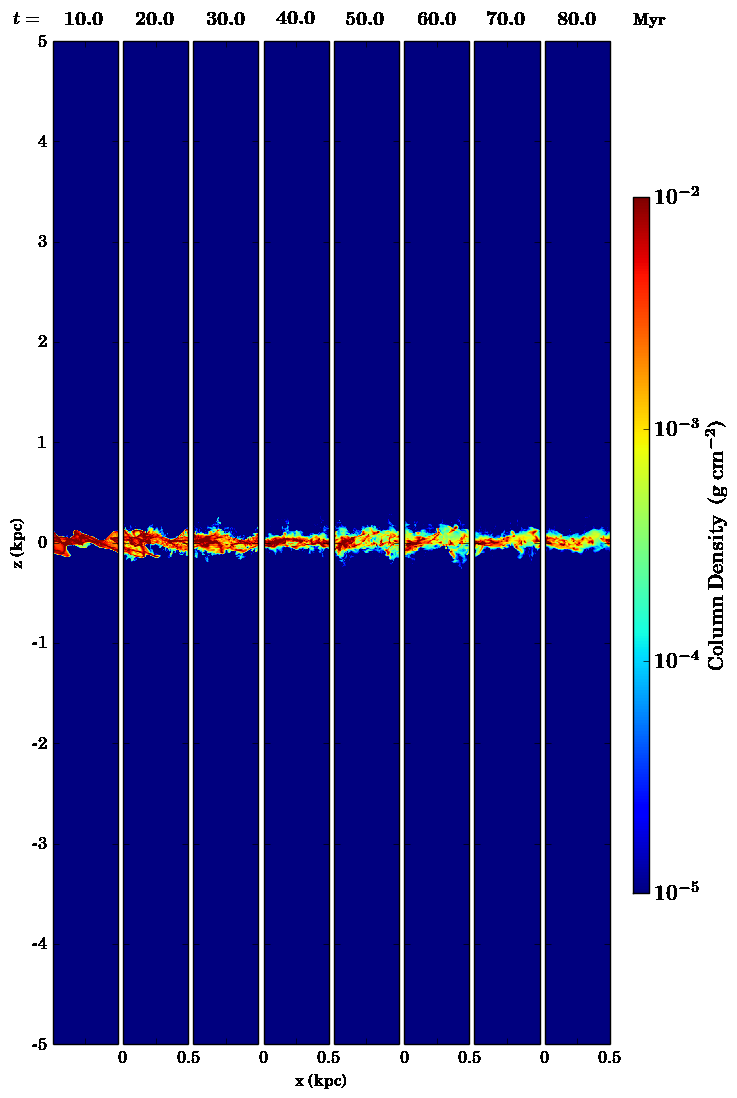} & 
 \includegraphics[width=0.5\textwidth]{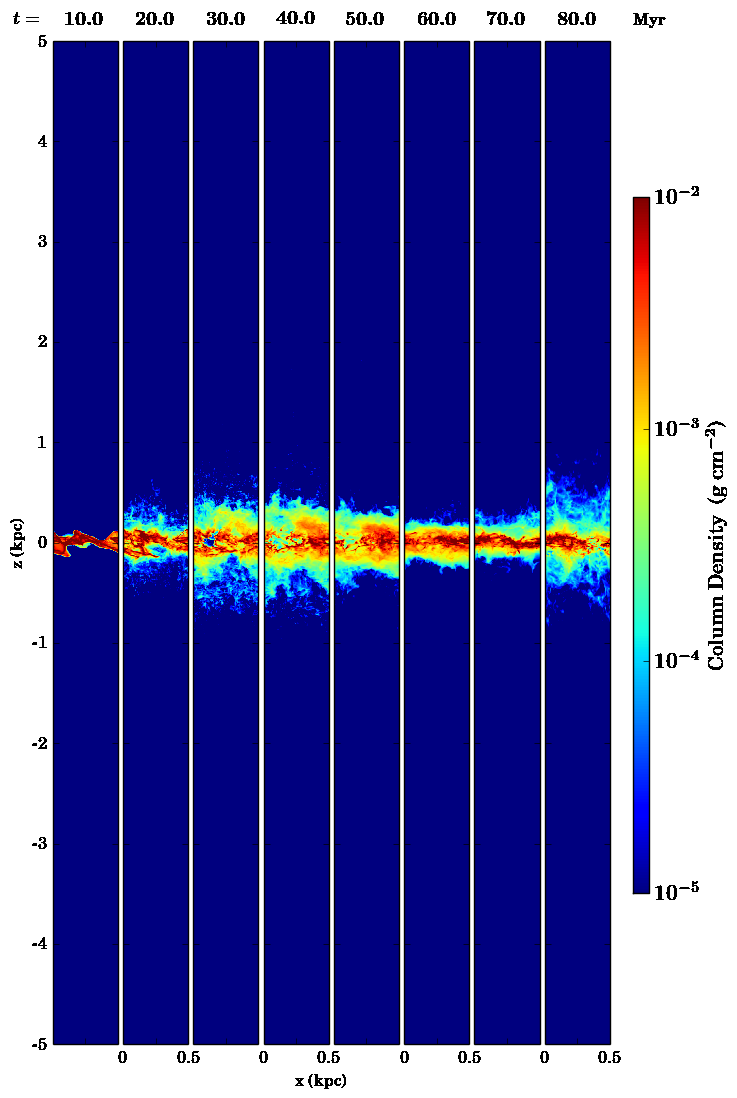}  
 \end{tabular}
\caption{Time evolution (from left to right) of the
  total gas column density. {\it Left:} Simulation {\it NoF-n1e2}
  without feedback. In this case no outflows are driven and the gas
  collapses to the mid-plane. {\it Right:} Simulation {\it FW-n1e2}
  with stellar winds (no supernovae) originating from the massive
  stars within the cluster sink particles.}\label{fig:hill1}   
\end{figure*}

\begin{figure*}
\centering
\begin{tabular}{cc}
 \textbf{Run {\it FSN-n1e2}}  & \textbf{Run {\it FWSN-n1e2}}\\
 \includegraphics[width=0.5\textwidth]{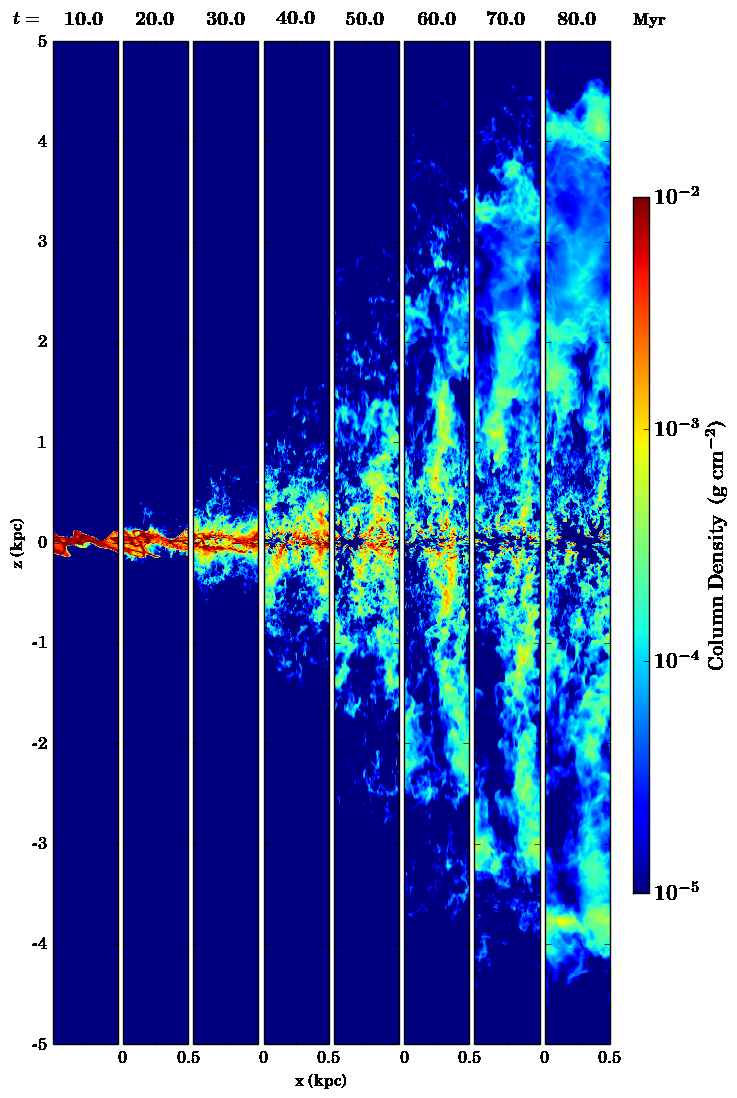}&
 \includegraphics[width=0.5\textwidth]{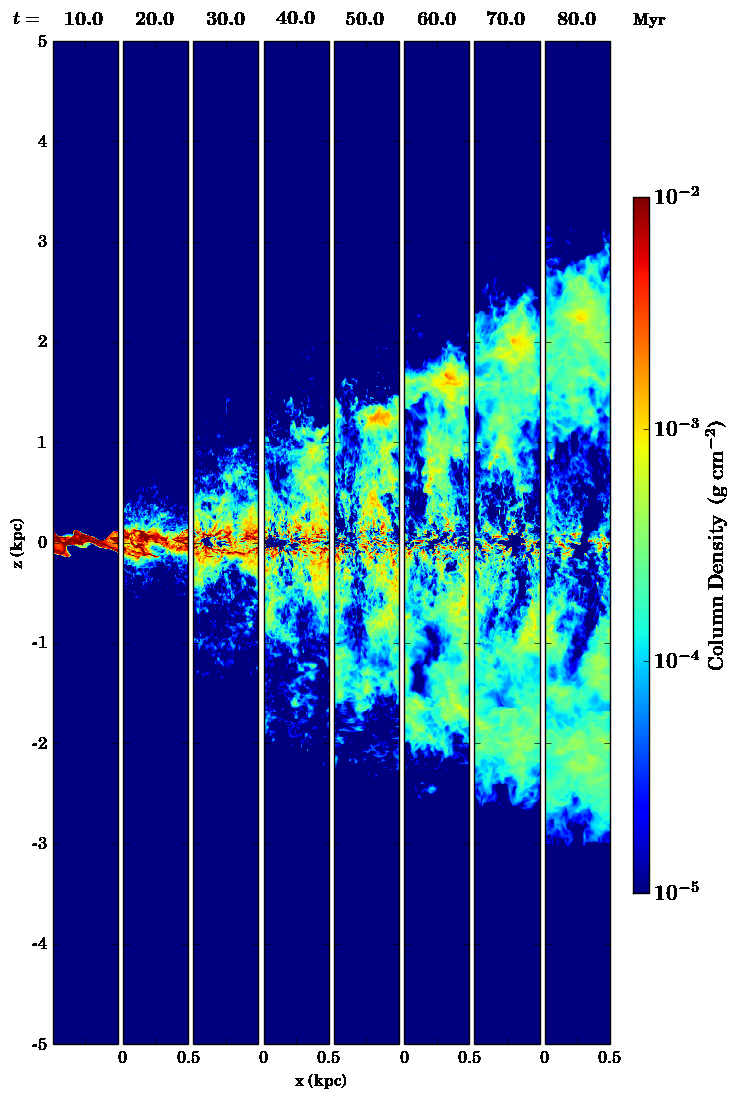}
 \end{tabular}
\caption{Same as Fig. \ref{fig:hill1} for simulation {\it FSN-n1e2}
  with just supernova explosions (left panel). This simulation has a
  high star formation rate and drives the strongest outflows. {\it
    Right:} Simulation {\it FWSN-n1e2} with both, stellar winds
  and supernova explosions. Here stellar winds reduce the star
  formation rate and outflow.}\label{fig:hill2}   
\end{figure*}

\begin{figure*}
\centering
\begin{tabular}{cc}
 \textbf{Run {\it FWSN-n1e3}}  & \textbf{Run {\it FWSN-n1e4}}\\
 \includegraphics[width=0.5\textwidth]{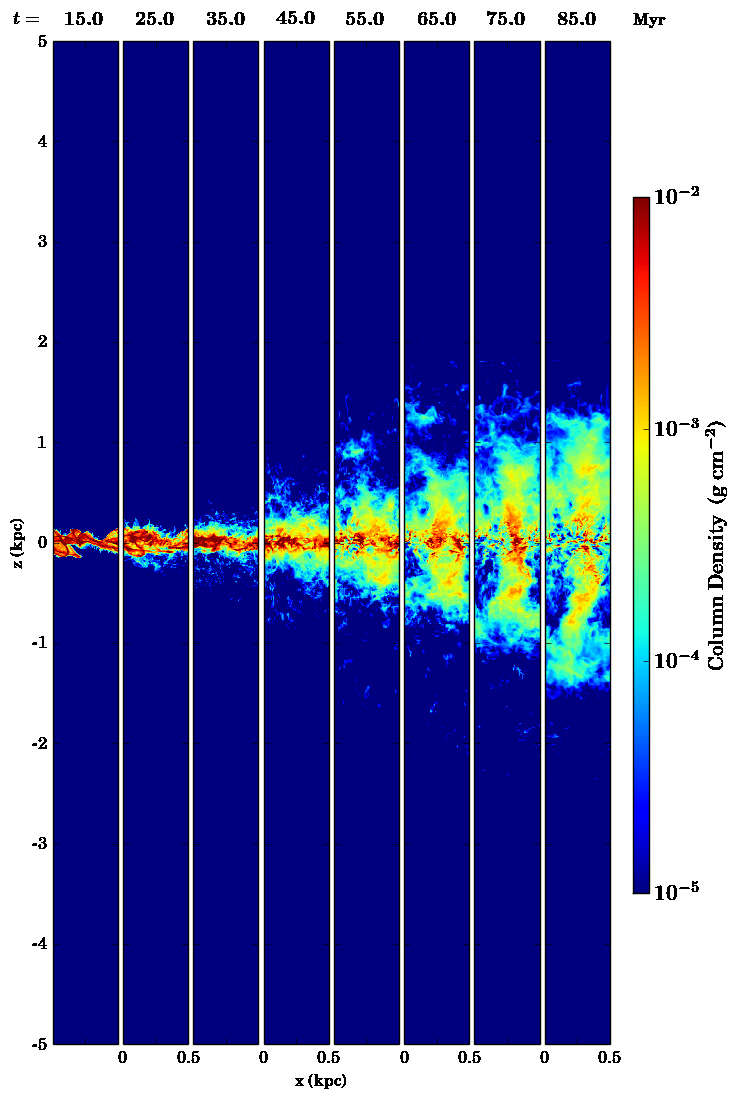} & 
 \includegraphics[width=0.5\textwidth]{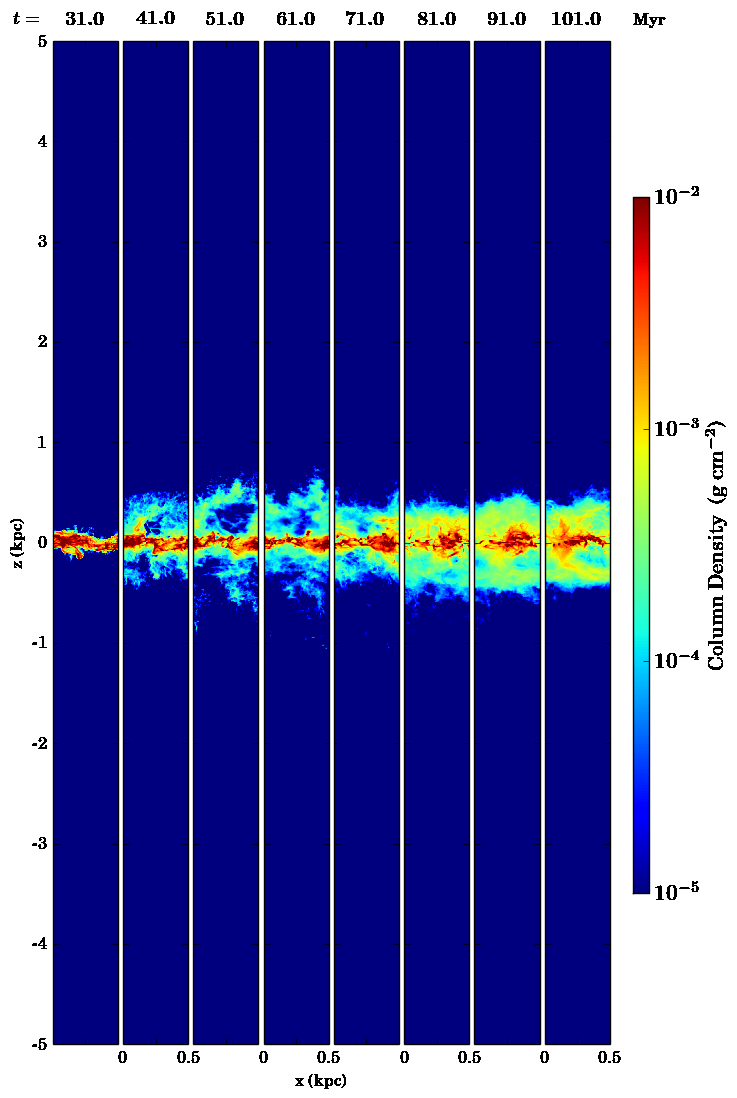}
 \end{tabular}
\caption{Same as Fig. \ref{fig:hill1} for simulations {\it FWSN-n1e3}
  (left panel) and {\it FWSN-n1e4} (right panel), both with stellar
  winds and supernovae. In these simulations the formation of cluster
  sink particles is enabled above $\rthr = 2\times 10^{-21} \;{\rm g\; cm}^{-3}$ and
  $\rthr = 2\times 10^{-20} \;{\rm g\; cm}^{-3}$, respectively. The star formation rate
  decreases with increasing $\rthr$ leading to smaller disc scale
  heights.}\label{fig:hill3}   
\end{figure*}

\section{Qualitative discussion of the simulations}\label{sec:results1}

In Fig. \ref{fig:snap1}, we give one example for the resulting temperature, density, and chemical structure of the
ISM\footnote{Movies of all simulations are available at \texttt{www.astro.uni-koeln.de/silcc}} for run {\it FWSN-n1e3} at $t=45$ Myr. From left to right we show a slice of the gas density at $y=0$ (top) and $z=0$ (bottom) and temperature followed by the column densities of all gas and the different species that we trace in the simulation, i.e. $\hp,\;\h,\ \htwo$, and CO, respectively. The filled, white circles show the position of the formed cluster sink particles. In this simulation the star formation rate is low and there are not many clusters. Similar figures for all simulations at comparable times ($t \approx t_\mathrm{sink,0} + 31$ Myr) after the formation of the first cluster at $t_\mathrm{sink,0}$ are shown in Appendix \ref{AppendixA}. The formation time of the first cluster in each simulation is listed in Table \ref{tab:tab2}. 

Run {\it FWSN-n1e4} features significant amounts of $\htwo$ and CO, which are organised in filamentary and clumpy structures near the disc mid-plane. The cluster sink particles form within the densest clumps and redistribute the surrounding gas by wind and supernova feedback. In particular stellar winds disperse the gas early during cluster formation and evolution. Supernovae heat the gas efficiently but their onset is delayed with respect to cluster formation.  

For comparison, we show the time evolution of the total gas column density for all six simulations (see Table \ref{tab:simlist}) in Figs. \ref{fig:hill1}, \ref{fig:hill2}, and \ref{fig:hill3}. At first we show the run without feedback for reference (Fig. \ref{fig:hill1}, left panel). The lack of pressure support from stellar feedback results in a compact configuration around the disc mid-plane. In the right panel we depict the evolution of run {\it FW-n1e2} with feedback from stellar winds, which are emitted by the massive stars within the forming stellar clusters. The cluster sinks are allowed to accrete throughout the simulation and a new massive star is formed every time a mass of $120 \mo$ has been accreted onto the cluster (see section \ref{sec:sub-grid}). We randomly assign a mass to each formed star (sampled from the high-mass stellar IMF), so most of the forming stars are B type stars and contribute only weak wind feedback which does not heat the gas efficiently.  

The disc scale height increases dramatically when supernova feedback is included (Fig. \ref{fig:hill2}). Here we note that the ISM in the run with only supernova feedback (left panel) appears to be more clumpy and structured than run {\it FWSN-n1e2} with wind and supernova feedback (right panel). Early feedback by stellar winds suppresses gas accretion onto young cluster sinks. Gas that is unbound by stellar wind feedback is available within the ISM, causing the ISM to be somewhat more diffuse. In particular, the outflowing gas is slightly colder in this simulation. 

Fig. \ref{fig:hill3} illustrates the impact of the sink density formation threshold, $\rthr$. For higher $\rthr$ the number of cluster sinks and hence the star formation rate is significantly reduced. Therefore we have less feedback, which results in a smaller disc scale height and less to no outflowing gas. In the following sections we discuss these findings quantitatively. 
  
\begin{figure*}
\begin{tabular}{ccc}
{\it NoF-n1e2} & {\it FW-n1e2} & {\it FSN-n1e2} \\
\includegraphics[trim = 0mm 0mm 0mm 10mm, clip, width=0.33\textwidth]{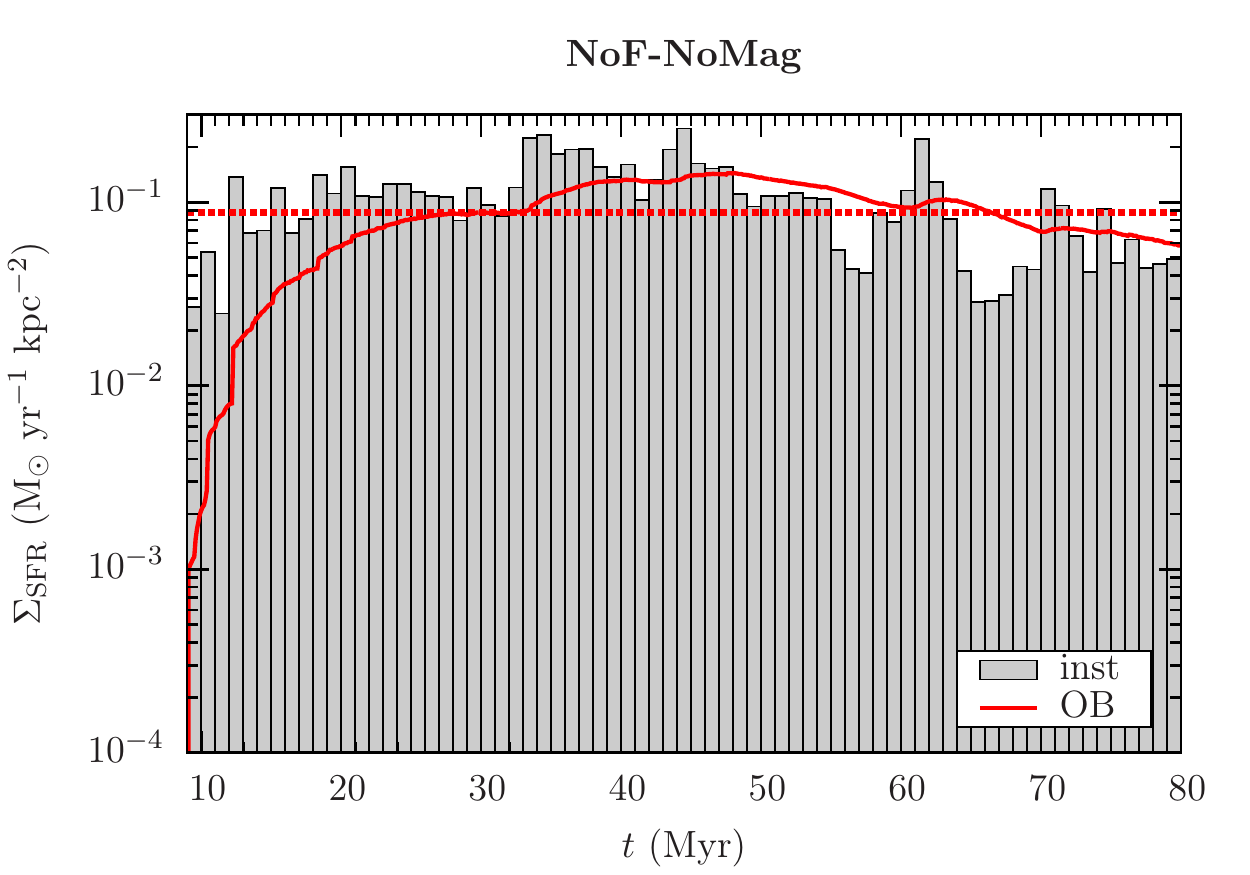} &
\includegraphics[trim = 0mm 0mm 0mm 10mm, clip,width=0.33\textwidth]{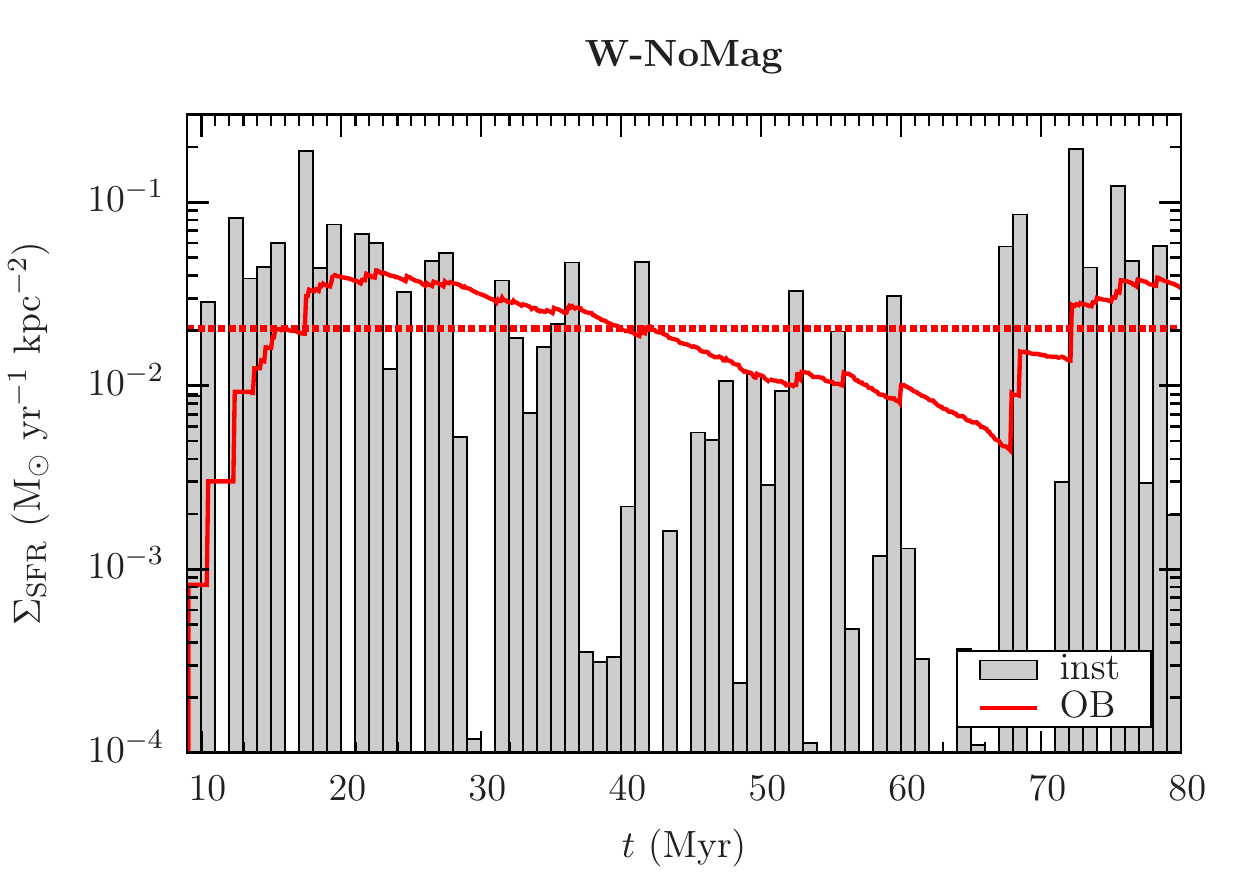} &
\includegraphics[trim = 0mm 0mm 0mm 10mm, clip,width=0.33\textwidth]{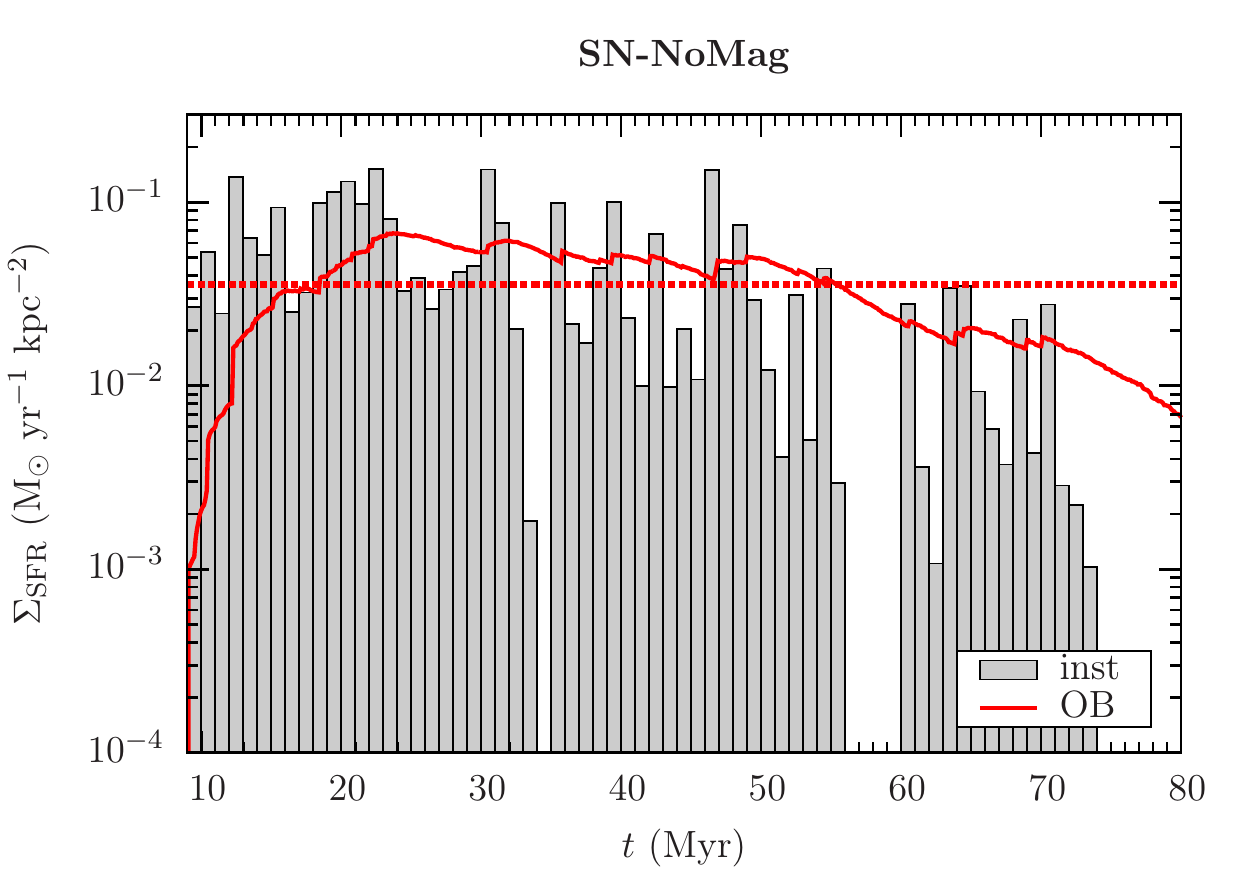}\\
{\it FWSN-n1e2} & {\it FWSN-n1e3} & {\it FWSN-n1e4} \\
\includegraphics[trim = 0mm 0mm 0mm 10mm, clip,width=0.33\textwidth]{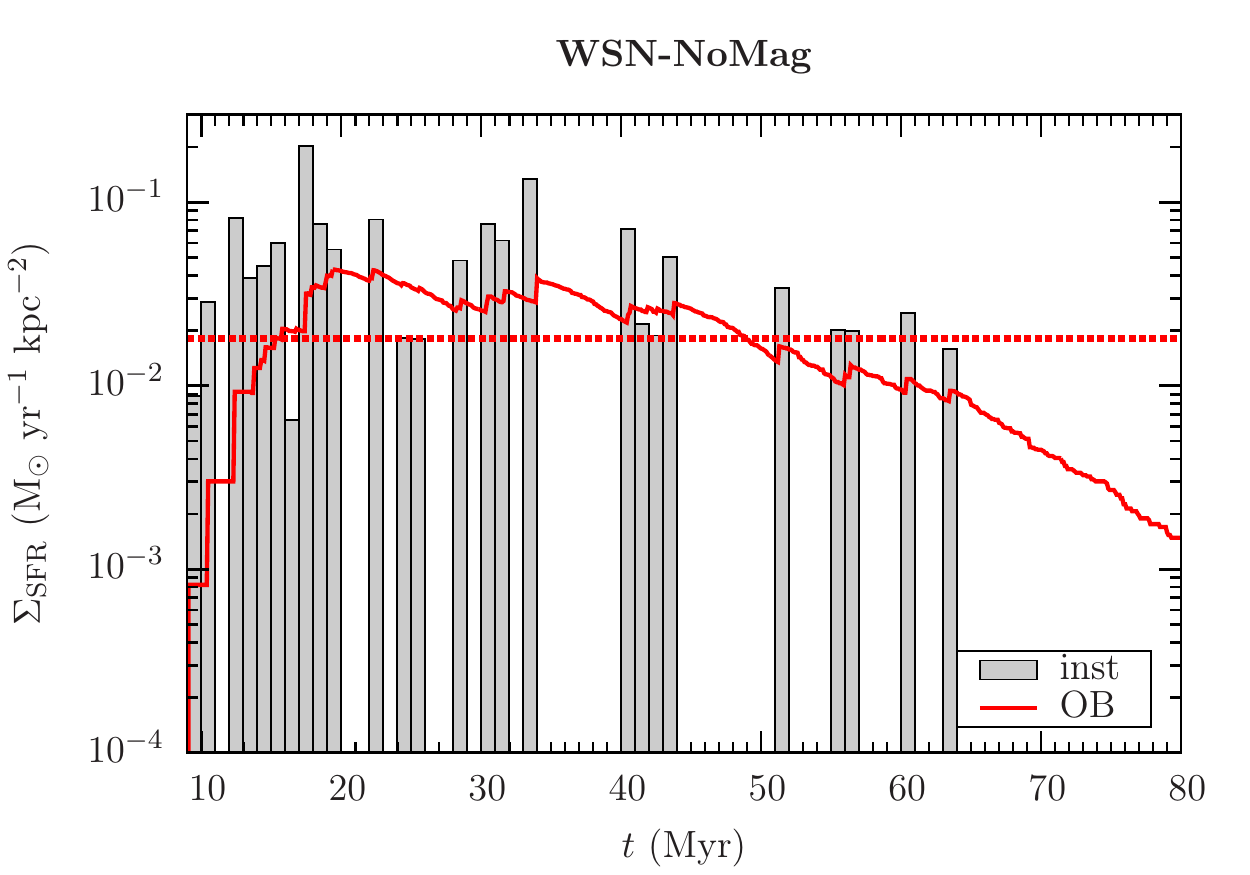} &
\includegraphics[trim = 0mm 0mm 0mm 10mm, clip,width=0.33\textwidth]{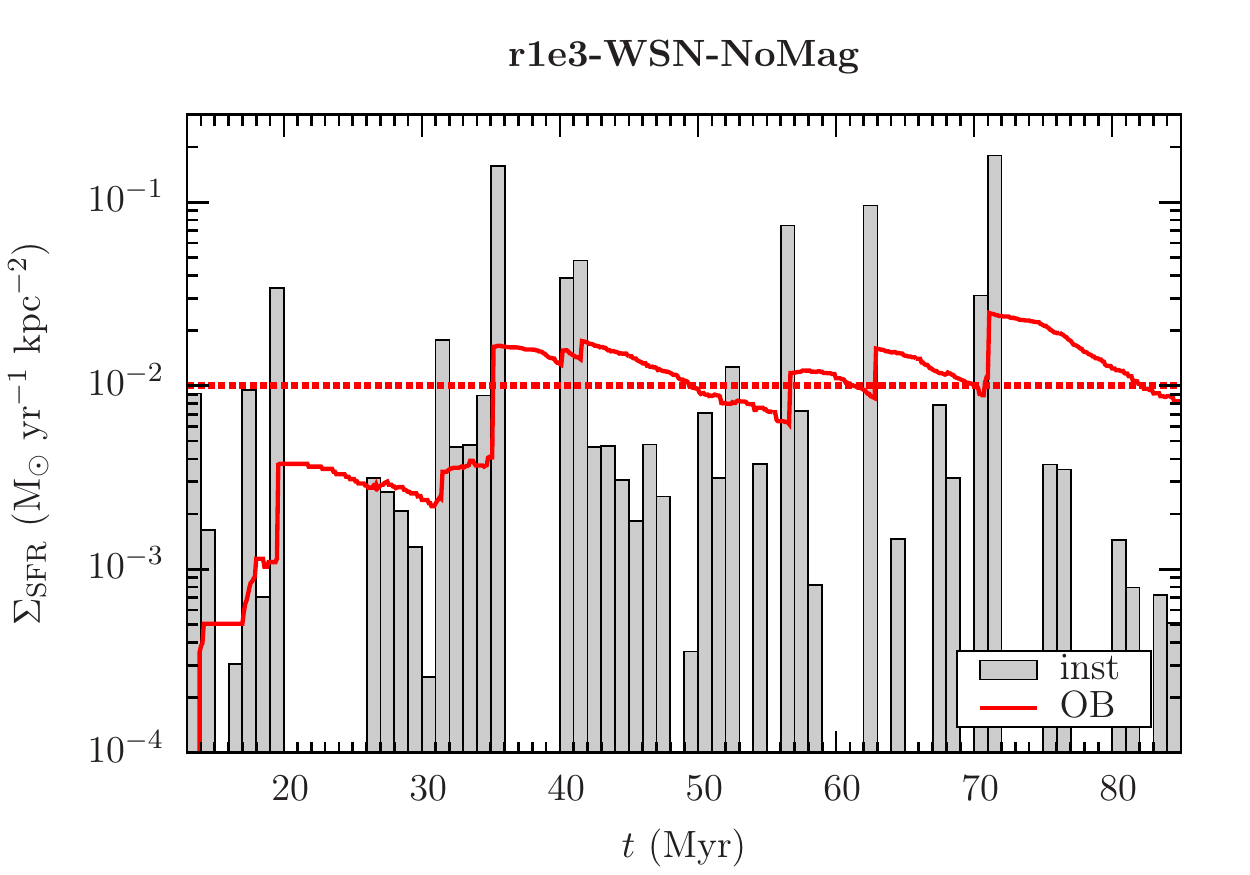} &
\includegraphics[trim = 0mm 0mm 0mm 10mm, clip,width=0.33\textwidth]{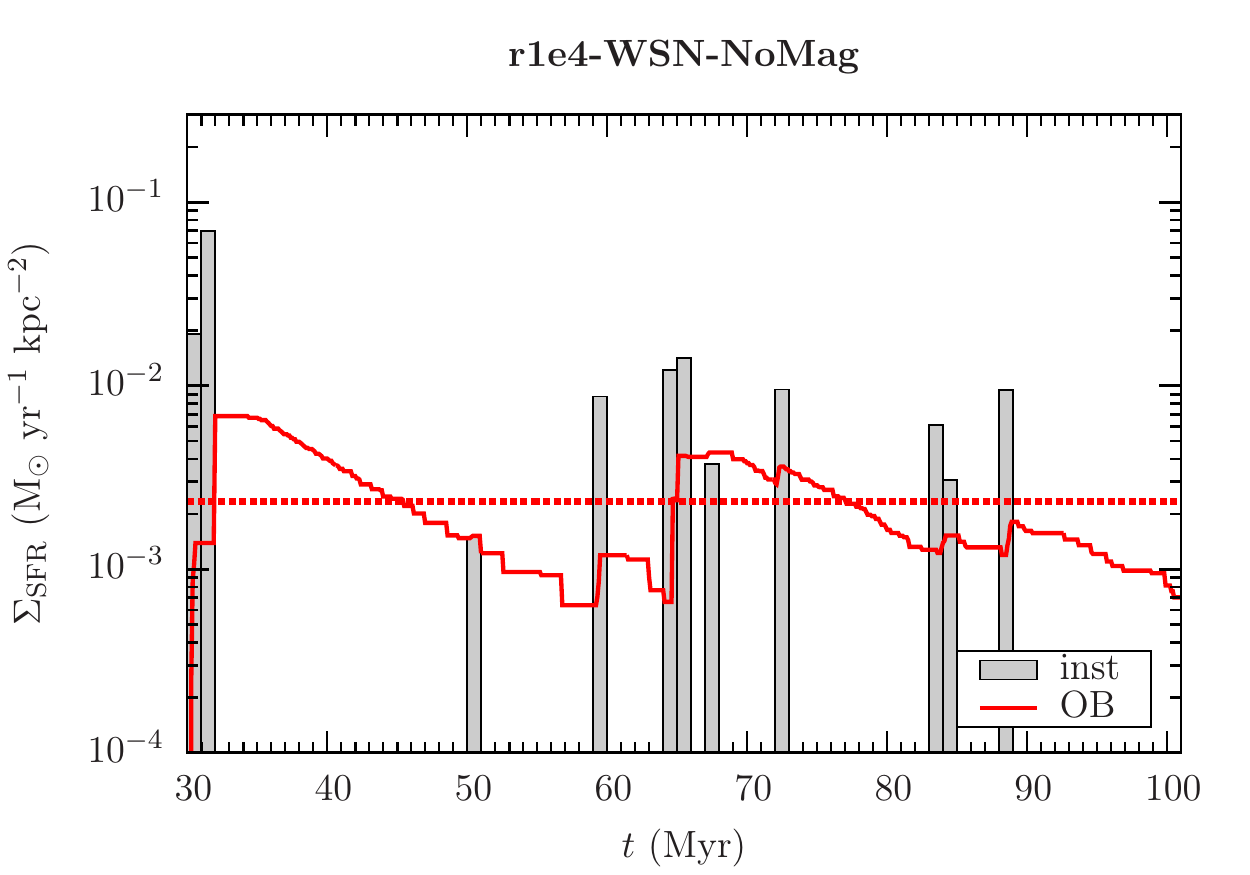}
\end{tabular}
\caption{Evolution of the $\sfrsd$ for all six simulations. The grey
  bins represent the instantaneous star formation rate surface
  densities, $\sfrsdinst$ (gas locked in sinks, see
  Eq. \ref{SFRinst}), while the red lines indicate the 'observed'
  values, $\sfrsdob$ as derived from the O- 
  and B type star lifetimes (see Eq. \ref{SFRob}). The horizontal, red
  dotted line is the average $\sfrsdob$ over 71 Myr of star formation 
  activity. In particular the wind feedback (the FW simulations)
  renders star formation more stochastic by early termination of
  cluster sink growth.}\label{fig:sfrsd}  
\end{figure*}
\begin{figure*}
\includegraphics[width=0.49\textwidth]{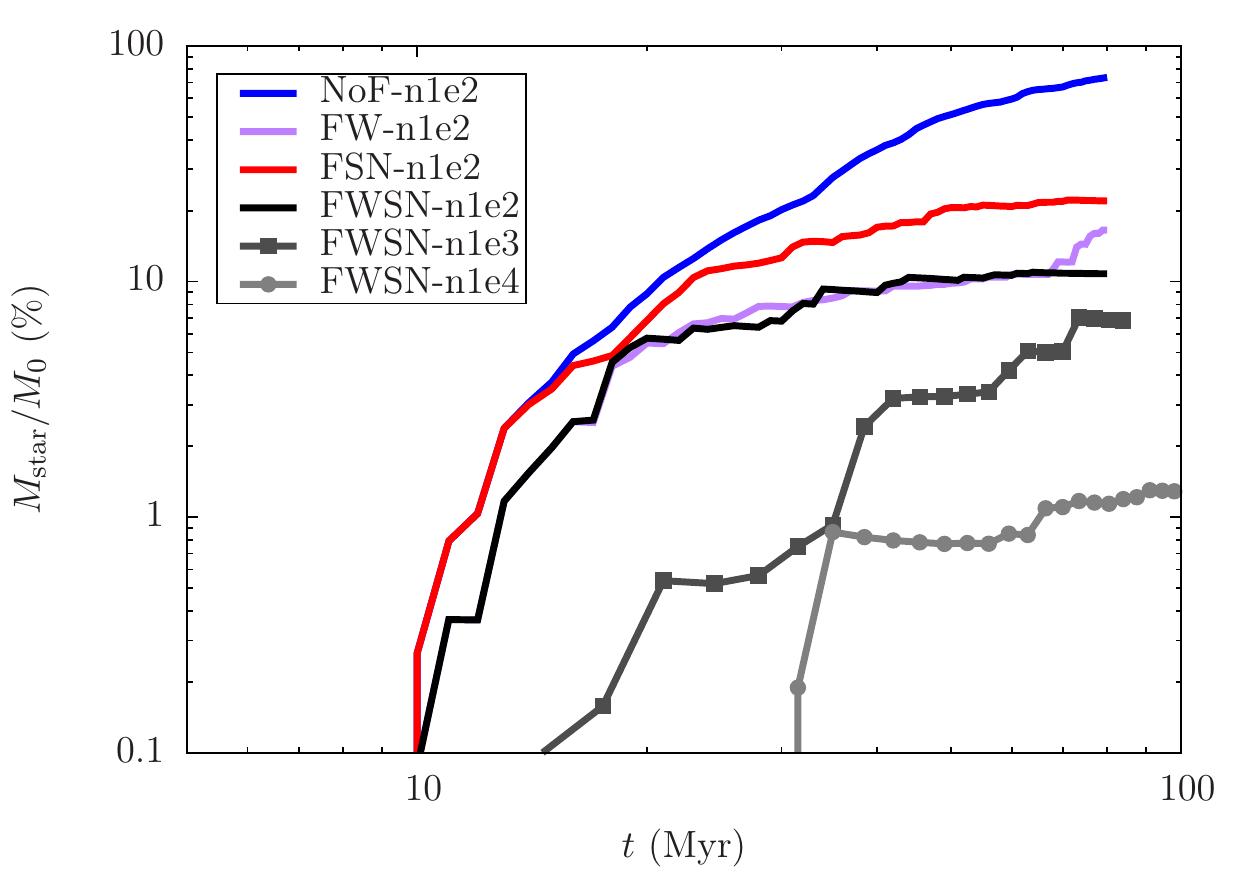}
\includegraphics[width=0.49\textwidth]{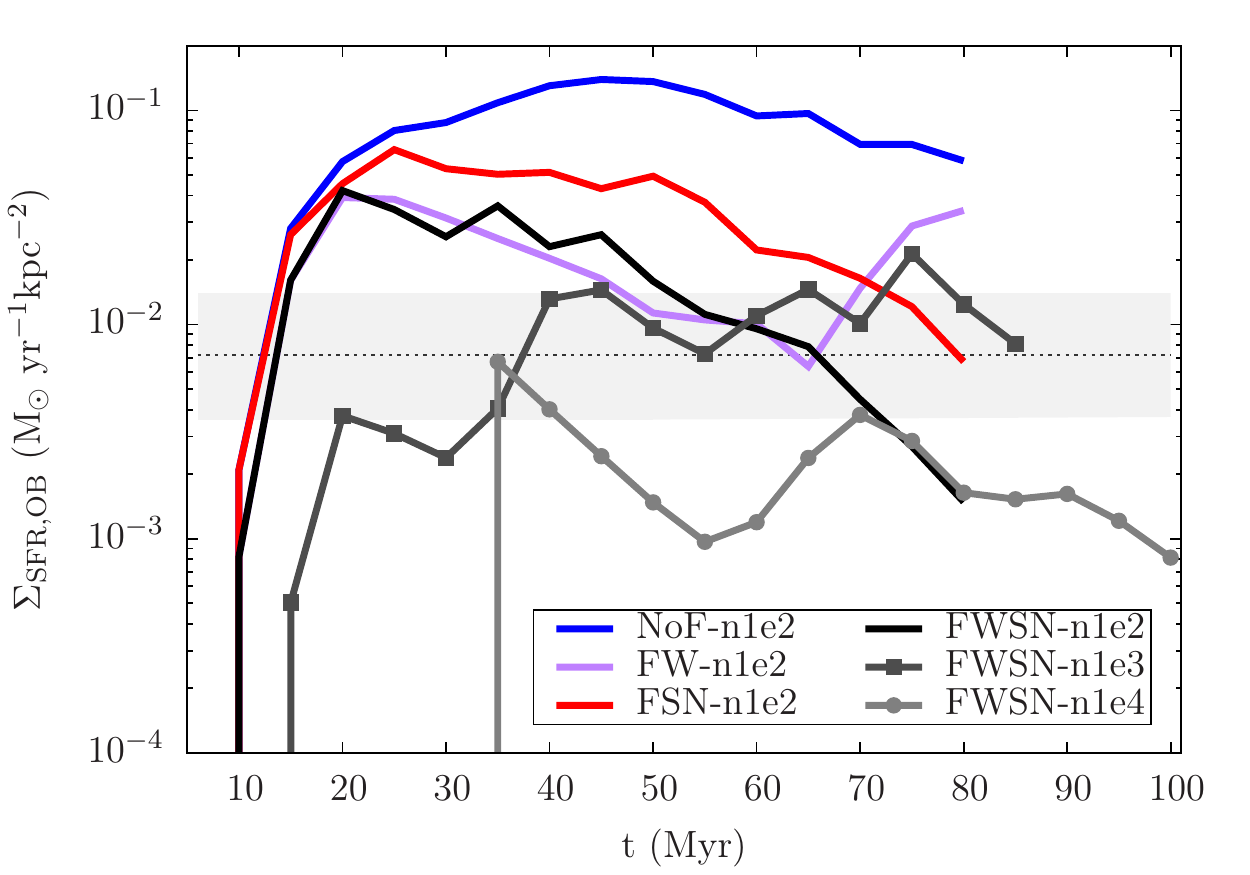}
 \caption{{\it Left panel:} Time evolution of the total mass in cluster sink particles, i.e. in stars, for the six simulations. The no feedback run (blue line) has the highest star formation rate and is shown for reference. Stellar winds suppress the accretion of gas onto the sinks immediately after the first massive stars form and efficiently limit sink formation (by at least a factor of two, see purple line), while supernova feedback is delayed and is therefore less efficient. Star formation is also reduced for higher sink density thresholds (grey lines). {\it Right panel:} Time evolution of the star formation rate surface density for the six simulations. The curves are slightly smoothed with respect to Fig. \ref{fig:sfrsd} to reduce noise. The grey dashed line indicates $\Sigma_{\rm SFR}$ as expected from the Kennicutt-Schmidt relation for $\Sigma_{\rm gas} = 10 \mopc$ and the grey band indicates a factor of two uncertainty. Only runs with stellar wind feedback lie in the observed range with the best fitting simulation being {\it FWSN-n1e3}. Supernova feedback alone is not efficient enough and acts too late to significantly limit the star formation rate (red line).}\label{fig:msink}  
\end{figure*}
\begin{figure}
 \includegraphics[width=0.5\textwidth]{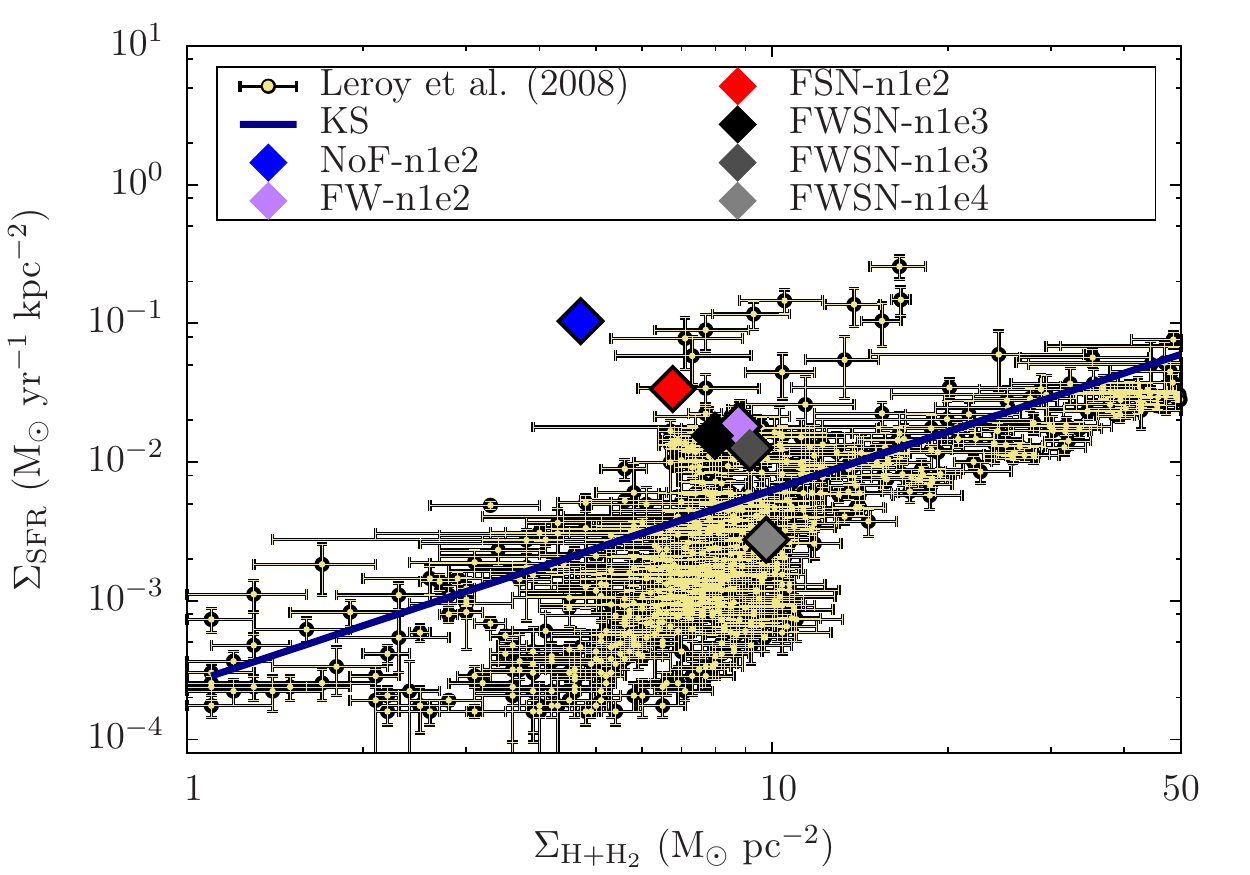}
\caption{Mean SFR surface density vs. mean total gas ($\h$ and $\htwo$) surface
  density for the different simulations. The means were computed between
  30 and 80 Myr. The simulations with wind feedback agree best
  with observations. The light yellow points with black contours show the observational data from
  \citet{Leroy+08}, while the dark blue line represents the KS relation.}\label{fig:obs} 
\end{figure}

\begin{figure*}
 \includegraphics[width=0.49\textwidth]{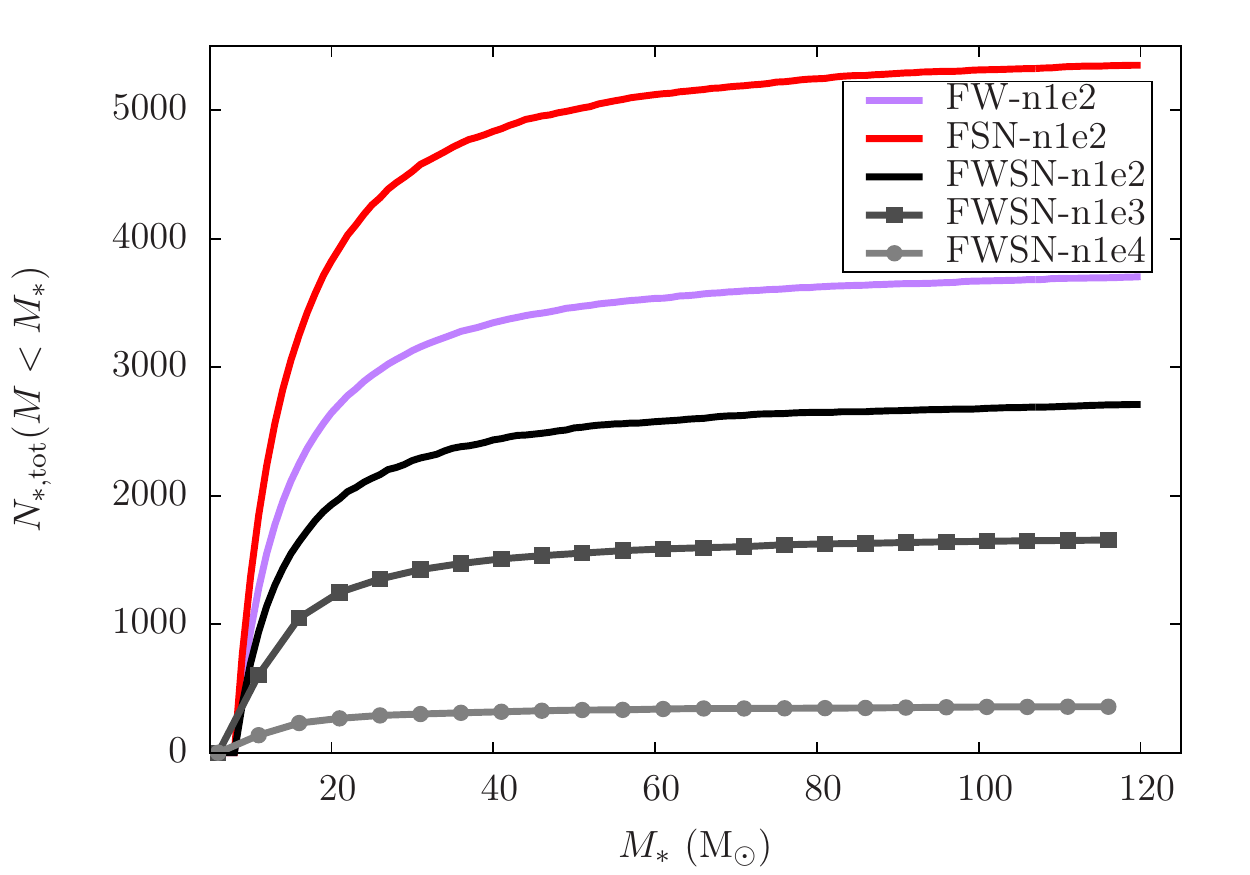}
 \includegraphics[width=0.49\textwidth]{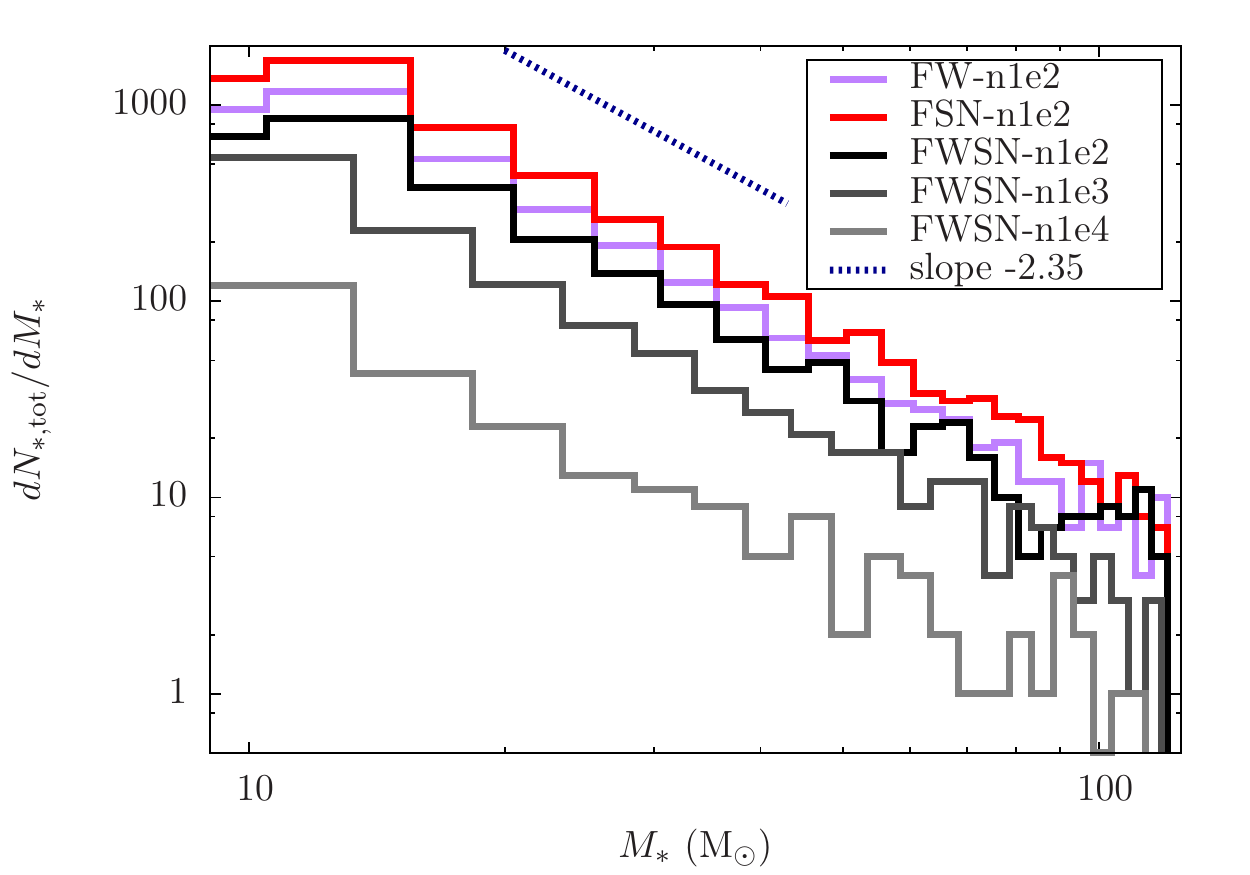}
 \includegraphics[width=0.49\textwidth]{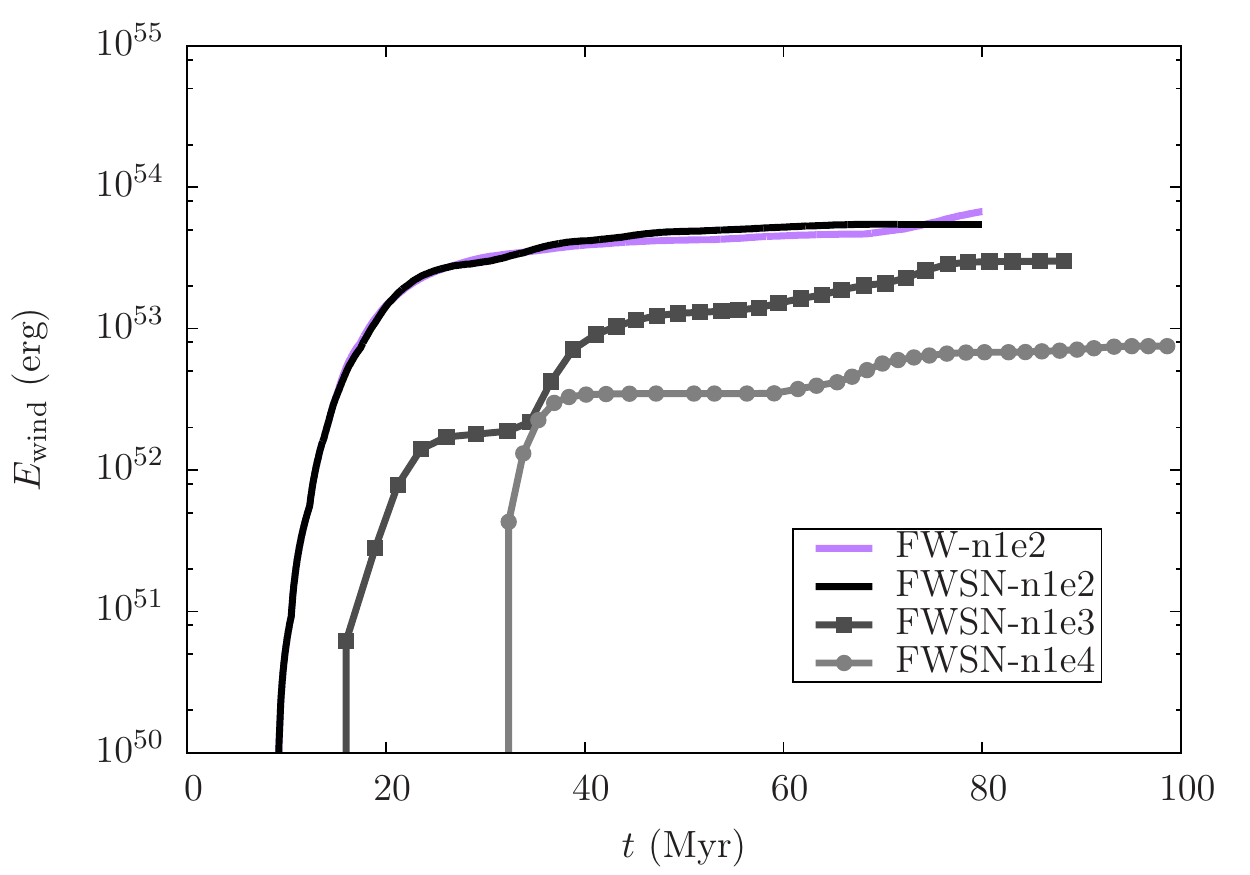}
 \includegraphics[width=0.49\textwidth]{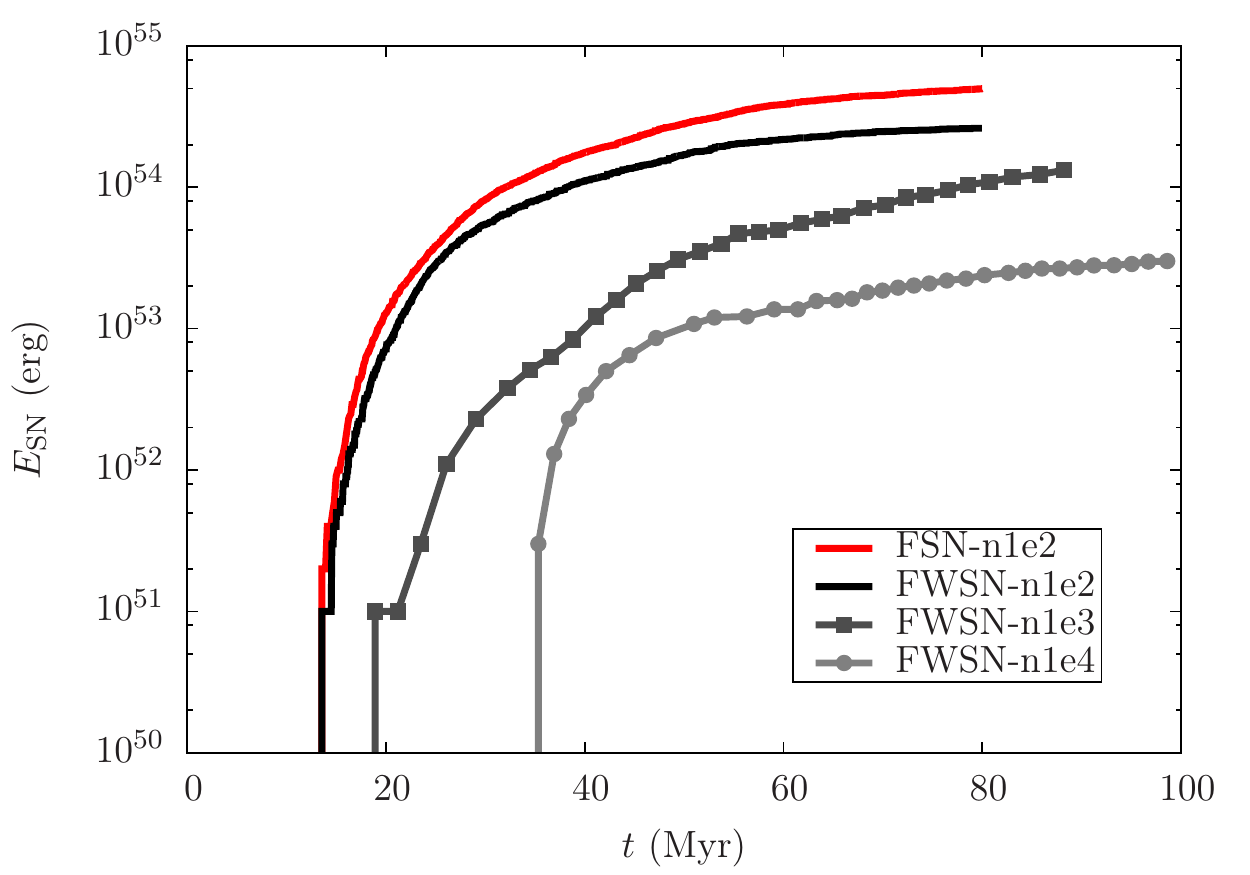}
 \caption{{\it Top left panel:} Cumulative distribution of all massive
   stars, $N_\mathrm{*, tot}$, that form in the simulations with feedback. Simulations with
   a higher mass in cluster sink particles also form a larger number of massive
   stars. {\it Top right panel:} The stellar IMF of the massive stars
   formed in these runs for a mass bin size of $5\;\mo$. All show a
   Salpeter slope (as indicated by the dotted, dark-blue line) modulo some noise from the random sampling. {\it
     Bottom panels:} Cumulative wind energy input (left) and supernova
   energy input (right) from all massive stars in the
   simulations. The energy input from SN explosions is only a factor
 of $\sim 3$ higher than from stellar winds. 
 }\label{fig:cmf}  
\end{figure*}

\begin{figure*}
 \includegraphics[width=0.49\textwidth]{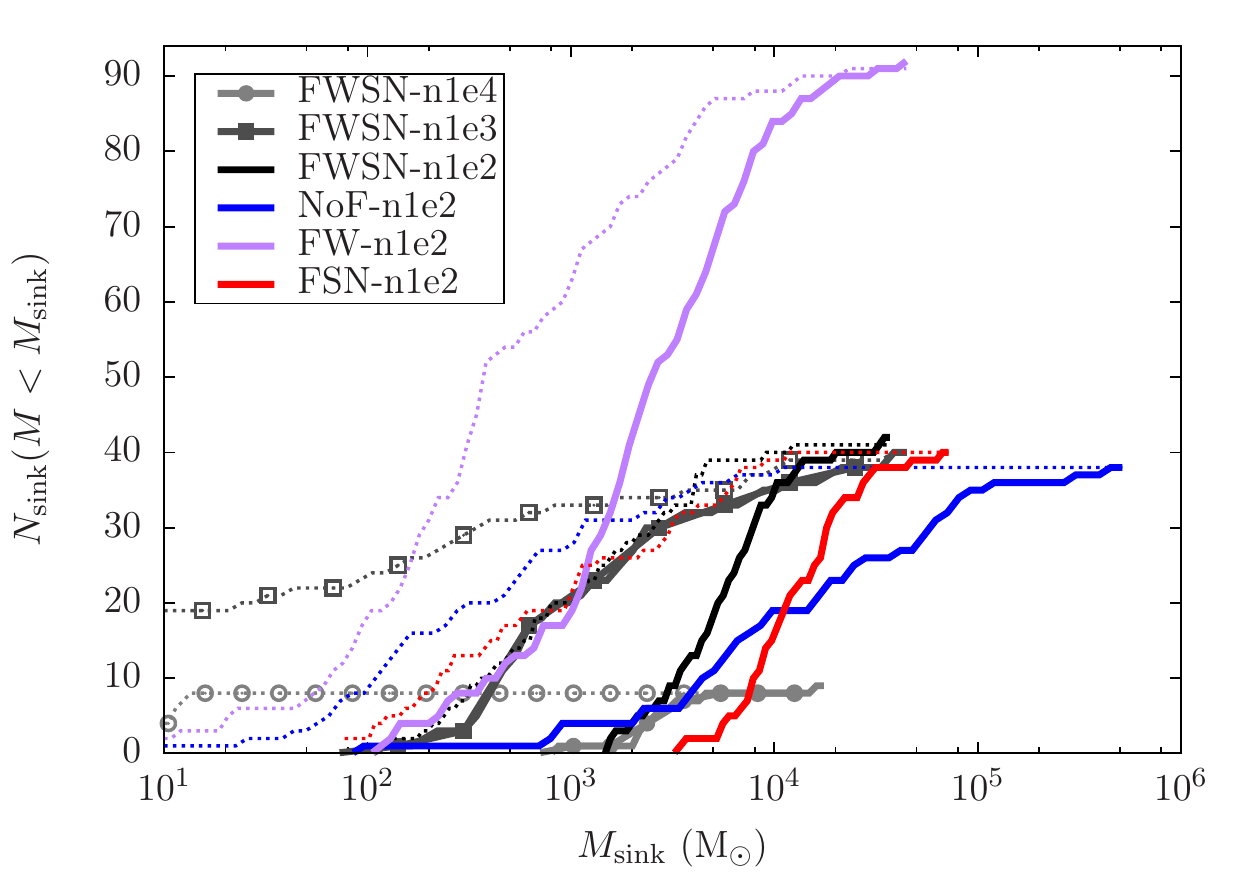}
 \includegraphics[width=0.49\textwidth]{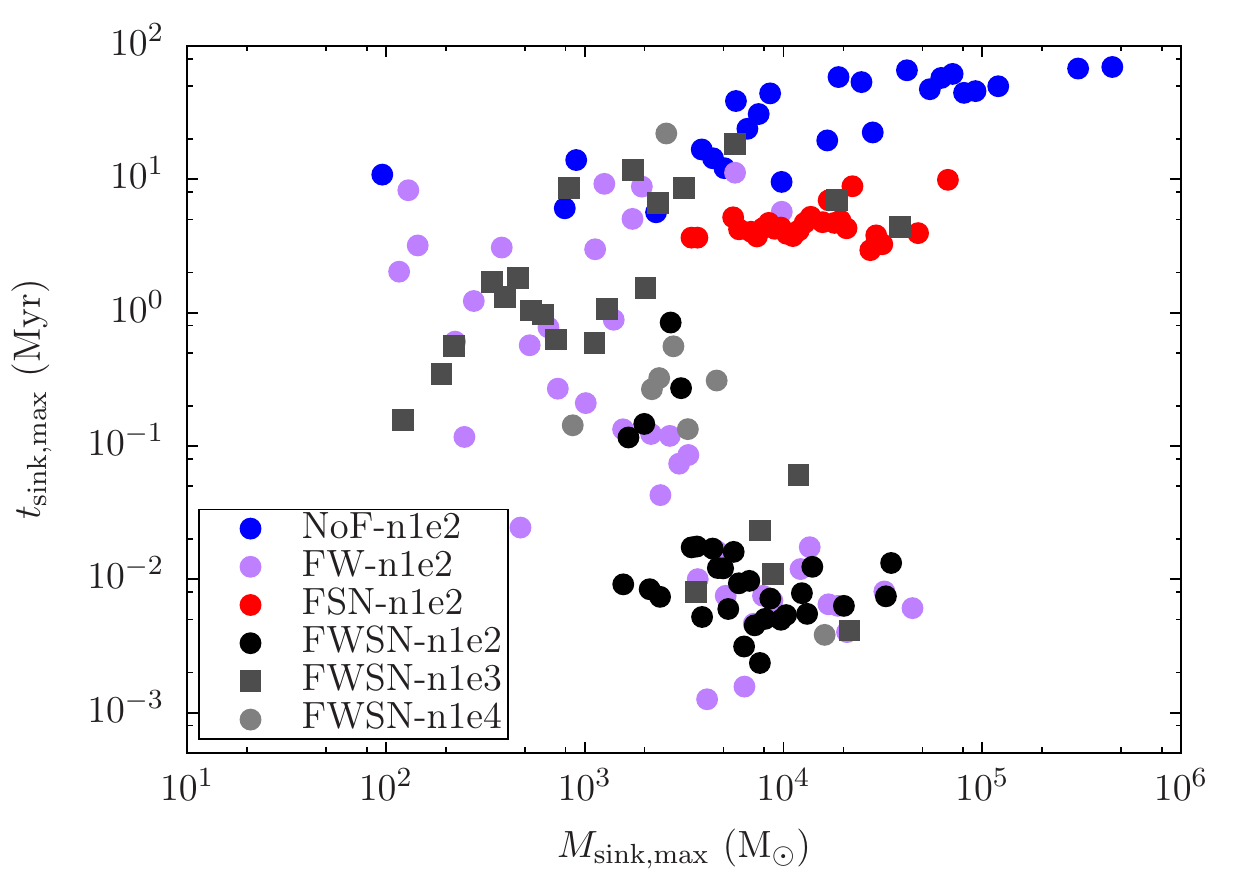}
 \caption{{\it Left panel:} Cumulative mass distribution for all
   cluster sinks in the simulations at the time of their formation (dotted
   lines) and at their maximum mass (solid lines). There is
   substantial gas accretion onto most of the cluster sink
   particles, which causes the distributions (dotted vs. solid at the same colour) 
   to shift towards higher masses. At a given threshold density supernovae (red line) reduce the
   cluster mass range but not their number (compare to the run without feedback,
   blue line). Wind feedback in addition reduces the cluster masses
   even more (black line). Wind feedback alone results in the
   formation of more clusters at lower masses (purple line). {\it
     Right panel:} Cluster growth timescales - the time it takes each
   sink particle to reach its maximum mass, $t_\mathrm{sink, max}$,
   plotted against the respective maximum sink mass. Runs with wind
   feedback form a  population of lower mass sinks with short accretion times, whereas
   runs without wind feedback (blue and red dots) accrete for longer
   and assemble higher maximum masses. }\label{fig:cmf2}  
\end{figure*}



\section{How feedback regulates star cluster formation}\label{sec:results2}
The presented set of simulations allows us to determine the relative importance of stellar wind and supernova feedback in terms of regulating the star formation rate in the simulated portions of the galactic disc.   

\subsection{Mass evolution and star formation rates}
\label{sec:massevol}
There are different ways to measure the star formation rate (SFR) surface density in our simulations. Naively, one could just count how much gas is collapsing into sink particles (representing the stellar population) within a given time bin $\Delta t$. We call this the {\it instantaneous SFR}, $\sfrsdinst$, which is computed as 
\begin{equation} \label{SFRinst}
\noindent\sfrsdinst(t) = \frac{1}{A}
\sum_{j=1}^{N_{\mathrm{sink}}}\dot{M}_{{\rm sink},j} (\Delta
t)~\;[\mo~{\rm yr}^{-1}~{\rm kpc}^{-2}], 
\end{equation}
for $t-\frac{\Delta t}{2}<t<t+\frac{\Delta t}{2}$ and the area of the computational domain in the disc mid-plane $A=(0.5\;{\rm kpc})^2$. 

However, the SFR derived in this way depends on $\Delta t$ and is not directly comparable to the SFR an observer would measure, e.g. when tracing the SFR with H$_\alpha$ emission. The H$_\alpha$ emission sensitively depends on the presence of OB and WR-stars, which have short lifetimes of $\sim 5 - 40$ Myr. Since we follow every massive star, $i$, in our simulation (1 massive star is formed for each 120 $\mo$ of gas that is turned into stars), we can use the current number of massive stars and their respective lifetime, $t_{{\rm OB},i}$, to estimate an {\it observable SFR surface density}, $\Sigma_{\sfrob}$, as 
\begin{equation} \label{SFRob}
\Sigma_{\sfrob}(t) = \frac{1}{A}\sum_{i=1}^{N_\mathrm{*}} \dfrac{120 \mo}{t_{{\rm OB},i}}\ ,
\end{equation}
for $t_{{\rm form},i}<t<t_{{\rm form},i}+t_{{\rm OB},i}$, where
$t_{{\rm form},i}$ is the formation time of  massive star $i$ and $N_\mathrm{*}$ is the number of 'active' massive stars at time t.  

In Fig. \ref{fig:sfrsd}, we show $\sfrsdinst$ for $\Delta t = 1$ Myr (grey bars) and $\Sigma_{\sfrob}$ (red lines), as well as the average $\Sigma_{\sfrob}$ (red dotted line) for the different simulations (different panels). A bin size of $\Delta t = 1$ Myr corresponds to $\sim 1000$ time steps in the simulations with feedback (the typical time step is $\sim 10^3$ yr). Young star clusters (with ages $\lesssim 5-10$ Myr, i.e. before the first SN explodes) have high accretion rates and contribute most to $\Sigma_{\sfrinst}$. We find that $\sfrsdinst$ becomes more bursty in the presence of stellar winds (e.g. {\it NoF-n1e2} vs. {\it FW-n1e2}) which truncate cluster growth, as well as for a lower total amount of star formation per unit area, $\Sigma_{\mathrm{SF,tot}} = \int_0^{t_{\rm stop}}\sfrsdinst {\rm d}t $. Overall the O- and B-type star lifetimes are still long enough to hide the time variation from an observer, who would measure $\Sigma_{\sfrob}$. The variation is significant and thus $\Sigma_{\sfrob}$ can be orders above the current star formation rate as well as up to a factor of 10 below it.   \\

In Fig. \ref{fig:msink} (left panel), we show the total mass in cluster sink particles as a function of time for all six simulations. In run {\it NoF-n1e2} most of the gas ($\sim$ 80\%) has collapsed into sinks by $t=80$ Myr, followed by run {\it FSN-n1e2} with $\sim$ 20\% in sinks. Until the very end of the simulation, runs {\it FW-n1e2} and {\it FWSN-n1e2} evolve similarly and $\sim$ 10\% of the gas is converted into sinks. This shows that stellar wind feedback efficiently regulates star formation right after the first massive star was born. Supernova feedback acts with a time delay and therefore allows for more star formation. Wind and supernova feedback together closely follow the case of only wind feedback because the gas that would be available to accrete onto formed sinks is already unbound by the stellar winds and supernovae have little additional effect. We note that this result might not be generally applicable with increasing gas surface density in the disc. 
When the sink density threshold is increased, the mass in sinks decreases to $\sim$ 1\% for {\it FWSN-n1e4}. In addition, star formation starts later in these simulations and we have therefore run them for longer (see $t_\mathrm{sink,0}$ as listed in table \ref{tab:tab2}).   

The later onset of star formation in simulations with higher $\nthr$ can be compared with the free-fall time at the given $\nthr$, $\tau_{\rm ff} = (3 \pi / (32 G m_p \nthr))^{1/2}$. For example, we have $\tau_{\rm ff}(\nthr = 10^2\;{\rm cm}^{-3}) \approx 5$ Myr, while we assume that star formation proceeds instantaneously within the cluster sinks formed at this density. In 5 Myr, the gas has quite some time to move around (a typical turbulent velocity of 10 km s$^{-1}$ roughly corresponds to 10 pc Myr$^{-1}$ and hence a distance of 50 pc can easily be crossed) and may not be accreted onto a cluster sink, which is introduced at a higher $\nthr$. In addition, fewer cells are filled with higher density gas in a turbulent environment (consider a lognormal structure of the volume-weighted density PDF) and therefore the sink formation becomes more stochastic as fewer cells meet the density formation criterion for higher $\nthr$. Formally, the high density thresholds are unresolved with respect to e.g. the Truelove criterion (see section \ref{sec:sinks}).

In Fig. \ref{fig:msink} (right panel), we show $\Sigma_{\sfrob}$ as a function of time for all simulations. The horizontal, grey, dashed line shows the SFR surface density corresponding to the Kennicutt-Schmidt value at $\Sigma_\mathrm{gas}=10 \mopc$ and the light grey band indicates an uncertainty of a factor of 2. Clearly, runs without feedback or with supernova feedback alone have too high $\Sigma_{\sfrob}$, while the value found for run {\it FWSN-n1e4} is a bit low. We note that runs with higher threshold densities ({\it FWSN-n1e3} and {\it FWSN-n1e4}) have relatively flat star formation rates and are missing the initial peak.     

We place our simulation results on the familiar Kennicutt-Schmidt diagram \citep{Kennicutt98a} in Fig. \ref{fig:obs}. Here, we plot the derived average surface mass density in atomic plus molecular hydrogen, $\Sigma_\mathrm{H+H_2}$, against the average $\Sigma_{\sfrob}$, where the averages were computed between $t_\mathrm{sink,0}$ and $t_\mathrm{stop}$. We also show the observations of 23 (11 dwarfs and 12 large spirals) nearby {\it normal} star-forming galaxies by \cite{Leroy+08} (yellow points). These are composed of hundreds of radial profiles of $\sfrsd$, $\Sigma_{\h}$ and $\Sigma_{\htwo}$ (only for spirals) at 800 pc (spirals) and 400 pc (dwarf) resolution. We multiply their SFRs by a factor of 1.59 in order to rescale them from a \cite{Kroupa01} to a \cite{Salpeter55} IMF. The thin, blue line again indicates the standard KS relation \citep[as in Fig. \ref{fig:msink}; ][]{Kennicutt98a}: 
\begin{equation} \label{KS}
\frac{\sfrsdks}{\mo \yr^{-1} \kpc^{-2}}=2.5 \times 10^{-4}\ \biggl(\frac{\Sigma_{\h+\htwo}}{\mo \pc^{-2}} \biggr)^{1.4}\ .
\end{equation}
Simulations without stellar wind feedback result in a $\sfrsd$  that is too high and do not agree well with observations. 


In Fig. \ref{fig:mf} (Appendix B) we show the corresponding time evolution of the total gas mass (top left panel), and of the mass fractions of atomic hydrogen (top right), ionized hydrogen (bottom left), and molecular hydrogen (bottom right), all normalised to the total gas mass at $t=0$, $M_0$ (see section \ref{sec:initial}). The total gas mass evolution is complementary to the sink mass evolution.  

\begin{table*}
\begin{tabular}{lccccccc}
Run name & $t_\mathrm{sink,0}$ & $t_\mathrm{stop}$ & N$_\mathrm{*, tot}$ &$\dot{N}_\mathrm{SN}$ & $M_\mathrm{sink, max}$ & $M_\mathrm{sink, med}$ & $<t_\mathrm{sink, max}>_{_\mathrm{log}}$  \\
& [Myr] & [Myr] & &[Myr$^{-1}$]& [$\mo$] &[$\mo$] &[Myr] \\
\hline
{\it NoF-n1e2} & 9.06 & 80.0 & 14082 & -- & $5.1 \times 10^5$ &$9.8 \times 10^3$ & 27.9\\
{\it FW-n1e2} & 9.06 & 80.0 & 3705 & -- & $4.4 \times 10^4$ &$2.0 \times 10^3$& 0.12 \\
{\it FSN-n1e2} & 9.06 & 80.0 & 5350 & 70.2 &$7.2 \times 10^4$&$1.2 \times 10^4$& 4.51\\
{\it FWSN-n1e2} & 9.06 & 80.0 & 2710 & 36.9 &$3.7 \times 10^4$& $5.6 \times 10^3$& 0.01\\
{\it FWSN-n1e3} &  13.9 & 85.0 & 1656 & 17.2 & $3.9 \times 10^4$&  $8.5 \times 10^2$& 0.69\\
{\it FWSN-n1e4} &   30.3 & 101.0 & 358 & 4.5 & $1.8 \times 10^4$& $2.5 \times 10^3$& 0.26\\
\hline
\end{tabular}
\caption{For each simulation (column 1) we list the time at which the
  first star cluster forms (column 2), the time at which we stop the
  simulation (column 3), and the total number of massive stars formed
  (column 4; see Fig. \ref{fig:cmf}). In column 5 we list the average supernova rate per
  Myr, where we average over $t_\mathrm{stop} - t_\mathrm{sink,0}$. 
  In column 6 and 7 we give the maximum and the median of
  the cluster sink mass distribution (see Fig. \ref{fig:cmf2}), and in column 8 we list the logarithmic mean accretion time of all clusters in the respective simulations. }\label{tab:tab2} 
\end{table*}
\subsection{Regulation of star formation by stellar winds}
\label{sec:star-formation}
Depending on the simulation, our cluster-sink sub-grid model results in a population of a few $10^2-10^3$ massive stars (see Fig. \ref{fig:cmf}, top left panel). Simulations with higher overall star formation rates also form more massive stars. With a few thousand massive stars, we achieve a good random sampling of the IMF for massive stars with $M_{*} \lesssim 80 \mo$. Therefore, the different simulations give the same slope of the IMF but a different $y-$axis offset (see top right panel of Fig. \ref{fig:cmf}, where we plot the massive star IMF using a bin size of $5\mo$). Since the slope of the IMF is very steep, we only form a small number of very massive stars ($\lesssim 10$ stars per bin at $\sim 100\mo$). Due to the low number statistics in the highest mass bins, all simulations with $\nthr=10^2\;{\rm cm}^{-3}$ have comparable numbers of very massive stars. Runs with higher $\nthr=10^3 - 10^4\;{\rm cm}^{-3}$ form fewer stars and consequently have fewer very high mass stars.   

In the lower panels of Fig. \ref{fig:cmf}, we show the cumulative energy input from stellar winds (left panel) and from supernovae (right panel). In runs with winds and supernovae, the supernova energy input is only a factor of $\sim 2.5 - 3$ larger than the cumulative wind energy input. Note that this applies for the solar metallicity case and that the ratio might be different in lower metallicity environments as lower metallicity stars have, during most of their evolution,
winds with lower mass-loss rates \citep[e.g.,][]{Kudritzki1987,Vink2001,Krticka2006,Mokiem2007,Grafener2008} and somewhat lower terminal velocities \citep{Leitherer1992, Krticka2006}. Although the very massive stars are so rare and only live for a very short time, they are the ones which contribute the most wind energy (see cumulative wind energy input shown in Fig. \ref{fig:tracks}). This renders the wind energy input to be quite stochastic for individual star-cluster forming regions, depending on the masses of the individual very massive stars. 

To quantitatively assess the differential impact of stellar winds and supernovae on star cluster formation, we investigate the accretion history of the forming cluster sink particles in Fig. \ref{fig:cmf2}. In the left panel, we show the cumulative mass distribution of all cluster sinks at the time of formation (thin dotted lines) and at their maximum mass (solid lines). We define the {\it cluster formation time}, $t_\mathrm{sink, max}$, as the time it takes each cluster sink to reach its maximum mass. The cumulative distribution of maximum cluster masses is shifted to higher masses for all simulations, which indicates that a significant amount of mass is gained by gas accretion. This subsequent gas accretion tends to steepen the cumulative mass distributions (in  the left panel of Fig. \ref{fig:cmf2}, the cumulative distributions of the maximum masses (solid lines) are steeper than the corresponding cumulative distributions of the initial masses shown by dotted lines), which means that the variance of the actual mass distributions, which have an approximately log-normal shape (the corresponding cumulative distributions can be represented with an error function), decreases with time.  

For runs {\it NoF-n1e2}, {\it FSN-n1e2}, and {\it FWSN-n1e2}, the number of formed cluster sinks is comparable ($\sim 40$) and also the initial cluster sink mass distributions are very similar. However, the maximum mass distributions are different, since the runs with more feedback subsequently accrete less mass. Therefore, the run without feedback forms the most massive clusters with up to $M_\mathrm{sink, max}\sim 5\times 10^5\mo$ and a median mass of $M_\mathrm{sink, med} \sim 10^4\mo$, the run with only supernova feedback forms somewhat lower mass clusters with up to $M_\mathrm{sink, max} \sim 7\times 10^4\mo$ and a median mass of $M_\mathrm{sink, med} \sim 10^4 \mo$, and the run with supernova and stellar wind feedback forms even lower mass clusters with a maximum mass of up to $M_\mathrm{sink, max}  \sim 3.7\times 10^4\mo$ and a median of $M_\mathrm{sink, med} \sim 5.6 \times 10^3\mo$.  

Interestingly, run {\it FW-n1e2} with just winds forms approximately twice as many cluster sinks, where most of the additional clusters have low masses ($\lesssim 2\times 10^3\mo$) and therefore do not contribute significantly to the total mass in cluster sinks. For higher sink density thresholds with wind and supernova feedback, the star formation rate is lower and  fewer clusters form in case of run {\it FWSN-n1e4}. The maximum masses are comparable to run {\it FWSN-n1e2} but the median masses are somewhat lower with $\sim 0.85 - 2.5 \times 10^3 \mo$ (we list $M_\mathrm{sink, max}$ and $M_\mathrm{sink, med}$ for all clusters in table \ref{tab:tab2}). For comparison, in the Milky Way there are only a handful of known star clusters with masses above $2\times 10^4\mo$ \citep[see e.g.][and references therein]{Piskunov2008, Fujii2016} and therefore the run without stellar feedback is in clear disagreement with observations. The runs including stellar wind feedback show the best agreement with the solar neighbourhood observations of young star clusters \citep{Lada2003}, where the solar neighbourhood motivates our initial conditions.   

In the right panel of Fig. \ref{fig:cmf2} we plot the cluster formation time $t_\mathrm{sink, max}$ as a function of maximum cluster mass. This is equivalent to the time scale on which a cluster accretes gas efficiently. Clearly, runs without stellar wind feedback accrete for a long time, from 5 Myr up to 71 Myr, which is the maximum possible
accretion time for run {\it NoF-n1e2}, $t_\mathrm{sink, max} = t_\mathrm{stop} - t_\mathrm{sink,0}$, where $t_\mathrm{stop}=80$ Myr and $t_\mathrm{sink,0} \approx 9$ Myr for the first cluster sink particle. For run {\it FSN-n1e2} (red points), $t_\mathrm{sink, max} \sim 4.51$ typically corresponds to the life time of the most massive star that first explodes as a supernova, i.e. the minimum supernova delay time. Overall, these clusters all accrete for about the same but relatively long timescale of $\sim 5$ Myr and thus, all become quite massive. All clusters have masses above $\sim 10^{3.3} \mo$. \\
On the other hand, all runs with stellar winds in addition to supernovae show a qualitatively different trend. The most massive clusters grow on the shortest timescales $\sim 10^4$ yr, much shorter than the shortest stellar lifetimes. Here the winds from the forming massive stars efficiently clear out the local environments and thus, terminate the gas accretion onto the cluster. This process is more efficient for more massive clusters and we see a clear anti-correlation between cluster formation time and cluster mass, ranging from $\sim 10^6$ yr for clusters with $M_\mathrm{sink, max} \sim 10^{2.7} \mo$ to $\sim 10^4$ yr for $M_\mathrm{sink, max} \sim 10^{4} \mo$. We note that the quoted values for the cluster accretion times are most likely underestimates, since we do not consider the dynamical evolution of gas within the sink particle.
The typical accretion rates onto the cluster sinks are $10^{-3} -10^{-2} \mo\;{\rm yr}^{-1}$. In case of winds, this is only valid within approximately the first Myr. The logarithmic mean of $t_\mathrm{sink, max}$ is also given in table \ref{tab:tab2}. All our models with stellar feedback are in agreement with the idea of the rapid removal of gas from the clusters on time scales $\lesssim 10-30$ Myr, which is observed in star clusters in the Milky Way (\citealt{Lada2003}; see also \citealt{deGrijs2010} and references therein for a summary).

We conclude that stellar winds regulate the accretion of gas onto the forming star cluster sink right after the first massive star(s) have been born, while supernovae explode only late (after $\gtrsim$ 5 Myr) and fail to regulate accretion in a way to produce enough lower mass clusters. 



\begin{figure}
\includegraphics[width=0.49\textwidth]{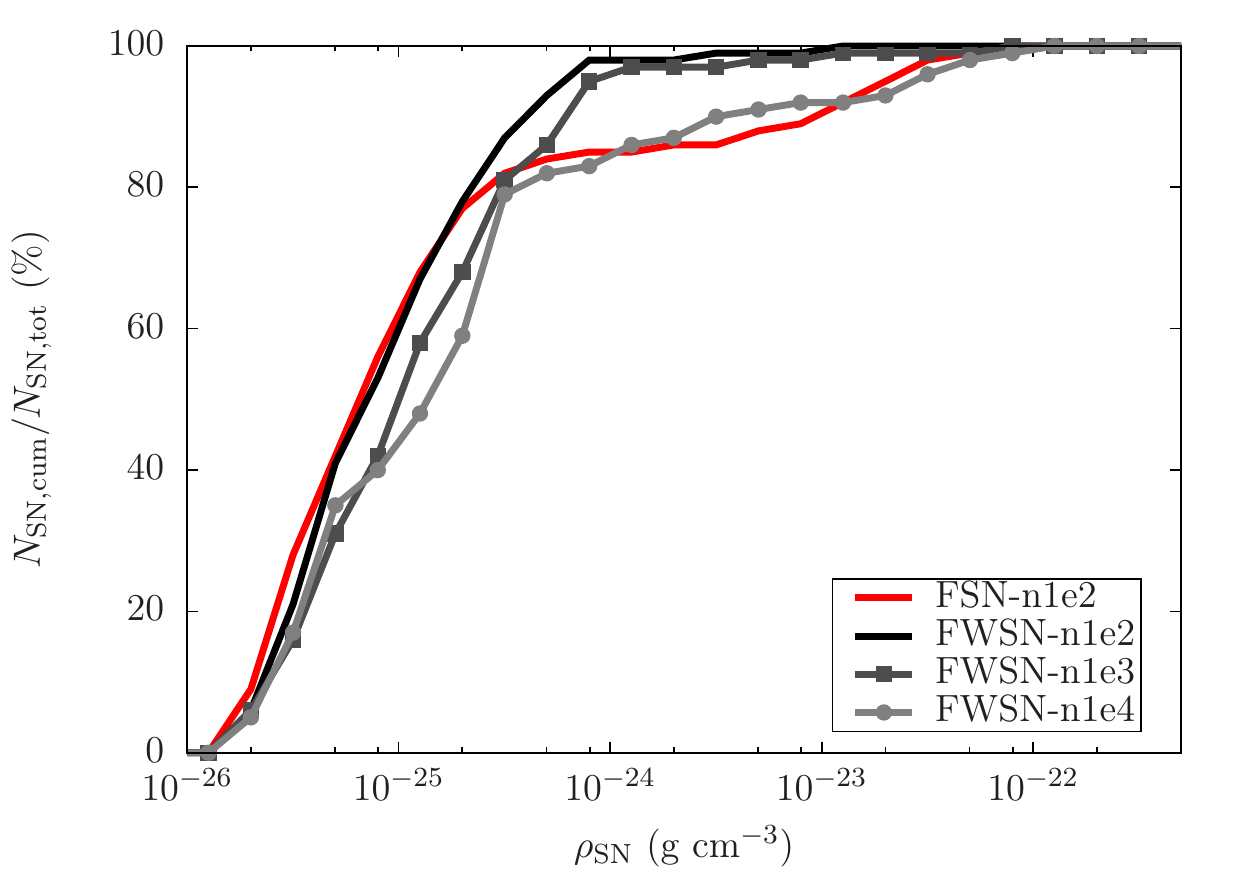}
  \caption{Normalised cumulative distribution of SNe as a function of
    the mean environmental gas density (in the region where the SN
    explodes), $\rho_{\rm SN}$, for all simulations with
    supernovae. For high enough star formation rates, 
    the stellar wind feedback clears out the SN environment before
    their explosion resulting in typical environmental densities of
    $\sim 10^{-25} \;{\rm g\; cm}^{-3}$ . But even the clustering of massive
    stars and their supernovae alone is sufficient to have 80\% of all
    supernovae explode in environments with reduced density.}\label{fig:rhoSN} 
\end{figure}

\section{How Supernova feedback drives galactic outflows}\label{sec:results3}

The impact of the SN explosions on the ISM depends on the structure and the density of the gas near the explosion centre \citep[see e.g.][for recent high-resolution numerical simulations]{KimOstriker15,WalchNaab15,Martizzi+15,IffrigHennebelle15}. SNe exploding in high density environments are subject to rapid radiative cooling and do not inject large amounts of radial momentum \citep[see][for studies of
the momentum injection  of SNe in different environments]{2016MNRAS.458.3528H, Haid2016}. The average SN rate per Myr in each simulation, $\dot{N}_\mathrm{SN}$, is listed in table \ref{tab:tab2}.  

We probe the mean density in each SN injection region, $\rho_\mathrm{SN}$, to understand which ambient conditions the SN explosions encounter in the different simulations. This is depicted in Fig. \ref{fig:rhoSN}, where we plot the cumulative distribution of all SNe as a function of $\rho_\mathrm{SN}$. We find that in run {\it FSN-n1e2} with only supernova feedback, $\sim 80$\% of all SNe explode in relatively low density gas with $\rho \sim 10^{-25}\;{\rm g\;cm}^{-3}$ and only $\sim 15$\% explode in higher density gas with $\rho \gtrsim 10^{-23}\;{\rm g\;cm}^{-3}$. The reason is the clustering of the massive stars and hence of the SN feedback. Only the first SN in a cluster interacts with a denser environment, while the following ones explode inside the low density
bubble \citep{MacLowMcCray88, ChuMacLow90}. Clustering can also lead to the formation of super bubbles \citep{Wunsch+08}. With wind feedback included, basically all SN environments are reduced in density before the first explosion. Only run {\it FWSN-n1e3}, which has a much lower star formation rate, also has a small fraction of SNe that interact with dense gas. 

Observations of OH maser emission also identify that $\sim 10$\% of all SN remnants in the Milky Way are interacting with dense gas \citep{HewittYusefZadeh09}. Furthermore, \cite{Elwood2016} have studied the distribution of environmental densities for SN remnants in M31 and M33. They derive a narrow lognormal distribution of environmental densities with a mean number density of $\bar{n}_\mathrm{SN}=0.07\;{\rm cm}^{-3}$ and a standard deviation of $\sigma_\mathrm{SN}=0.7$. To compare with the observed results, we fit a lognormal distribution for the environmental densities of the two runs that can be fitted with a single component. The following values give the best fit to the respective distribution.
\begin{itemize}
\setlength{\itemindent}{0.3cm}
\item For run {\it FWSN-n1e2}, we find a mean density of $\sim 8\times 10^{-26}\;{\rm g\;cm}^{-3}$, which corresponds to $\bar{n}_\mathrm{SN}\approx 0.07\;{\rm cm}^{-3}$, and a standard deviation of $\sigma_\mathrm{SN}=0.9$.
\item For run {\it FWSN-n1e3}, we find a mean density of $\sim 10^{-25}\;{\rm g\;cm}^{-3}$, which corresponds to $\bar{n}_\mathrm{SN}\approx 0.09\;{\rm cm}^{-3}$, and a standard deviation of $\sigma_\mathrm{SN}=1.1$.
\end{itemize}
Overall, the SN remnants in the simulations presented here encounter low density environments, and are therefore well resolved. In this case, thermal energy input can be safely used without the problem of numerical over-cooling \citep{Gatto+15}. 

 
\begin{figure}
\includegraphics[width=0.49\textwidth]{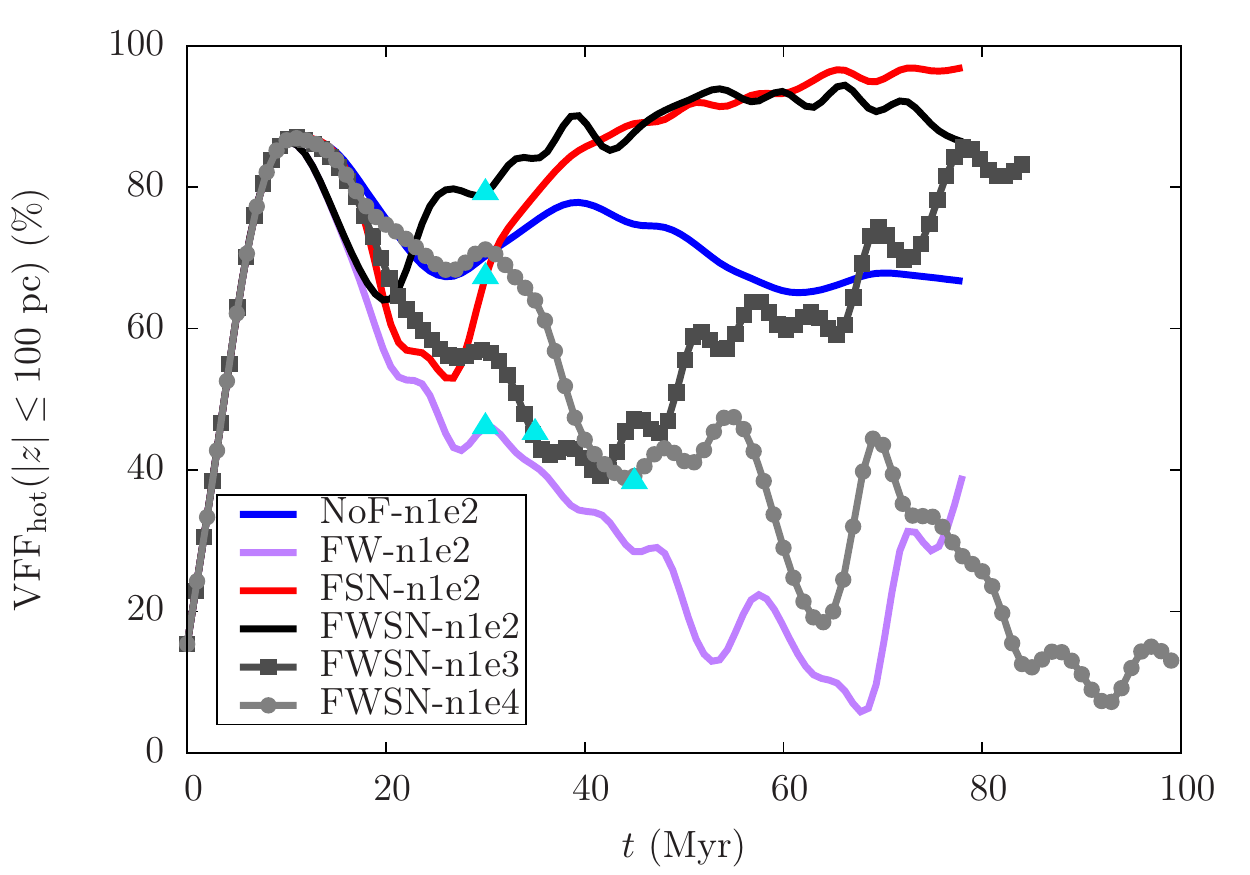}
 \includegraphics[width=0.49\textwidth]{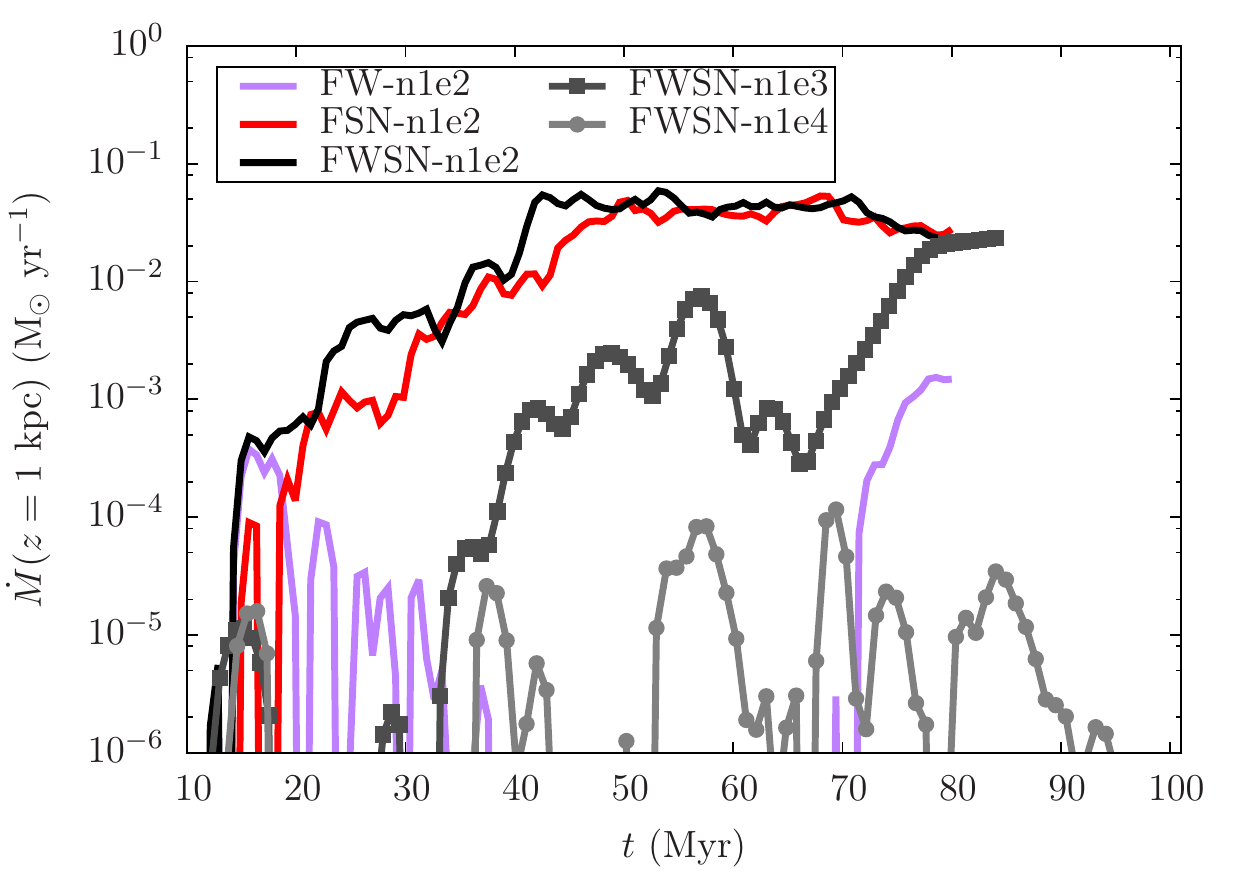}
 \caption{{\it Top:} Time evolution of the hot gas volume-filling fractions (all gas with $T > 3 \times 10^5$ K) within $z=\pm 100$ pc around the disc mid-plane. The cyan triangles show the time from which onward we compute the mass loading factors for each run as shown in Fig. \ref{fig:massload}, which corresponds to $t_\mathrm{sink,20} = t_\mathrm{sink,0} + 20$ Myr. {\it  Bottom:} Time evolution of the total gas outflow rates at 1 kpc above and below the disk mid-plane. Only simulations with high enough supernova rates ({\it FSN-n1e2, FWSN-n1e2, FWSN-n1e3}) develop high volume-filling fractions of hot gas and significant outflows.}\label{fig:mdot1kpc-tot}  
\end{figure}
We define gas as {\it hot gas} if it has a temperature $T > 3 \times 10^5$ K. This gas is in the {\it thermally stable}, hot phase \citep{Dalgarno1972}. In the following we show that, in case enough supernovae explode in low density environments, a hot, volume-filling phase is developed \citep[see e.g.][for this process in regions with periodic boundaries]{Gatto+15, Li+15} and galactic outflows can be launched. In the upper panel of Fig. \ref{fig:mdot1kpc-tot}, we plot the time evolution of the hot gas volume-filling fraction (VFF) within $z=\pm 100 \pc$ around the disc mid-plane. The  cyan triangles mark the time of the formation of the first cluster plus 20 Myr ($t_\mathrm{sink, 20} = t_\mathrm{sink,0} + 20$ Myr). At this point feedback from the first massive stars had enough time to change the structure of the surrounding ISM. All runs with supernova feedback and high star formation rates (runs {\it FSN-n1e2}, {\it FWSN-n1e2}), and 
{\it FWSN-n1e3}) develop volume-filling hot gas with VFF $\gtrsim 50$\% at $t_\mathrm{sink, 20}$ and more than 80\% towards $t_\mathrm{stop}$. Run {\it NoF-n1e2} also seems to have a fairly high hot gas VFF, but this is caused by the collapse of the disc into a thin sheet and the accretion of most of the gas into sink particles. However, run {\it FW-n1e2} without supernova feedback and run {\it FWSN-1e4}, which has a star formation rate surface density that is below the KS relation, do not form a hot volume-filling phase and the hot gas VFF stays below 40\% after $t_\mathrm{sink, 20}$. 

\noindent In total we define four temperature regimes (see Paper I):
\begin{itemize}
\setlength{\itemindent}{0.3cm}
 \item hot: $T > 3 \times 10^5$ K,
 \item warm-hot: $8000 < T \leqslant 3 \times 10^5$ K,
 \item warm: $300 < T \leqslant 8000$ K,
 \item cold: $30 < T \leqslant 300$ K.
\end{itemize}
For completeness, the evolution of warm-hot, warm, and cold gas is shown in Fig. \ref{fig:vff2} (see Appendix B).\\   

Furthermore, we show the time evolution of the total outflowing gas mass through surfaces at $z=\pm 1$~kpc in the lower panel of Fig. \ref{fig:mdot1kpc-tot}. All runs with a hot volume-filling phase also have relatively high outflow rates of $\dot{M}(z=\pm1\;{\rm kpc}) \gtrsim 2\times10^{-2}\mo\;{\rm yr}^{-1}$ at $t_\mathrm{stop}$. The two runs with a low hot gas VFF also have significantly lower outflow rates.\\ 

\begin{figure}
\includegraphics[width=0.49\textwidth]{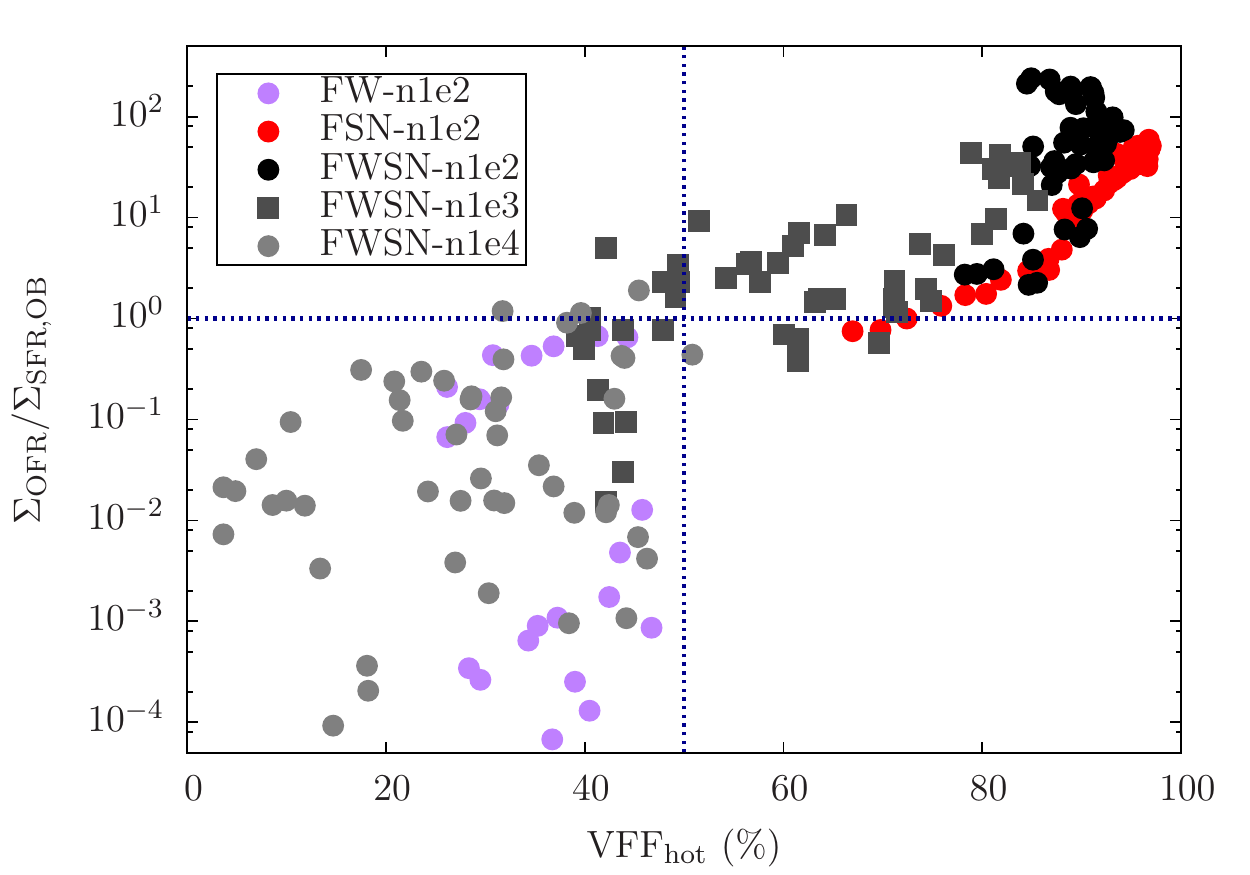}
 \caption{Mass loading, measured from the
   ratio of the surface density of the outflowing gas at $z = \pm 1$
   kpc and the star formation rate surface density, as a function of
   the hot gas volume-filling  fraction within $\pm 100$ pc for all simulations with feedback (the
   run without feedback does not develop any outflows and is therefore
   not shown). The data is binned in time with $\Delta t= 1$
   Myr. Only simulations with a hot gas filling fraction of more than
   $\sim 50$\%, which has been created due to supernova feedback, are
   associated with outflows with mass loading $\gtrsim 1$ (as indicated by the dotted, dark blue lines).
 }\label{fig:massload}  
\end{figure}

A better quantity to define the efficiency of galactic outflows that are driven by thermal pressure is the so-called {\it mass loading factor}, which is defined as the total gas outflow rate surface density, $\Sigma_\mathrm{OFR}$ in $[\mo {\rm yr}^{-1}\;{\rm kpc}^{-2}]$, over the star formation rate surface density. We note that this quantity only becomes meaningful in combination with the distance (from the star formation event) where the outflow rate is measured. We correlate the hot gas VFF and the mass loading factor in Fig. \ref{fig:massload}. Each point represents the average over a time bin of $\Delta t=1$ Myr as calculated for all times starting from $t_\mathrm{sink, 20}$ up to $t_\mathrm{stop}$. The dotted horizontal line indicates a mass loading factor of 1, and the dotted vertical line shows the VFF of 50\%. There is a very clear trend that simulations with hot gas VFFs that are higher than 50\% have mass loading factors above 1, while the two runs with low hot gas VFFs (run {\it FW-n1e2} and run {\it FWSN-n1e4}) also have mass loading factors smaller than 1. The runs with low mass loading and low hot gas VFF actually have some points missing where there is no outflow at all. This is also the case for run {\it NoF-n1e2}, which has a mass loading factor equal to zero at all times. While there seems to be an exponential correlation above for the high mass loading factors, there is a large scatter for low mass loading factors and therefore we do not provide fits to the distributions. 

In summary, we clearly find that outflows from the galactic disc can be launched by the thermal pressure of the supernova-driven, hot gas, in cases where the star formation rate (i.e. the supernova rate) is high enough to cause a hot volume
filling phase \citep[see][for wind launching mechanisms driven by non-thermal cosmic rays]{Girichidis+16}. Simulations with stellar winds alone fail to produce significant outflows. 

This is in qualitative agreement with investigations by \cite{Gatto+15} and \cite{Li+15} who study the SN-driven ISM in periodic setups. They show that the thermal pressure becomes very high in cases where the SNe can drive the hot gas VFF above 50\%. In periodic setups, where the gas is confined and the pressure cannot be released by outflows, this leads to a thermal runaway, where most of the gas mass is compressed into small clumps and most of the volume is filled with hot gas. A high hot gas VFF is reached for high enough SN rates, in which case the bubbles start to overlap. 



\section{Conclusions}\label{sec:conclusions}
We study the impact of stellar winds and SNe on the multi-phase ISM in a representative piece of a galactic disc with $\Sigma_\mathrm{gas} = 10 \mopc$ and a size of $(500 \pc)^2 \times \pm\ 5 \kpc$. We include an external, static, stellar potential as well as gas self-gravity, radiative cooling and diffuse heating, sink particles and stellar feedback in the form of stellar winds and SN explosions. We take into account dust and gas (self-~)shielding and we track the distribution of molecular gas using a chemical network that allows us to follow the formation, evolution and destruction of $\h,\hp,\htwo,\co,\cp$. 

Star formation is modelled via cluster sink particles, which are allowed to accrete throughout the simulation. We implement a sub-grid model for the feedback from massive stars, where we randomly sample massive stars from the IMF and follow the wind feedback of each single massive star using the latest Geneva stellar tracks. The injected wind luminosity corresponds to the total wind luminosity of all massive stars in the cluster.
At the end of their lifetime the stars undergo a type II supernova explosion. We switch on and off wind and SN feedback to study how each feedback mechanism affects the multi-phase ISM structure.  

We find that 
\begin{itemize}
\setlength{\itemindent}{0.3cm}
 \item For a given stellar population, the energy injected by stellar
   winds is mostly dominated by short-lived very high-mass stars,
   while the majority of the energy injected as SNe comes from
   long-lived progenitors with lower masses. Compared to stellar
   winds, SNe dominate the total injected energy, but only by a factor
   of $\sim 3$. 
 
\item Models with stellar winds and SNe show the best agreement with
  observations of nearby {\it normal} star-forming galaxies.   
 
 \item Stellar winds regulate the growth of young cluster sinks by
   quenching the gas accretion onto them shortly after the first
   massive star has been born. SN feedback is significantly delayed
   and thus, allows for longer time scales of efficient gas accretion
   (up to the first SN explosion after $\sim 5 \Myr$). Stellar winds
     qualitatively change cluster formation timescales. More massive
     clusters have shorter formation timescales. In simulations
     without winds such an anti-correllation does not exist.
 
  
 \item Strong shock-heating by SN explosions and possibly overlapping
   SN remnants creates a hot volume-filling gas phase near the disc mid-plane. Stellar winds are
   less energetic and convert most of the cold and warm gas into a
   warm and warm-hot gas.  
 
 \item Thermal pressure of the hot gas can drive outflows with
   significant mass loading factors as measured at $z=\pm 1$ kpc. This
   is  possible if the star formation rate and hence the SN rate in
   the discs is high enough to produce a hot gas VFF of more than $\sim
   50$\%.  
 


\end{itemize}

\section*{Acknowledgements}

All simulations have been performed on the Odin and Hydra clusters hosted by the Max Planck Computing \& Data Facility (http://www.mpcdf.mpg.de/). We thank M. Anderson, C. Federrath, J. Mackey, M. M. Mac Low, E. Pellegrini, and X. Shi for useful discussions and \cite{Leroy+08,Ekstrom+12} for making their data publicly available. AG, SW, TN, PG, SCOG, RSK, and TP acknowledge the Deutsche Forschungsgemeinschaft (DFG) for funding through the SPP 1573 ``The Physics of the Interstellar Medium''. SW acknowledges funding by the Bonn-Cologne-Graduate School, by SFB 956 "The conditions and impact of star formation", and from the European Research Council under the European Community's Framework Programme FP8 via the ERC Starting Grant RADFEEDBACK (project number 679852). TN acknowledges support by the DFG cluster of excellence 'Origin and structure of the Universe'. RW acknowledges support by the Czech Science Foundation project 15-06012S and by the institutional project RVO:~67985815.
RSK and SCOG acknowledge support from the DFG via SFB 881 ``The Milky Way System'' (sub-projects B1, B2 and B8). RSK acknowledges support from the European Research Council under the European Community's Seventh Framework Programme (FP7/2007-2013) via the ERC Advanced Grant STARLIGHT (project number 339177). The software used in this work was in part developed by the DOE NNSA-ASC OASCR Flash Center at the University of Chicago. We thank C. Karch for the program package \texttt{FY} and M. Turk and the \texttt{yt} community for the \texttt{yt} project \citep{yt}. 

\appendix
\section{Simulation snapshots}\label{AppendixA}

\begin{figure*}
 \centering
 \textbf{Run {\it NoF-n1e2}, $t=40 \Myr$}\\
 \includegraphics[width=0.7\textwidth]{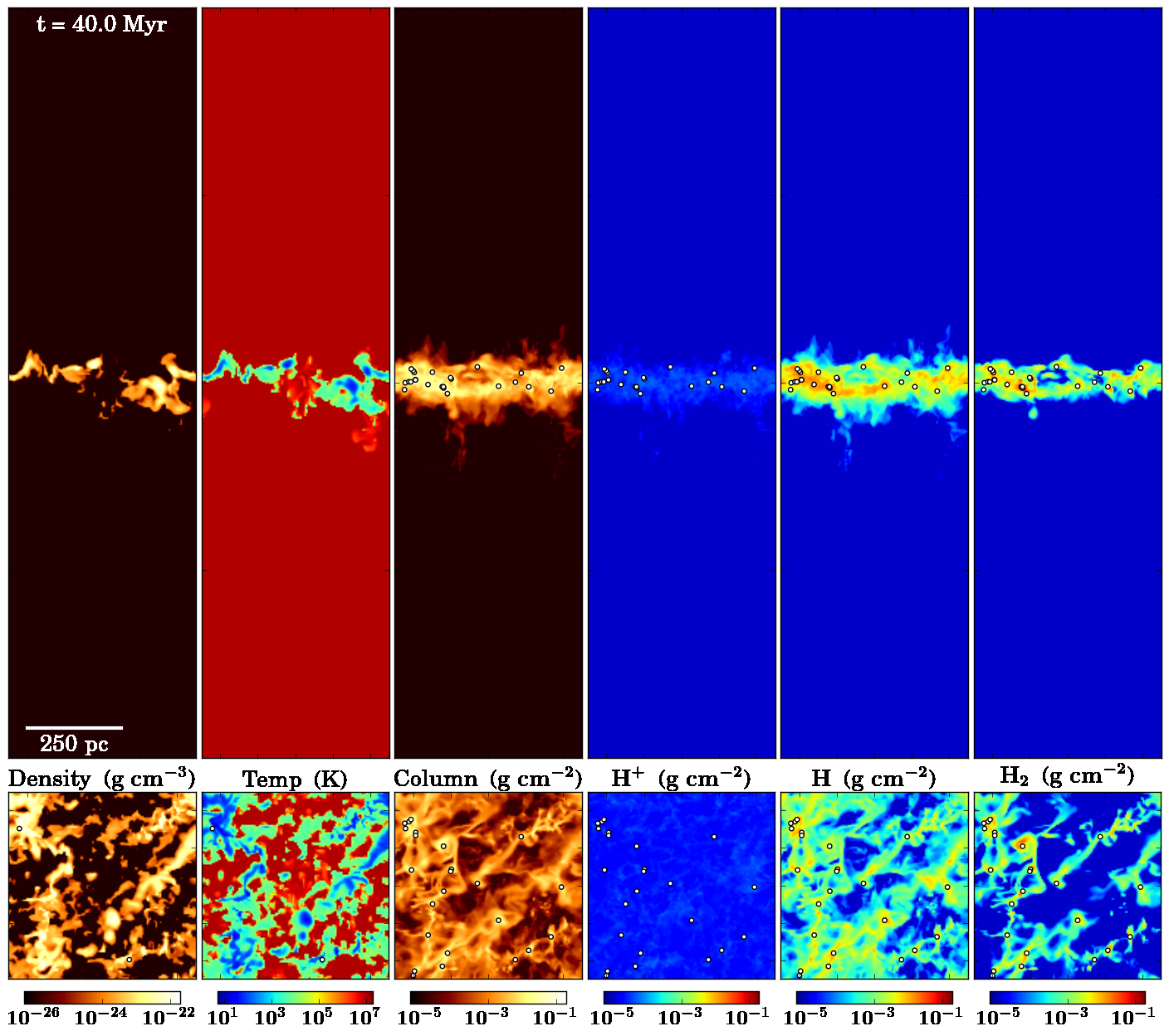}\\
 \textbf{Run {\it FW-n1e2}, $t=40 \Myr$}\\
 \includegraphics[width=0.7\textwidth]{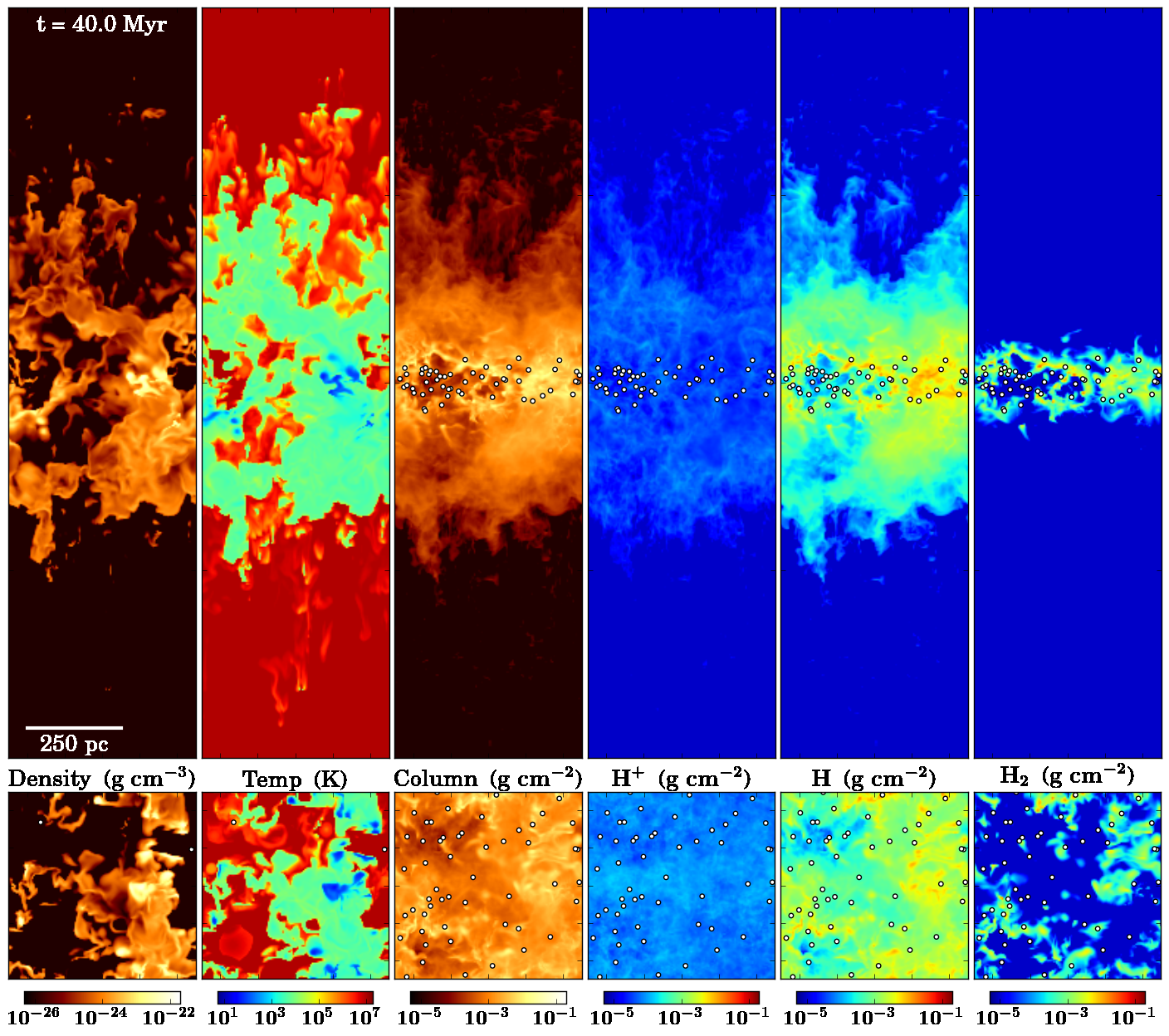}\\
\caption{Same as Fig. \ref{fig:snap1} for simulations {\it NoF-n1e2} without feedback from the cluster sinks (top) and run {\it FW-n1e2} with stellar wind feedback (bottom) at $t=40 \Myr$  
}\label{fig:snap1a}
\end{figure*}

\begin{figure*}
\centering
 \textbf{Run {\it FSN-n1e2}, $t=40 \Myr$}\\
 \includegraphics[width=0.7\textwidth]{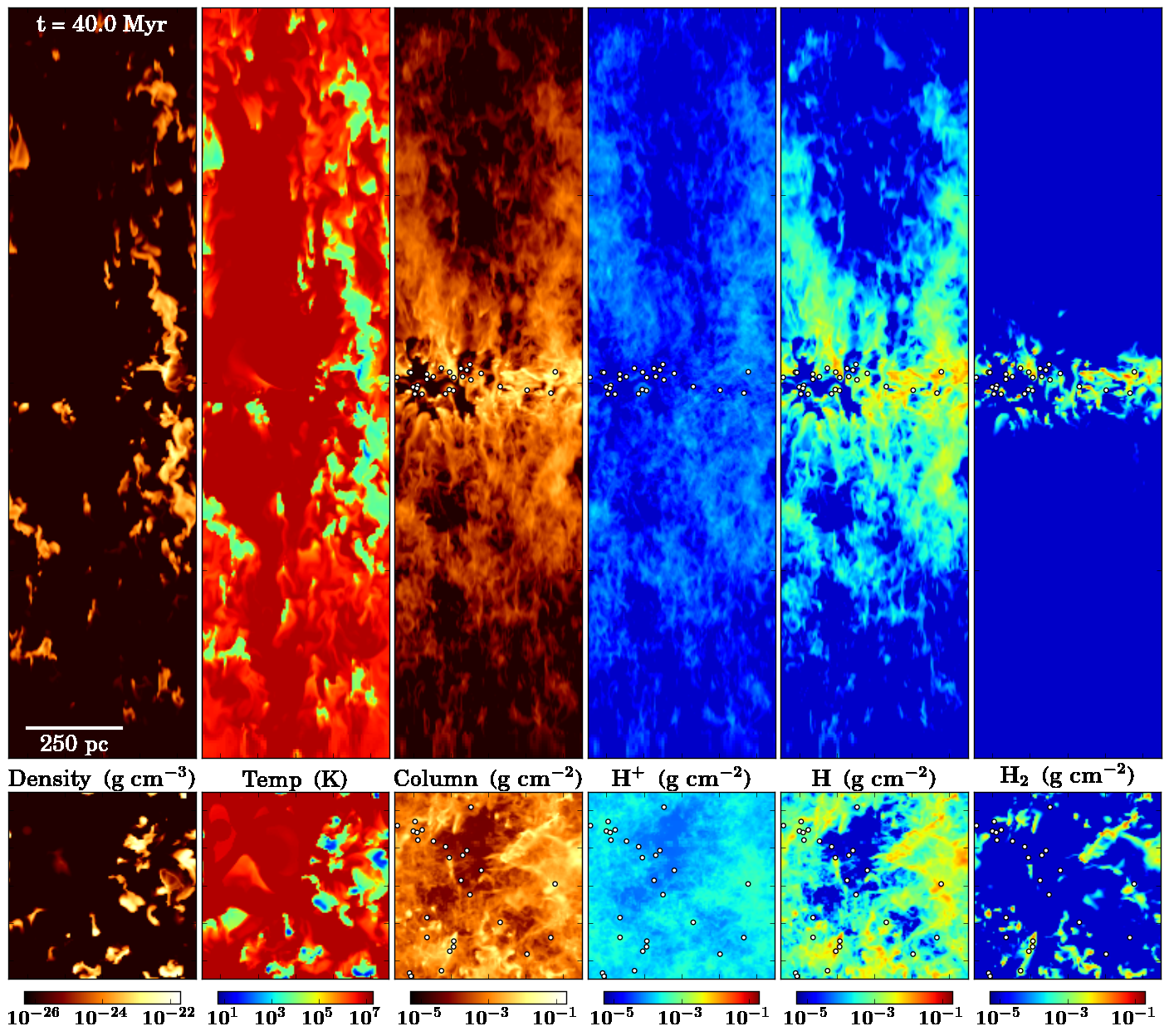}\\
 \centering
 \textbf{Run {\it FWSN-n1e2}, $t=40 \Myr$}\\
 \includegraphics[width=0.7\textwidth]{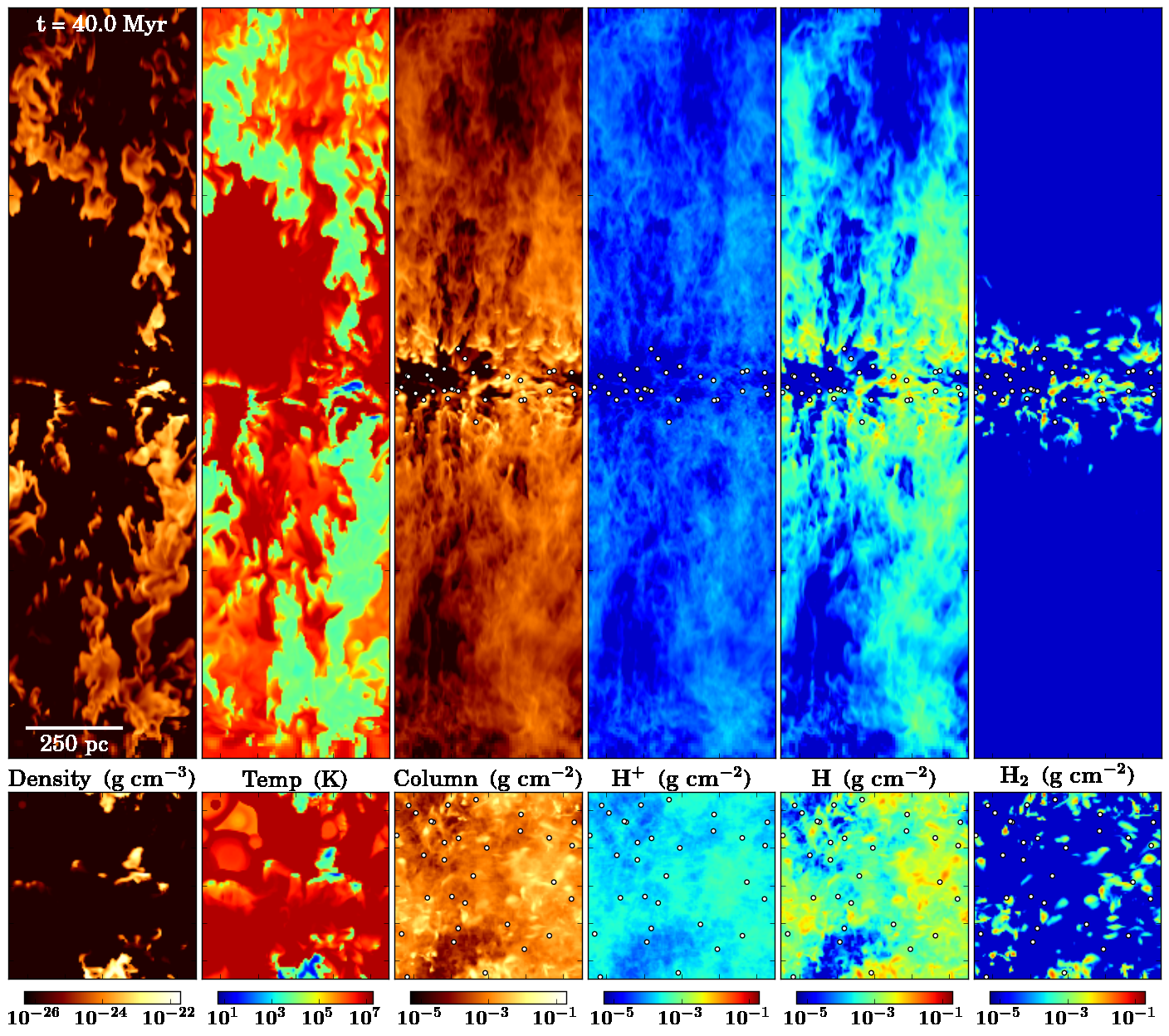}
\caption{Same as Fig. \ref{fig:snap1} for simulations {\it FSN-n1e2} with supernova feedback alone (top) and {\it FWSN-n1e2} with stellar winds and supernovae (bottom).}\label{fig:snap2a} 
\end{figure*}

\begin{figure*}
 \centering
 \textbf{Run {\it FWSN-n1e4}, $t=61$ Myr} \\
 \includegraphics[width=0.7\textwidth]{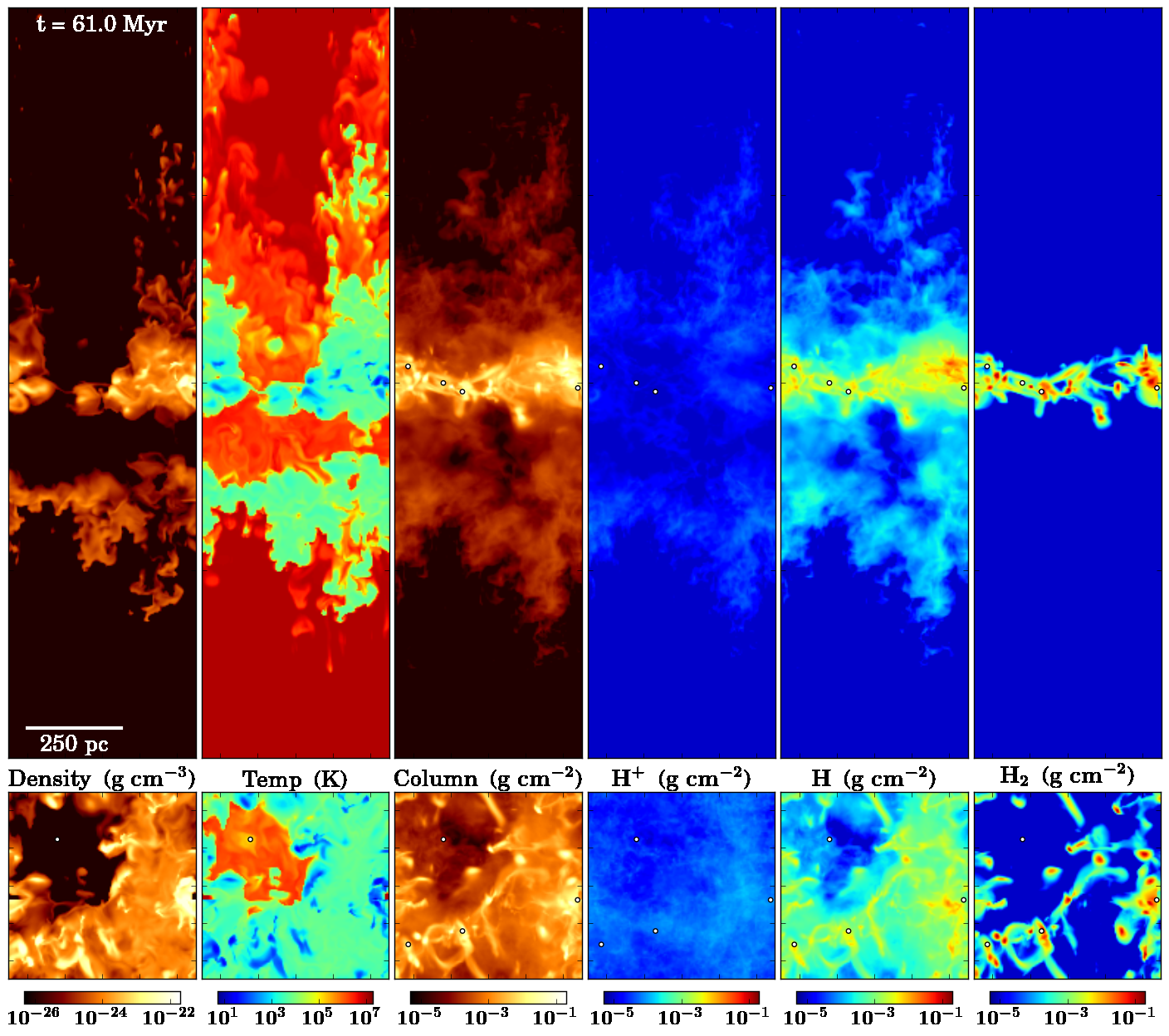}
\caption{Same as Fig. \ref{fig:snap1} for simulation {\it FWSN-n1e4} at $t=61$ Myr, where the formation of cluster sink particles is enabled above $\rthr = 2\times 10^{-20} \;{\rm g\; cm}^{-3}$. We show a snapshot at a somewhat later time because the first cluster sink is formed later in these simulations. The higher sink density threshold leads to smaller star formation rates and smaller outflow rates (see Fig. \ref{fig:mdot1kpc-tot}).}\label{fig:snap3a}
\end{figure*}

In Figs. \ref{fig:snap1a}, \ref{fig:snap2a}, and \ref{fig:snap3a} we show snapshots of simulations {\it NoF-n1e2} and {\it FW-n1e2} (Fig. \ref{fig:snap1a}), {\it FSN-n1e2} and {\it FWSN-n1e2} (Fig. \ref{fig:snap2a}), and {\it FWSN-n1e4} (Fig.\ref{fig:snap3a}) at $\sim t_\mathrm{sink,0} +31$ Myr. As described in section \ref{sec:results1} for Fig. \ref{fig:snap1}, the different panels show (from left to right) a slice of the total density and of the gas temperature at $y=0$ (top) and at $z=0$ (bottom), the total gas column density, and the column densities of $\hp$, H, and $\htwo$. The location of the cluster sink particles is indicated by the small white circles. 

Runs with supernova feedback show larger disc scale heights, which is clearly visible in the projections of the total column density or the atomic hydrogen column density. Only run {\it FWSN-n1e4}, which has a low star formation rate, has a smaller scale height but more molecular hydrogen. 

\section{Evolution of mass and volume-filling fractions}\label{AppendixB}

\begin{figure*}
 \includegraphics[width=0.49\textwidth]{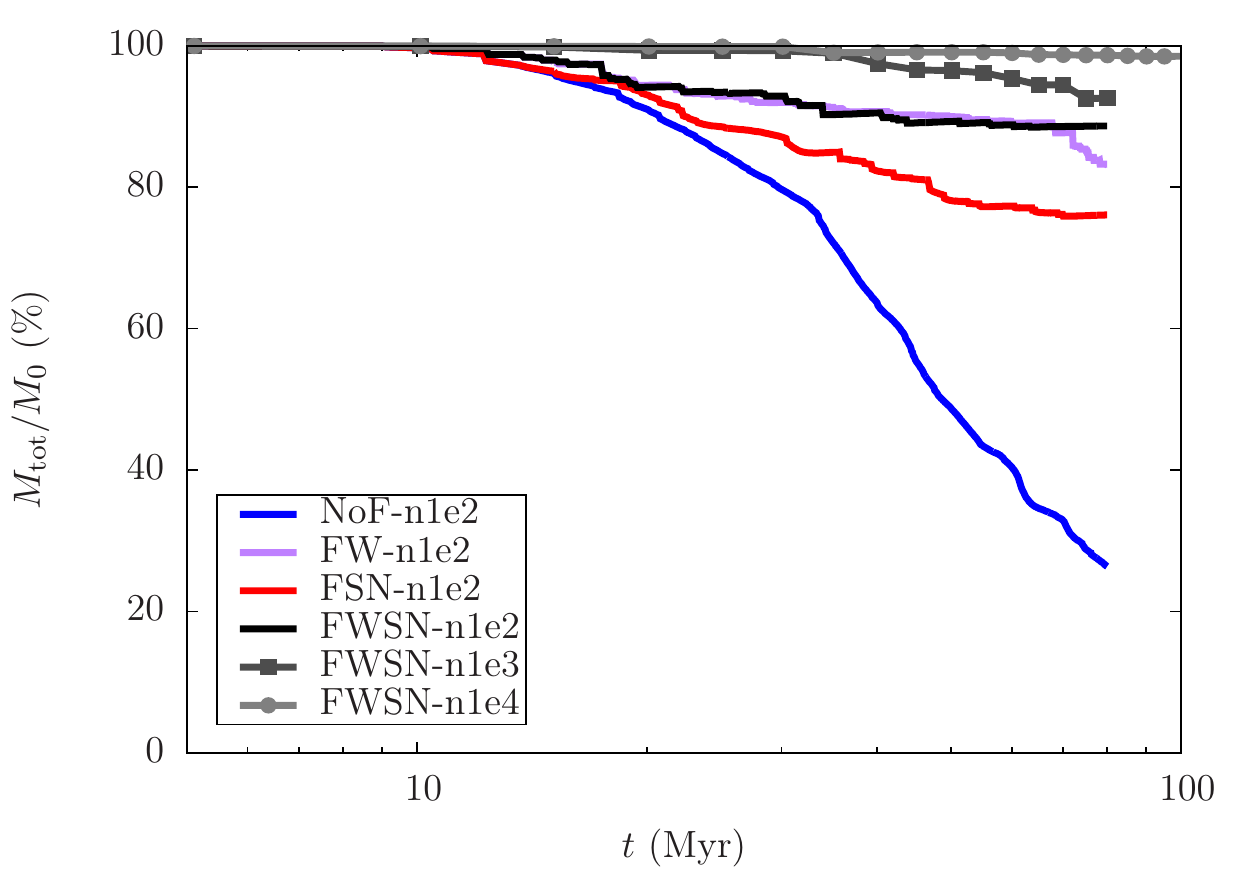}
  \includegraphics[width=0.49\textwidth]{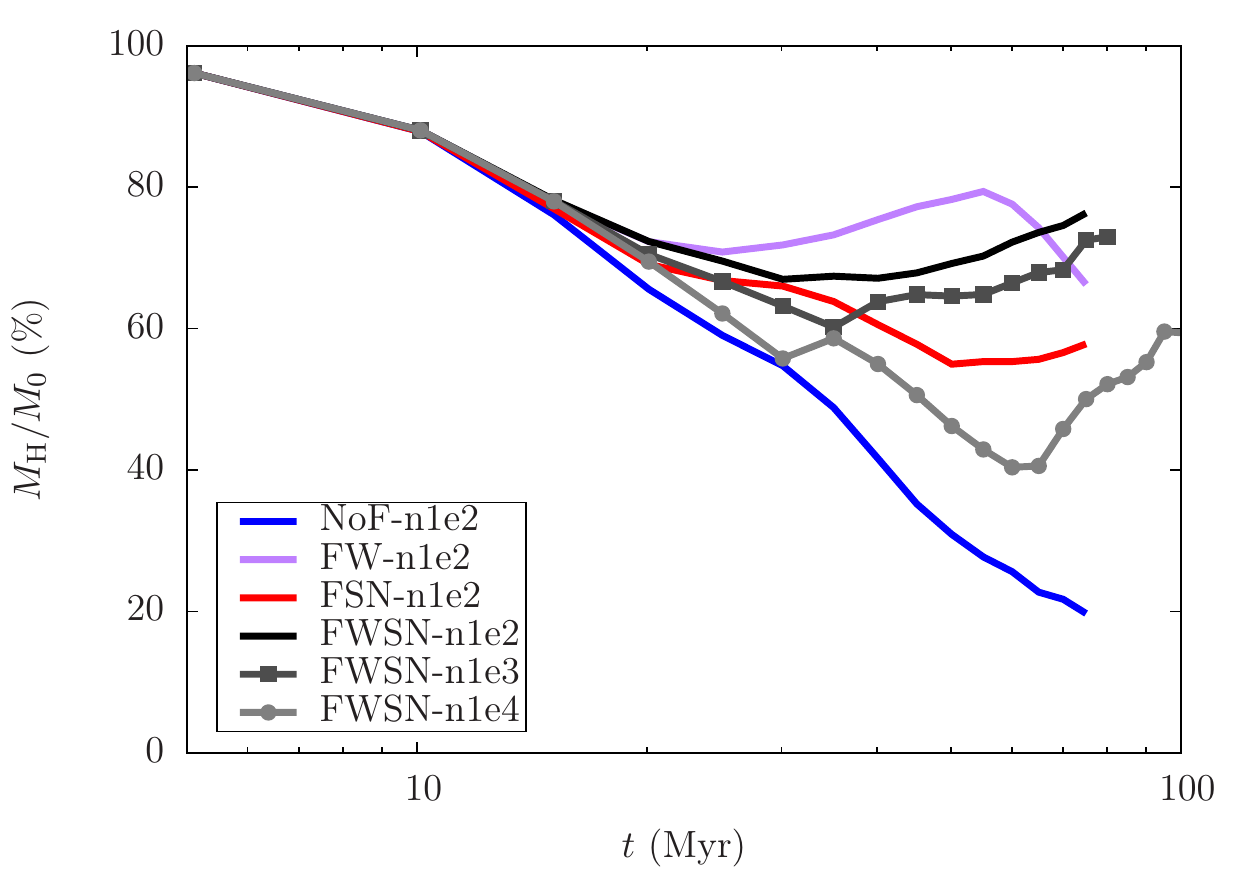}
  \includegraphics[width=0.49\textwidth]{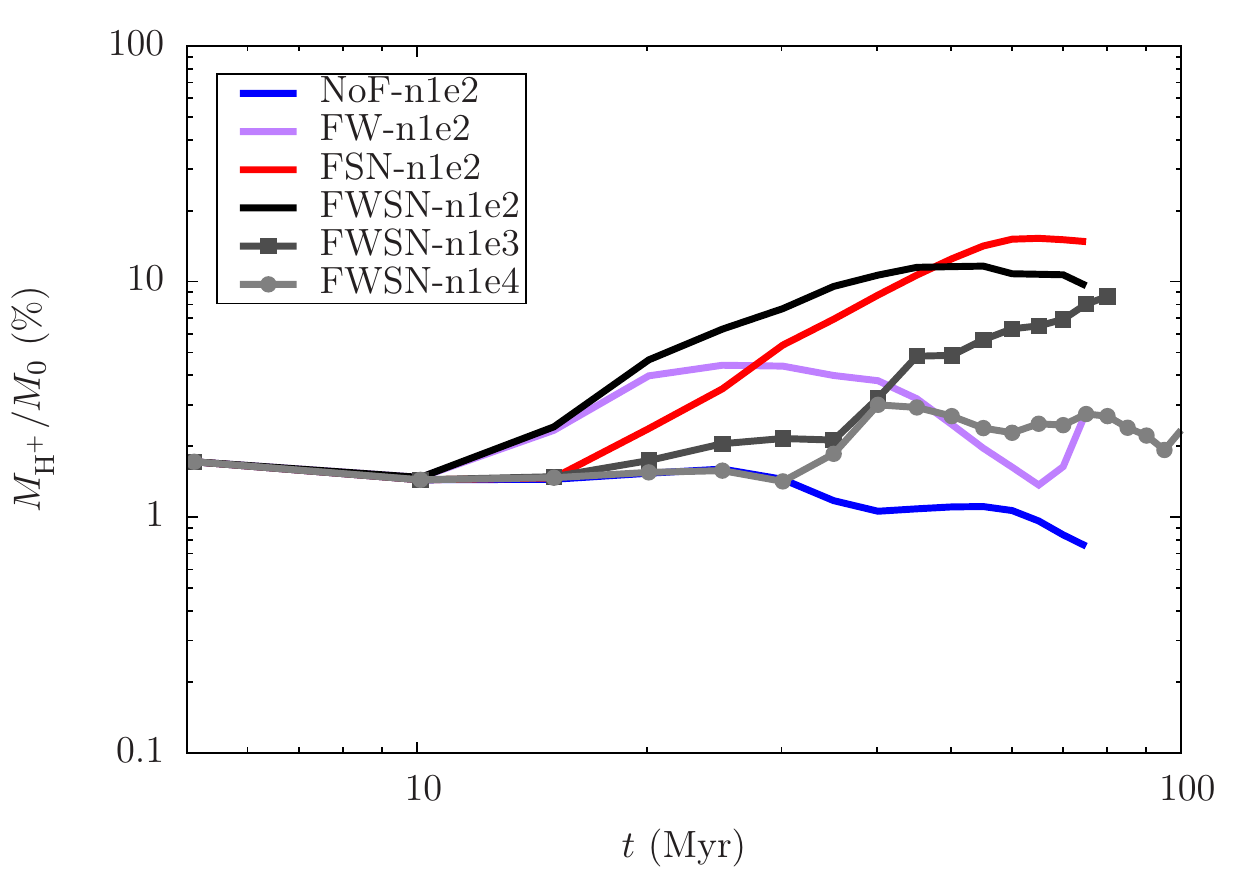}
 \includegraphics[width=0.49\textwidth]{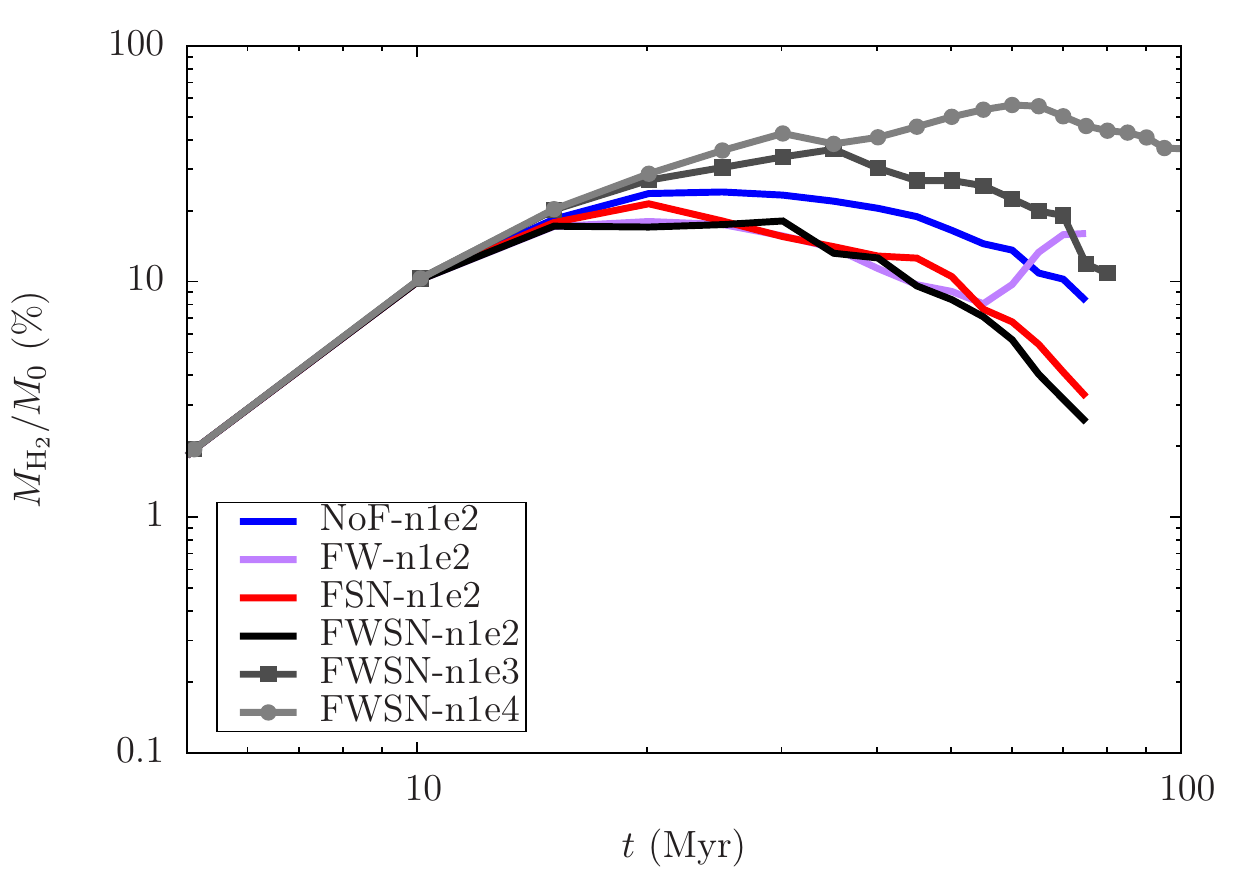}
\caption{Time evolution of the total gas mass ({\it top left}), and the mass fractions of atomic hydrogen ({\it top right}), ionized hydrogen ({\it bottom left}), and molecular hydrogen ({\it bottom right}), all normalised to the total mass in hydrogen at $t=0$. Overall the total gas mass decreases as star formation proceeds. At any time most of the gas mass is in atomic hydrogen (top right panel), although we caution that our neglect of radiative feedback from massive stars means that we overproduce warm neutral atomic gas at the expense of warm ionized gas. Most of the hot, ionized gas is caused by supernova feedback. For molecular hydrogen, the sink density threshold plays an important role. For the low density sink threshold, much of the molecular hydrogen gas is accreted onto the sink particles and is assumed to form stars. This leads to an underestimation of the H$_2$ mass fractions in runs with $\nthr=10^2 {\rm cm}^{-3}$.}
\label{fig:mf}   
\end{figure*}
In Fig. \ref{fig:mf} we show the time evolution of the total gas mass (top left panel), and of the mass fractions of atomic hydrogen (top right), ionized hydrogen (bottom left), and molecular hydrogen (bottom right), all normalised to the hydrogen gas mass at $t=0$, $M_0$ (see section \ref{sec:initial}). The total gas mass is complementary to the sink mass evolution. In the run without stellar feedback the star formation rate is so high (see Fig. \ref{fig:msink}) that only $\sim$ 25\% of the total gas mass is left after 80 Myr, while in run {\it FWSN-n1e4} only little mass has collapsed into stars. 

At any time most of the hydrogen mass is in atomic form (top right panel). Overall, runs with a lower star formation rate have less mass in hot, ionized gas (bottom left panel) but more mass in atomic and molecular gas (bottom right panel). We find that supernova feedback is needed to produce hot, ionized gas. When comparing the two simulations {\it FW-n1e2} and {\it FWSN-n1e2}, which have a very similar evolution of the total gas mass (and of star formation), we find that the latter has significantly more ionized hydrogen but less atomic hydrogen. In addition, run {\it FWSN-n1e4} has a significantly lower star formation rate than {\it FW-n1e2} but a comparable amount of ionized gas.

For molecular hydrogen (lower right panel), the sink density threshold also plays an important role. For the low sink density threshold, much of the molecular hydrogen gas is accreted onto the sink particles and is assumed to form stars. This leads to an apparently smaller H$_2$ fraction but a larger $M_{\rm sink}$. This result shows that we cannot use runs with $\nthr=10^2 {\rm cm}^{-3}$ to study the $\htwo$ content that develops within the galactic disc as most of the dense gas is accreted. For $\nthr=10^4 {\rm cm}^{-3}$ we begin to see similar $\htwo$ mass fractions as in runs without sink particles that we presented in Paper I, but the star formation rate is too low in this simulation. Therefore, the usefulness of runs with cluster sink particles to study the molecular gas content in a disc galaxy simulation is limited.

\begin{figure}
 \includegraphics[width=0.49\textwidth]{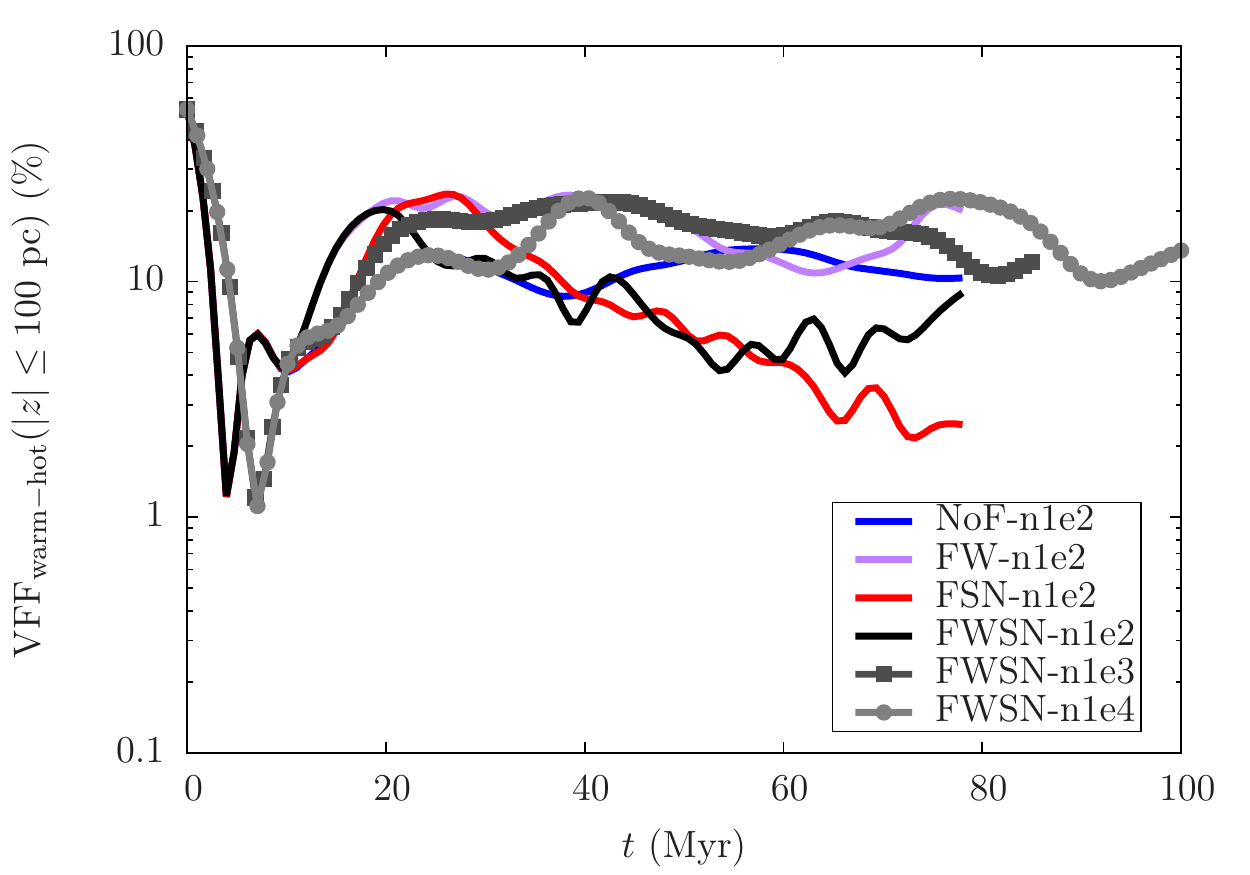}
 \includegraphics[width=0.49\textwidth]{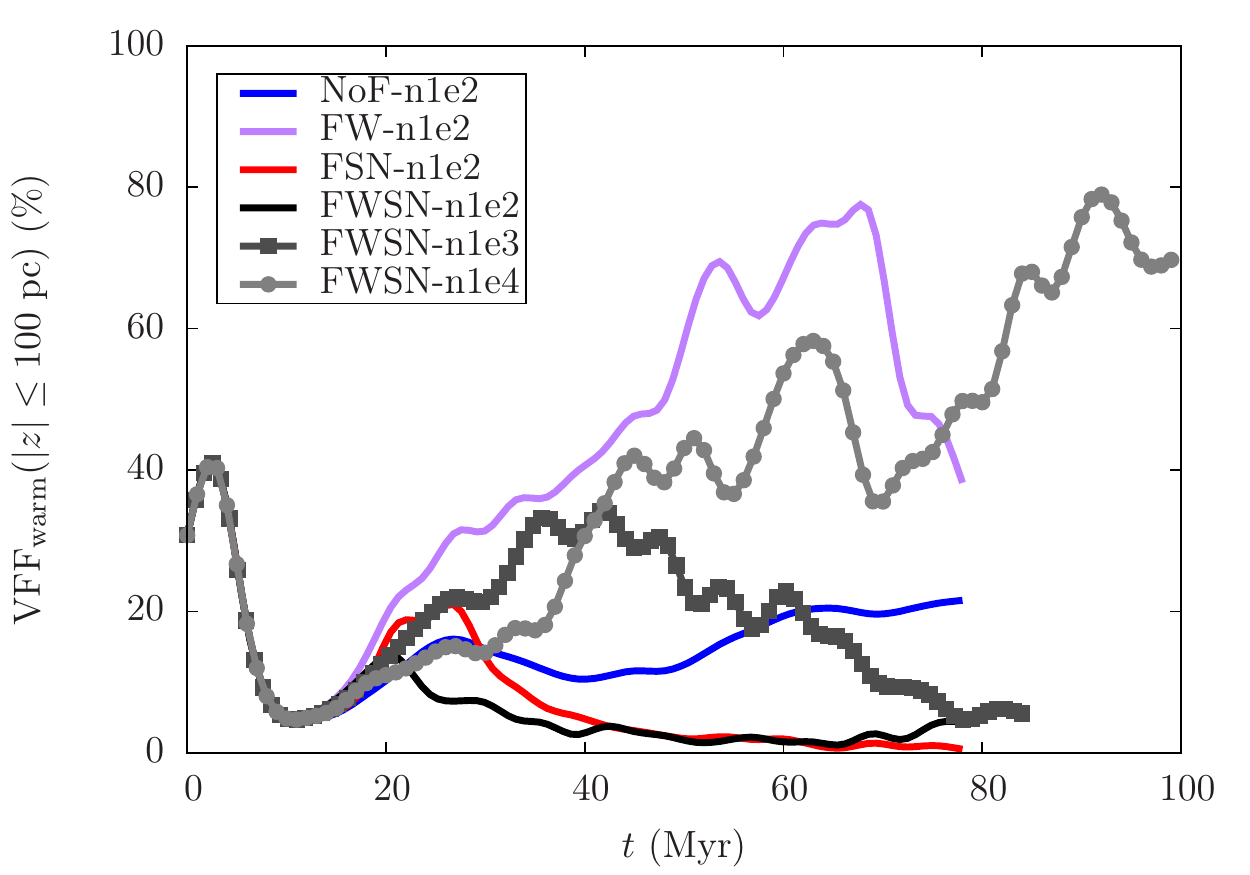}
 \includegraphics[width=0.49\textwidth]{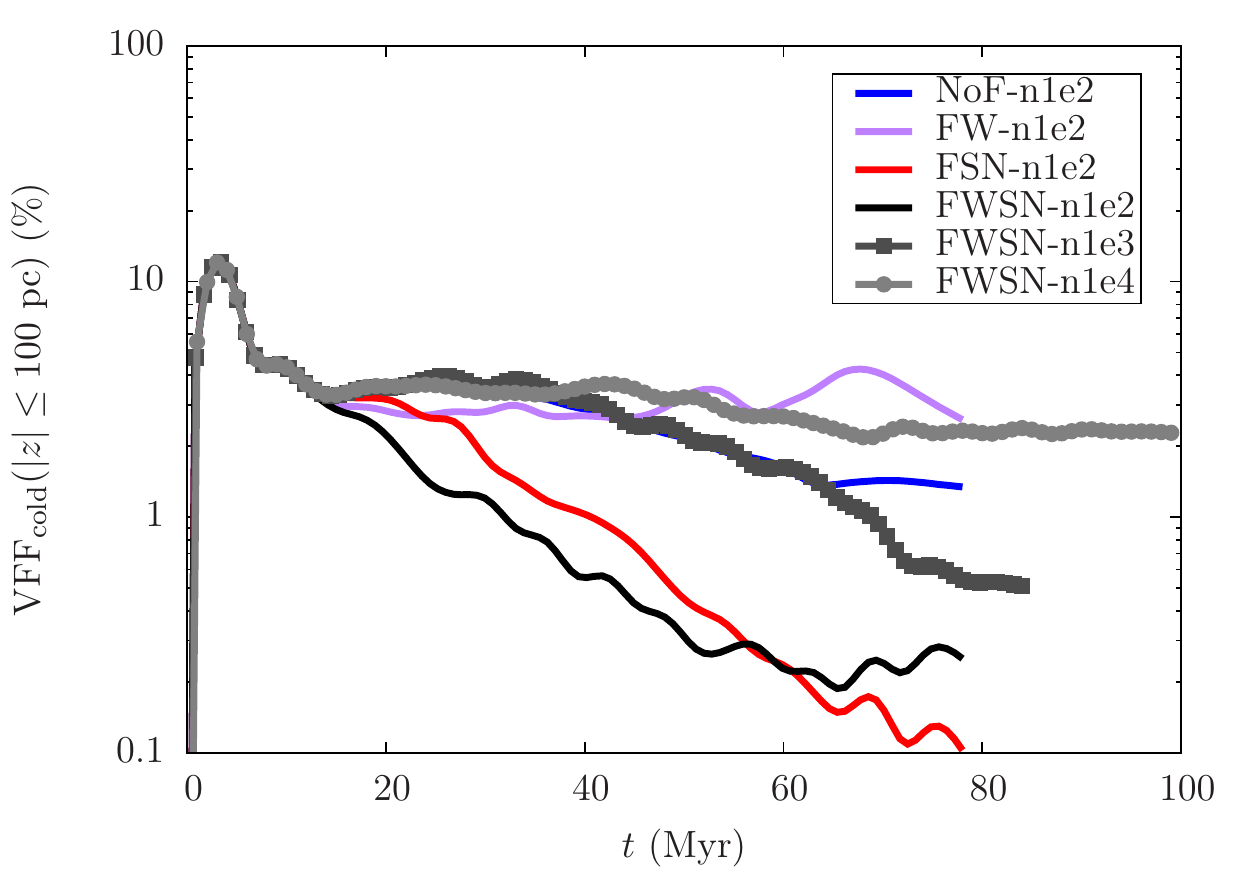}
\caption{Evolution of the volume-filling fractions of warm-hot ($8000 < T \leqslant 3 \times 10^5$ K), warm ($300 < T \leqslant 8000$ K), and cold gas ($30 < T \leqslant 300$ K) within $z=\pm 100 \pc$ from the disc mid-plane.}\label{fig:vff2} 
\end{figure}

In Fig. \ref{fig:vff2} we show the volume-filling fractions of the warm-hot (top), warm (middle), and cold gas (bottom) as a function of time within $z=\pm 100 \pc$ of the disc mid-plane. The hot gas VFF is shown in Fig. \ref{fig:mdot1kpc-tot}. Run {\it FW-n1e2} with stellar wind feedback alone has a low hot gas VFF but therefore a high warm and warm-hot VFF compared to the other simulations with SN feedback, which have a large fraction of the volume filled with hot gas but not much with warm and warm-hot gas. The same applies  for run {\it FWSN-n1e4} which has a SN rate that is too low to produce a large hot gas VFF. Also the cold gas VFFs follow this order: runs without SN feedback and/or with a lower star formation rate are generally colder and have a higher cold gas VFF. 


\bibliographystyle{mn2e}
\bibliography{references}

\label{lastpage}

\end{document}